%% file: main.tex
\documentclass[journal, 10pt]{IEEEtran}
\pdfoutput=1
\normalsize

\ifCLASSINFOpdf                                                                                                                                                                                                                                                                                                                                                                                                                
 \usepackage[pdftex]{graphicx}
\else
 \usepackage[dvips]{graphicx}
\fi

\DeclareGraphicsExtensions{.eps}
\usepackage[cmex10]{amsmath}
\usepackage{bm}
\usepackage{footnote}
\usepackage{epsfig}
\usepackage{latexsym}
\usepackage{enumerate}
\usepackage{multirow}
\usepackage{epstopdf}
\usepackage{longtable}
\usepackage{stfloats}
\usepackage{color} 
\usepackage{amssymb}
\graphicspath{{./Figures/}}
\usepackage{color}
\usepackage{tikz}
\usepackage{bbm}
\usepackage{authblk}
\usepackage{verbatim}
\usepackage{url}
\usepackage[figuresright]{rotating}

\usepackage{mathtools}

\usepackage[tight,footnotesize]{subfigure}
\usepackage{balance}

\usepackage[utf8]{inputenc}
\usepackage[T1]{fontenc}

\usepackage{siunitx}
\DeclareSIUnit\bps{bps}
\DeclareSIUnit\Torr{Torr}
\DeclareSIUnit\torr{Torr}
\DeclareSIUnit\sample{Sa}
\newcommand*{\circled}[1]{\lower.7ex\hbox{\tikz\draw (0pt, 0pt)%
  circle (.5em) node {\makebox[1em][c]{\small #1}};}}

\usepackage{cite}

\usepackage{color, soul}

\begin{document}
% TO COUNT LINES PER PAGE
%\setpagewiselinenumbers
%\pagewiselinenumbers 

\title{Terahertz Wireless Channels: A Holistic Survey on Measurement, Modeling, and Analysis} 
\author{Chong~Han,~\IEEEmembership{Member,~IEEE,} Yiqin~Wang, Yuanbo~Li, Yi~Chen,
Naveed~A.~Abbasi,~\IEEEmembership{Member,~IEEE},
Thomas~K\"urner,~\IEEEmembership{Fellow,~IEEE}
and Andreas~F.~Molisch,~\IEEEmembership{Fellow,~IEEE}

\thanks{
Chong Han acknowledges the support by National Key R\&D Program of China under Project No. 2020YFB1805700.
\par
Thomas K\"urner acknowledges the support by the Federal Ministry of Education and Research (BMBF, Germany) within the 6G Research and Innovation Cluster 6G-RIC under grant 16KISK031.
\par
Chong Han, Yiqin Wang, Yuanbo Li, and Yi Chen are with the Terahertz Wireless Communications (TWC) Laboratory, Shanghai Jiao Tong University, Shanghai, China, 200240. E-mail: \{chong.han, wangyiqin, yuanbo.li, yidoucy\}@sjtu.edu.cn.
\par
Naveed A. Abbasi and Andreas F. Molisch are with Wireless Devices and Systems Group (WiDeS), University of Southern California, Los Angeles, CA 90007 USA. E-mail: \{nabbasi, molisch\}@usc.edu.
\par
Thomas K\"urner is with the Institut f\"ur Nachrichtentechnik, Technische Universit\"at Braunschweig, 38106 Braunschweig, Germany. E-mail:  kuerner@ifn.ing.tu-bs.de.
}
}
\maketitle

\input{v1/abstract}

\input{v1/introduction}

\input{v1/sounding}
\input{v1/modeling}

\input{v1/simulator}

\input{v1/characterization}

\input{v1/openproblems}
\input{v1/conclusion}

\bibliographystyle{IEEEtran}
\bibliography{bibliography}

\begin{IEEEbiography}
[{\includegraphics[width=1in,height=1.25in,clip,keepaspectratio]{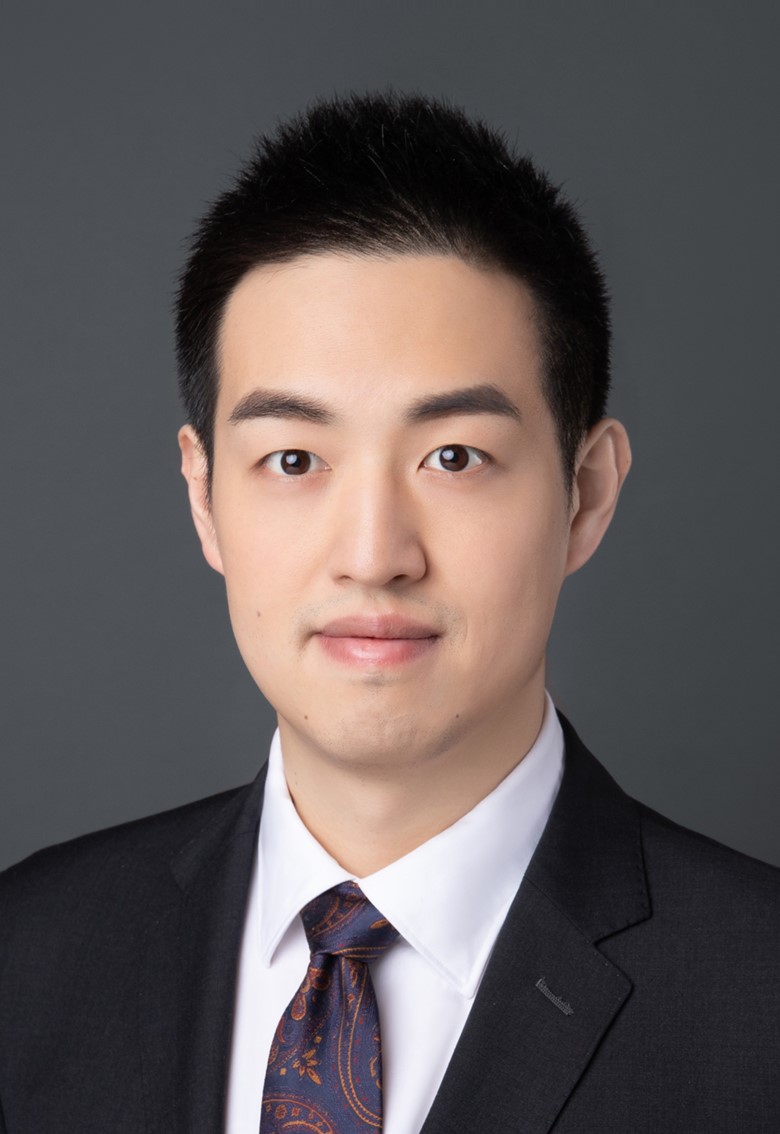}}]{Chong Han} \input{v1/1}
\end{IEEEbiography}
\begin{IEEEbiography}
[{\includegraphics[width=1in,height=1.25in,clip,keepaspectratio]{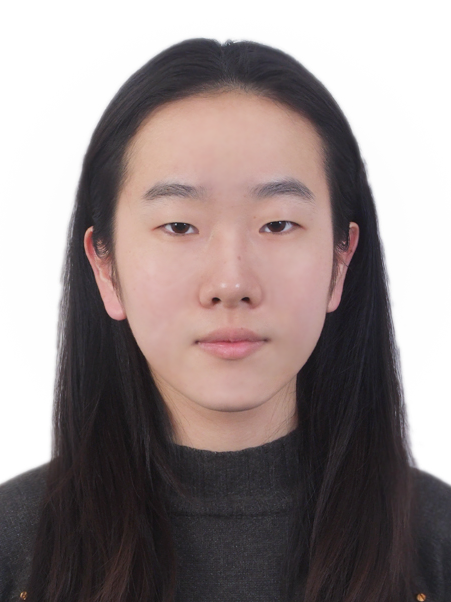}}]{Yiqin Wang} \input{v1/2}
\end{IEEEbiography}
\vspace{-10 mm}
\begin{IEEEbiography}
[{\includegraphics[width=1in,height=1.25in,clip,keepaspectratio]{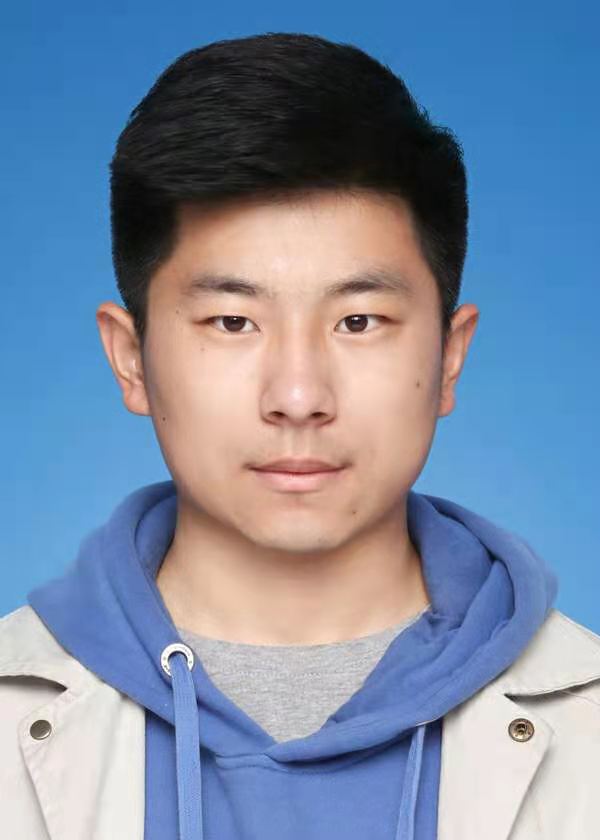}}]{Yuanbo Li} \input{v1/3}
\end{IEEEbiography}
\vspace{-10 mm}
\begin{IEEEbiography}[{\includegraphics[width=1in,height=1.25in,clip,keepaspectratio]{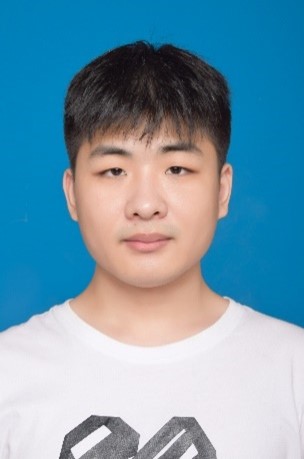}}]{Yi Chen} \input{v1/4}
\end{IEEEbiography}
\vspace{-10 mm}
\begin{IEEEbiography}
[{\includegraphics[width=1in,height=1.25in,clip,keepaspectratio]{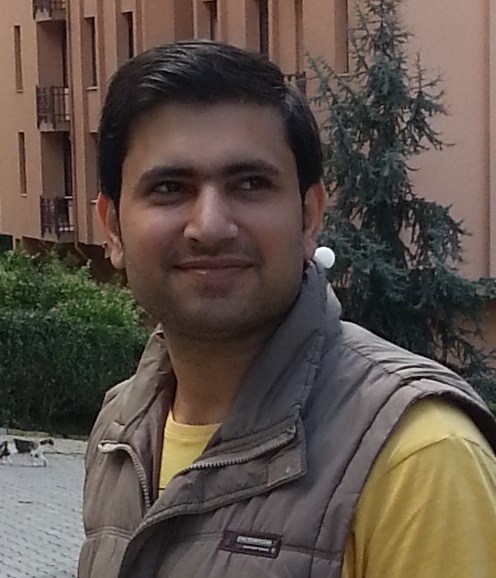}}]{Naveed A. Abbasi} \input{v1/5}
\end{IEEEbiography}
\vspace{-10 mm}
\begin{IEEEbiography}[{\includegraphics[width=1in,height=1.25in,clip,keepaspectratio]{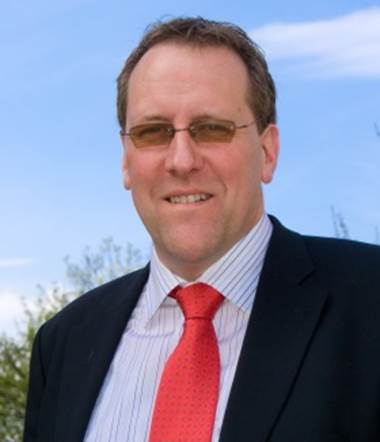}}]{Thomas K{\"u}rner} \input{v1/6}
\end{IEEEbiography}
\begin{IEEEbiography}[{\includegraphics[width=1in,height=1.1in,clip]{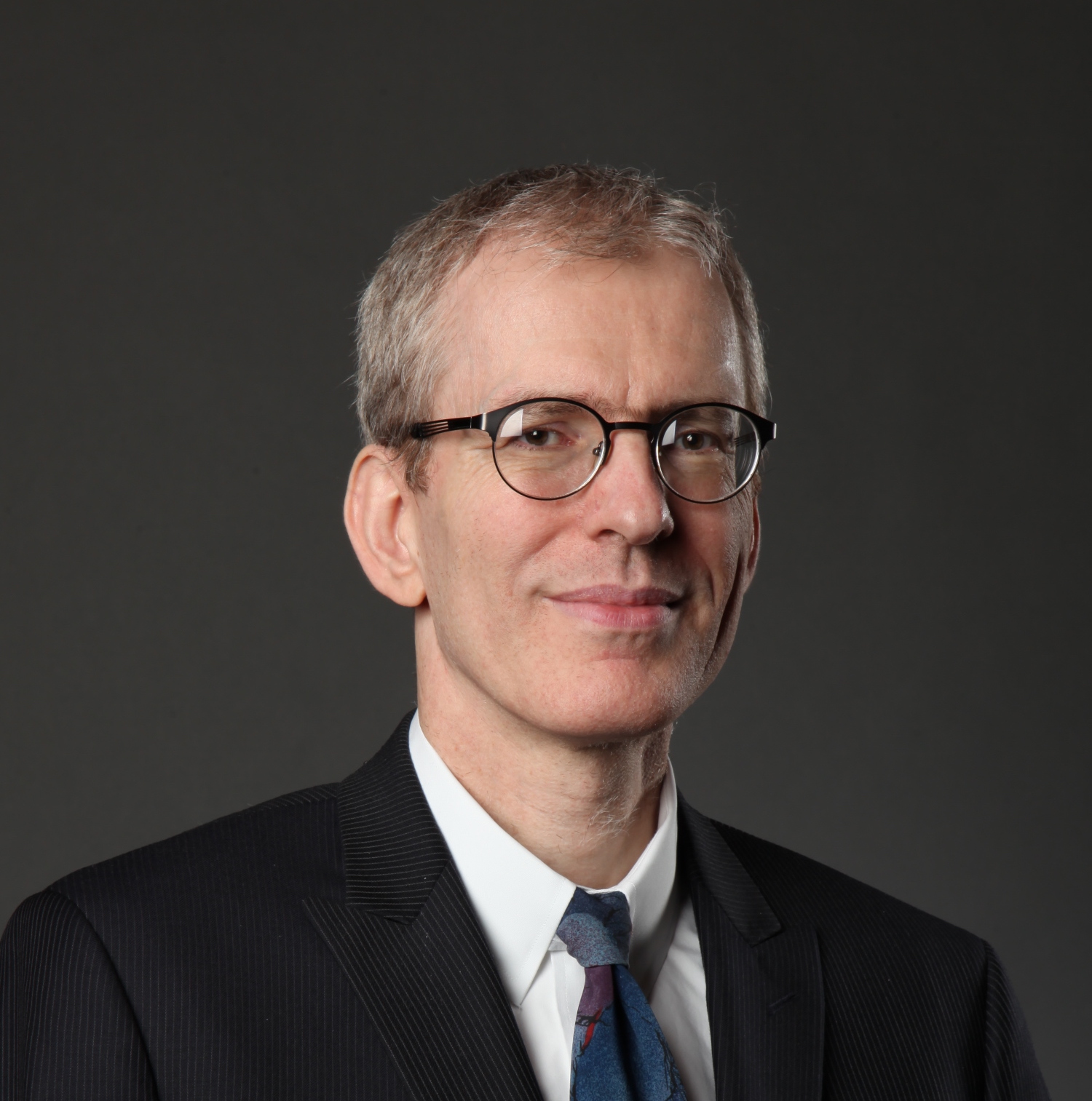}}]{Andreas F. Molisch} \input{v1/7}
\end{IEEEbiography}

\vfill

\end{document}

%% file: v1/abstract.tex
\begin{abstract}
Terahertz (0.1-10~THz) communications are envisioned as a key technology for sixth generation (6G) wireless systems. The study of underlying THz wireless propagation channels provides the foundations for the development of reliable THz communication systems and their applications. This article provides a comprehensive overview of the study of THz wireless channels. First, the three most popular THz channel measurement methodologies, namely, frequency-domain channel measurement based on a vector network analyzer (VNA), time-domain channel measurement based on sliding correlation, and time-domain channel measurement based on THz pulses from time-domain spectroscopy (THz-TDS), are introduced and compared. Current channel measurement systems and measurement campaigns are reviewed. Then, existing channel modeling methodologies are categorized into deterministic, stochastic, and hybrid approaches. State-of-the-art THz channel models are analyzed, and the  channel simulators that are based on them are introduced. Next, an in-depth review of channel characteristics in the THz band is presented. Finally, open problems and future research directions for research studies on THz wireless channels for 6G are elaborated.

%\boldmath
\end{abstract}

%% file: v1/introduction.tex
\section{Introduction}\label{sec:intro}

Revolutionary enhancement of data transmission rates in current and future wireless communication systems are required to meet the challenge of increasing communication traffic due to the massive exchange of information.  
Looking to 2030 and beyond, sixth generation (6G) and beyond systems are expected to achieve a peak data rate of terabits per second (Tbps), along with various further disruptive metrics, including experienced data rates ($R$) for users on the order of 10 to 100~Gbps, end-to-end latency ($L$) shorter than 0.1~ms and $\mu$s-level jitter performance, 100~bps/Hz spectrum efficiency, 99.99999\% data transmission reliability (i.e. transmission at $R$ data rate and $L$ latency with 99.99999\% success probability), enormous connectivity density of more than $10$ million devices per 1~$\rm km^2$, and 100$\times$ improved (compared to 5G) network energy efficiency (by reducing transmit power and circuit power in usage scenarios like smart city/factory/home), positioning and sensing resolution and accuracy\footnote{``Resolution'' is a measure of the distance in amplitude beyond which two objects are able to be resolved/identified/distinguished. ``Accuracy'' defines how close or far of a given set of measurements (observations or readings) are to their true values.} at millimeter level, among others~\cite{tataria20216g,saad2020vision,zhang20196g,akhtar2020shift,series2015IMTV}.

Though millimeter-wave (mmWave) communications (30-100 GHz)\footnote{In strict terminology, mmWave extends from 30-300 GHz. However, in the 5G/6G literature, it has become common to use mmWave for the 24-100 GHz range, and call the 100 GHz-10 THz range ``THz''. We will follow the latter convention in this paper.} have been officially adopted in recent fifth generation (5G) cellular systems~\cite{dahlman20205g}, it is difficult for mmWave systems to support Tbps-level data rates, since the total available bandwidth is only about 20~GHz~\cite{chen2021terahertz}.
Among the available frequency bands, the Terahertz (THz) band nominally occupies the region of 0.1-10~THz in the electromagnetic spectrum, corresponding to wavelengths from 30~$\mu$m to 3~mm, though current interest mostly concentrates on the 0.1-0.5 THz range. Main features include the broad bandwidth ranging from tens of GHz to possibly hundreds of GHz, the narrow beamwidth and associated high directivity, high propagation loss, and the vulnerability to line-of-sight (LoS) blockage, among others. The THz band has been identified as the most promising one to address the spectrum scarcity and capacity limitations of current wireless systems, and realize long-awaited applications~\cite{federici2010review,kleine2011review,akyildiz2014terahertz,huq2019terahertz,rappaport2019wireless,saad2020vision,schmidt2002THz,mahmoud20216G,you2021towards,huang2019survey,akyildiz2021terahertz,nasrallah2018ultra,han2019terahertz}, extending to wireless cognition~\cite{chinchali2021network}, sensing~\cite{schmidt2002THz}, spectroscopy and imaging~\cite{mittleman2018twenty}, and localization/positioning~\cite{kanhere2021position}.

Applications of the THz band in wireless communications are envisioned, which include but not limited to, Tbps WLAN system (Tera-WiFi), Tbps Internet-of-Things (Tera-IoT) in wireless data centers, Tbps integrated access backhaul (Tera-IAB) wireless networks, ultra-broadband THz space communications (Tera- SpaceCom), nano-networks, wireless networks-on-chip communications (WiNoC), among many other emerging applications to appear in the coming future~\cite{akyildiz2021terahertz}.
Though THz communications are envisioned as a key technology to fulfill future demands for 6G and beyond~\cite{petrov2020IEEE, akyildiz20206G, han2019terahertz}, the advancement of practical and commercial products is still ongoing. First, the design of THz transceivers mainly follows one of the following three approaches, namely, electronic-based (e.g. systems utilizing frequency up-conversion), photonics-based (e.g. systems based on optical signal generator and photomixer), and plasmonic-based (e.g. systems based on new materials like graphene). Second, the use of antennas depends on the requirement of the usage scenario. While open-ended waveguides, horn antennas, and parabolic antennas provide different options of directivity, advanced antenna and array techniques like electronically steerable antennas, beamforming, and beam switching are still under requisition for seamless and/or dynamic services. Third, wideband transceivers are developed with high output power, low phase noise and high sensitivity, for end-to-end communications in the THz band, among others~\cite{chen2019survey,akyildiz20206G,polese2020toward}. As a result, thanks to the progress in efficient THz transceivers and antennas in the last ten years~\cite{Koch2007Terahertz,huang2011terahertz,song2011present,crowe2017terahertz,rappaport2019wireless}, THz communication systems are becoming feasible~\cite{song2010terahertz,kallfass2011all,Wang2013b,wang2014340}.

The wireless propagation channel is the medium over which transportation of the signals from the transmitter (Tx) to the receiver (Rx) occurs, and as a consequence, channel properties determine the ultimate performance limits of wireless communications, as well as the performance of specific transmission schemes and transceiver architectures~\cite{molisch2011wireless,zhang2020channel}. Since wireless channels are the foundation for designing a wireless communication system in the new spectrum and new environments, it is imperative to study the THz radio propagation channels for 6G future wireless communications~\cite{tataria20216g,wang20206G,rappaport2021radio}. Studying wireless channel features relies on the physical channel measurements with channel sounders. The characteristics of the wireless channel are then analyzed based on the measurement results for developing channel models. Channel models capture the nature of wave propagation with reasonable complexity and allow  the fair comparison of different algorithms, designs and performance in wireless networks.

Several survey papers have focused on mmWave channels for 5G systems, since these channels have significantly different characteristics from the traditional centimeter-wave (cmWave) cellular bands~\cite{shafi2018microwave}.
For example, \textit{Rappaport et al.} summarized mmWave channel measurements conducted by NYU Wireless in 2015~\cite{rappaport2015wideband}.
\textit{Salous et al.} presented a review of propagation characteristics, channel sounders, and channel models for mmWave~\cite{salous2016millimeter}.
In~\cite{xiao2017millimeter}, a survey of mmWave communications for 5G networks was presented, including channel measurement campaigns and modeling results. \textit{Shafi et al.} present a survey of mmWave propagation channels as well as their effect on deployment and testbed results~\cite{shafi20175g}.
Later, \textit{Wang et al.} summarized the requirements of the 5G channel modeling including mmWave channels, provided a review of channel measurements and models, and discussed future research directions for channel measurements and modeling~\cite{wang2018survey}.
In 2019, \textit{Huang et al.} investigated the developments and future challenges in the study of 5G mmWave channel~\cite{huang20195G}. In particular, the article classified existing channel sounders, summarized recent channel measurement campaigns, and discussed channel modeling approaches, for mmWave communications. The monograph \cite{rappaport2021radio} extensively discusses channel sounders, best practices for measurements, and generic channel models for mmWave and sub-THz channels. 

However, channel characteristics in the THz band might be different from those of the mmWave and cmWave bands. Due to the smaller wavelength, waves in the THz band interact through reflection, scattering and diffraction with smaller environmental structures, such as surface roughness with sub-millimeter dimensions, and suffer from absorption and scattering by molecules and tiny particles in the atmosphere. They therefore exhibit unique behaviors compared with those in lower frequency bands. Though methodologies of channel modeling in the THz band inherit from those in the mmWave band, a specific modeling and parameterization of THz channels is still required in order to accurately characterize THz wave propagation for diverse application scenarios for 6G.
Furthermore, the smaller wavelength enables larger-scale antenna arrays. Though this is a distinct advantage for increasing capacity, modeling the channel with ultra-large-scale antenna arrays becomes more challenging on account of the higher accuracy requirement and non-stationarity over the array~\cite{gao2013massive,zhang2020channel}. Non-stationarity denotes the variation of channel statistics in temporal, spatial, and frequency domains, which results from the environment change, the large bandwidth, and the large antenna size~\cite{matz2005non}.
Besides, although channel measurement approaches for lower frequencies have been well explored, these methodologies cannot be directly applied to THz waves due to the constraint on the hardware of the system components~\cite{rappaport2021radio}.

While there are a number of surveys on 6G networks~\cite{rappaport2019wireless, huang2019survey, zhu2020towards, zhang2020channel, tataria20216g, mahmoud20216G, you2021towards} and THz communications~\cite{federici2010review, kleine2011review, akyildiz2014terahertz, ohara2019perspective, chen2019survey, han2019terahertz, elayan2019terahertz, THzSpringer, akyildiz2021terahertz}, those papers deal with various system innovations, and thus treat THz channels briefly. For example, \textit{Wang et al.} surveyed channel measurements, characteristics, and models for all application scenarios in all frequency bands (mmWave, THz, and optical)~\cite{wang20206G}.
Furthermore, several papers have investigated specific topics of THz communications, such as THz integrated technology~\cite{sengupta2018terahertz}, THz signal processing techniques~\cite{sarieddeen2021overview}, THz beamforming~\cite{headland2018tutorial}, THz localization~\cite{chen2021tutorial}, THz imaging~\cite{mittleman2018twenty}, THz nanocommunication and nanonetworking~\cite{lemic2021survey}, etc. However, none of these focus on the THz propagation channel.
Existing surveys and tutorials from the literature on THz channel studies only focus on certain aspects, without providing a systematic and thorough analysis of THz channel characteristics, measuring and modeling methodologies of THz channels. For instance, \textit{Han et al.} provided an in-depth view of channel modeling in the THz band~\cite{han2018propagation}. \textit{Al-Saman et al.} reviewed the radio propagation characteristics in the THz band and radio propagation measurements in the sub-THz band in indoor industrial environments~\cite{alsaman2021wideband}. \textit{He et al.} offered a tutorial on the design and applications of a ray-tracing simulation platform~\cite{he2019design}. Therefore, a comprehensive and detailed analysis of measuring methods, measurement campaigns, modeling methodologies, channel models, and characteristics for THz channels is still missing.

This article presents a comprehensive overview and analysis  of studies of THz wireless channels. State-of-the-art developments in measurement, modeling, and characterization of wireless channels in the THz band are reviewed and analyzed. The major contributions of this article include:
\begin{itemize}
    \item A review of THz channel measurement systems and measurement campaigns is given. Three measuring methodologies, namely, frequency-domain channel measurement based on vector network analyzer (VNA), time-domain channel measurement based on correlation, and time-domain channel measurement based on THz time-domain spectroscopy (THz-TDS), are introduced and compared.
    \item Channel modeling methodologies, that is, deterministic, stochastic, and hybrid approaches, are introduced and discussed. Corresponding channel models and channel simulators for the THz band are surveyed.
    \item THz channel characteristics and non-stationary properties are thoroughly analyzed, which reveal the unique features of THz wireless channels.
    \item Open problems remaining in research studies on THz wireless channels are highlighted, which presents opportunities for future research efforts on THz wireless channels for 6G and beyond communications, localization, and imaging.
\end{itemize}

\begin{figure*}
    \centering
    \includegraphics[width=0.9\linewidth]{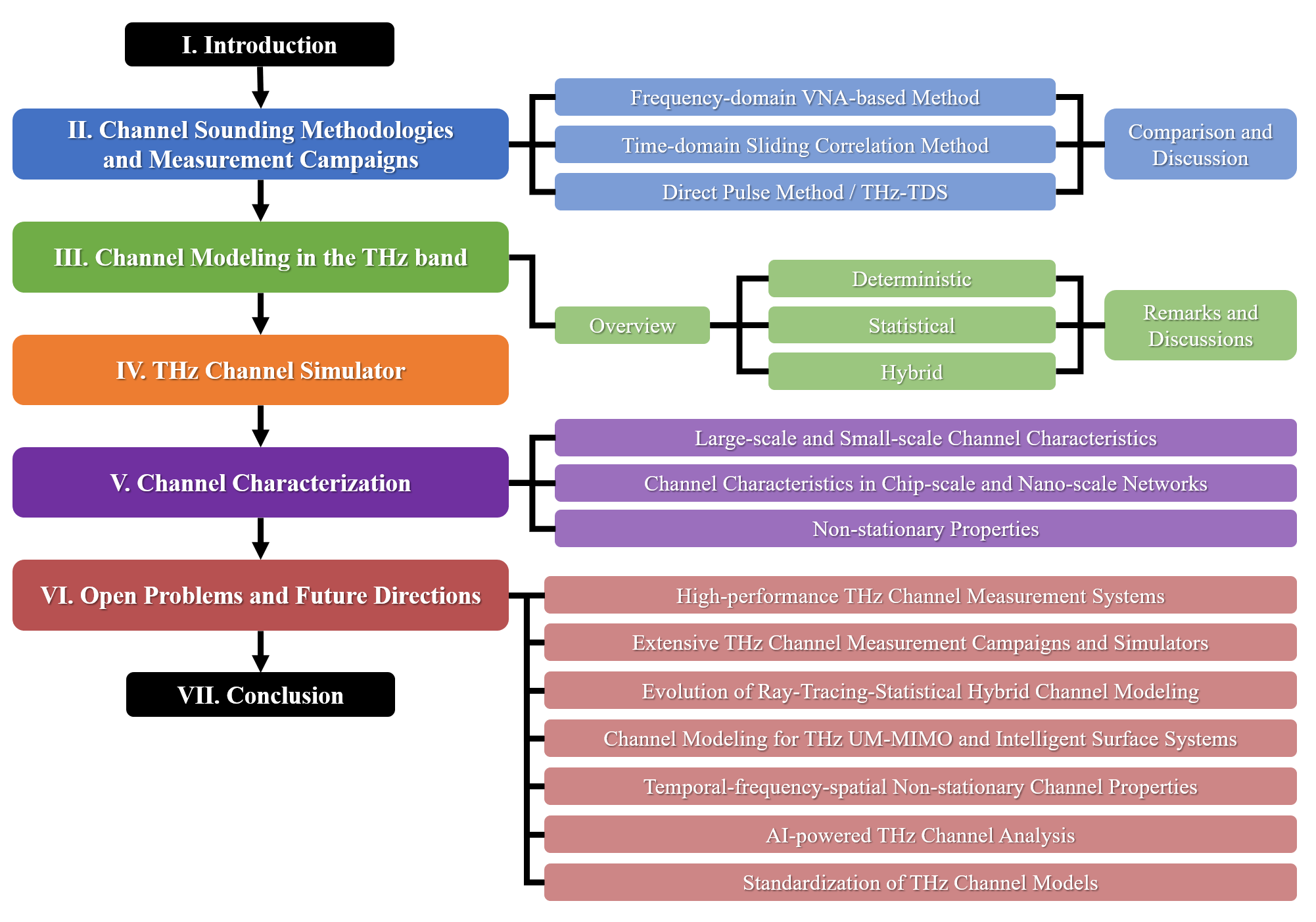}
    \caption{The structure of this article.}
    \label{fig:overview}
\end{figure*}

The remainder of this article is organized as shown in Fig.~\ref{fig:overview}.
In Section~\ref{sec:measurement}, channel measurement methodologies are first introduced and compared, and recent THz channel measurement systems and campaigns are surveyed.
Section~\ref{sec:modeling} reviews three THz channel modeling methodologies, namely deterministic, stochastic, and hybrid models.
Based on this, THz channel simulators are developed, which are introduced in Section~\ref{sec:simulator}.
Furthermore, Section~\ref{sec:characterization} specifies and summarizes THz channel statistics and non-stationary properties.
Finally, open problems and future research directions for studies on THz channels are elaborated in Section~\ref{sec:openproblem}.

%% file: v1/sounding.tex
\section{Channel Sounding Methodologies and Measurement Campaigns} \label{sec:measurement}

Any insights into propagation channels, as well as any channel models, must be based on, or verified by, measurements.
Basic channel measurement (or sounding) is based on the transmission of a known signal, (the input to the channels) and the observation of the received signal (the output of the channel), which thus contains information about the channel response~\cite{salous2013radio,rappaport2021radio}.
Channel sounders can be classified into narrowband and wideband. Narrowband sounders are useful to obtain partial characteristics of the channel, for instance, attenuation and distortion resulting from multipath propagation and fading. The Tx of a narrowband sounder consists of the signal generator transmitting a sinusoidal continuous wave (CW) at the carrier frequency and the transmit antenna, while the Rx is composed of a receive antenna and a spectrum analyzer that can capture the received signal spectrum within an observation window spanning a few hundred of hertz~\cite{alsaman2021survey}.

In contrast, the bandwidth of THz signals usually exceeds the coherence bandwidth of the channel due to the multi-path propagation. Therefore, frequency variations of the channel characteristics emerge and need to be investigated through wideband sounders in the THz band.
To support the frequency range of interest, measurement systems are adapted for the THz band by frequency conversion from electronic or photonic measurement equipment. To make up for the higher isotropic free-space path loss, as well as the frequency conversion loss from harmonic mixers and lower maximum amplifier output power, horn antennas are typically used at both ends of the transceiver.
	
The choice of the sounding technique is a trade-off among various factors, including measured bandwidth, speed, distance, power consumption, cost and complexity of the measurement system.
In this section, we analyze three feasible wideband channel measuring methodologies in the THz band, namely, frequency-domain channel measurement based on VNA, time-domain channel measurement based on sliding correlation, and time-domain channel measurement based on THz-TDS.
State-of-the-art measurement campaigns are summarized, and a comparison among the three measurement approaches is presented as well.

\input{v1/measurement_VNA-based}

\input{v1/measurement_Corr-based}

\input{v1/measurement_Pulse-based}

\input{v1/measurement_discussion}

%% file: v1/measurement_VNA-based.tex
\subsection{Frequency-domain VNA-based Method}

\subsubsection{Method}
VNA is an instrument that measures the response (outgoing wave) at one port of a component or a network to an incoming wave at another port of this component or network. VNA-based frequency-domain channel measurements are based on the characteristic of linear signal systems. To be specific, if a single carrier $x(t)=e^{j2\pi f_ct}$ is transmitted, the frequency response of the linear system at the carrier frequency $f_c$ can be calculated as the ratio of the received signal $y(t)$ to the transmitted signal.
Hence, for any two-port device under test (DUT), the response at one port due to a signal from any other port of the DUT is represented as the S-parameter, which describe the ratios of incoming and outgoing waves at the ports of a multi-port network~\cite{pozar2000microwave}. In particular, in the case that the DUT is the radio channel (propagation channel plus antennas), the frequency-dependent response at port 2 due to the input signal at port 1 (i.e. $S_{21}$) is the frequency response $H(f)$ of the radio channel. From this, the channel impulse response in the time domain, $h(t)$, can be obtained through the inverse Fourier transform.

VNAs are popular for channel measurements because they are well-calibrated precision measurement instrumentation that is available in many laboratories. Their main drawbacks include (i) low output power (which can be compensated by external power amplifiers), (ii) high noise figure of the Rx (which can be mitigated by an external low-noise amplifier), and (iii) long measurement duration. The choice of the intermediate frequency (IF) bandwidth allows to trade off, to a certain extent, the received noise power with the duration for a frequency sweep, yet in all practically relevant cases the coherence time of a THz channel, during which the channel can be considered as essentially time-invariant, is shorter than the measurement duration, thus allowing only measurements in quasi-static scenarios\cite{molisch2011wireless,salous2013radio}.

\begin{figure}
    \centering
    \includegraphics[width=0.95\linewidth]{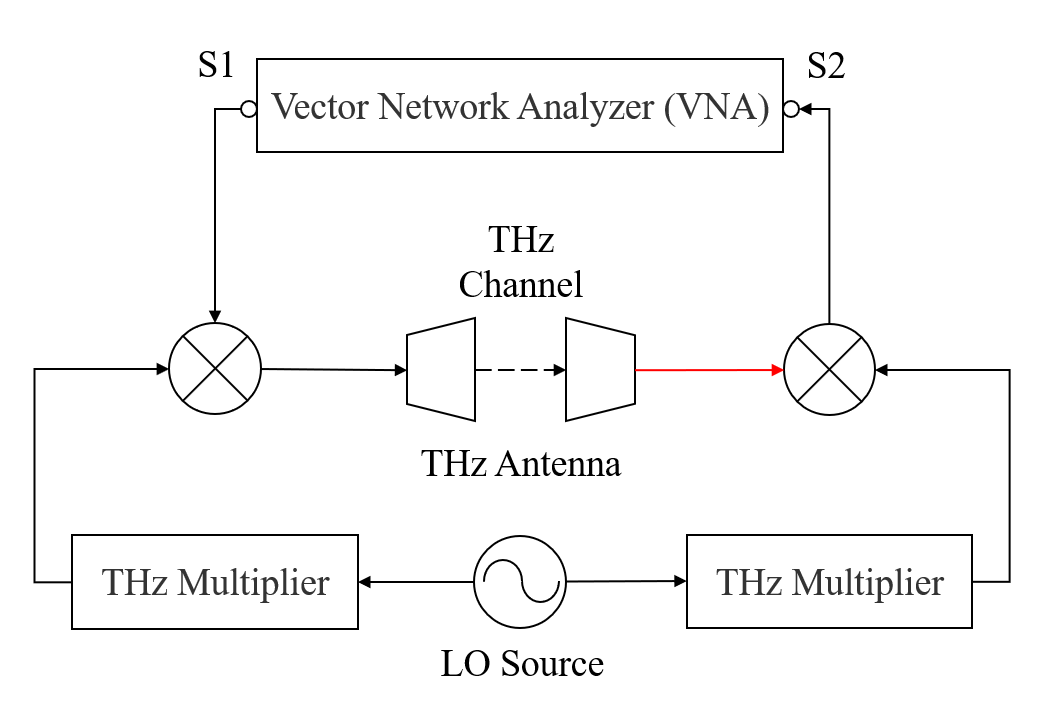}
    \caption{The architecture of a frequency-domain VNA-based measurement system.}
    \label{fig:VNA-based measurement}
\end{figure}

Commercial VNAs are typically limited to $<67$ GHz, so that THz measurements require upconversion modules. Those can either use commercial VNA frequency extenders for the desired frequency range~\cite{VDI,RohdeSchwartz}, or be created from basic components such as a precision oscillator, frequency multiplier, and THz mixer, as shown in Fig.~\ref{fig:VNA-based measurement}. The local oscillator signals can be supported by two additional ports of a four-port VNA (besides the original input/output ports). The devices in the frequency conversion module would bring extra loss and noise across the radio frequency (RF) frequency range, which reduces the dynamic range of the measurement system and thereby constrains the measuring distance.
According to the datasheet of the Virginia Diodes Inc. (VDI) mmWave and THz extenders~\cite{VDI}, THz frequency extenders have empirically worse magnitude and phase stability, smaller port power, and less dynamic range with increasing frequency, compared with the mmWave extenders.
At the Tx, the signal from the S1 port of the VNA is multiplied to the THz frequency through the multiplier and sent to the antenna. At the Rx, the received signal is down-converted to the IF after passing through the harmonic mixer, and is output to the S2 port of the VNA.

A further challenge is the feasible distance between transmit and receive antennas. Since Tx and Rx are housed in the same unit, cable losses from this unit to the antennas limit the range that can be covered, which becomes less than 10 m for THz and shrinks with increasing frequency. Radio-over-fiber connections are a possible remedy, which adopts the method of optical fiber expansion in the measurement system. In the RF-over-fiber (RFoF)-extended VNA-based system, either Tx or Rx is connected to the VNA by an optical fiber, instead of traditional coaxial cables, to transmit local oscillator signals and possibly IF signals. The much lower losses of optical fibers compared to co-axial cables greatly increases the feasible separation between Tx and Rx, and thus overcomes the disadvantage of the VNA-based method in this respect; this principle has been applied in the mmWave band~\cite{Nguyen2016Dual,Nguyen2018Comparing,mbugua2020phase,rappaport2021radio} as well as at THz~\cite{abbasi2020double}.

While VNA-based channel sounders apply  single-tone excitation and frequency sweeping over the desired spectrum, an alternative is channel sounding with multi-tone signal excitation, which can significantly reduce the measurement time and possibly capture the evolution of time-varying channels. However, subject to the linearity constraint of devices in the measurement system (e.g. the power amplifier), the multi-tone signal generator should carefully design phases of individual tones, so as to achieve a low peak to average power ratio (PAPR). Multi-tone-based channel sounding was suggested already 20 years ago~\cite{thoma2000identification} and has been used extensively since then for measurements in lower frequency bands. More recently it has been applied to mm-Wave band measurements~\cite{konishi2011channel,bas201728}. The same method could be adapted for the THz band, though this to our knowledge has not yet been done yet. 

To overcome fading margins in the link budget and increase the measurement distance, at least one directional antenna is required to be used at the transceiver. Hence, to capture multipath components in space, directional-scan-sounding (DSS) measurements need to be conducted. One common implementation is to use a rotating horn antenna, namely, to mount Tx or Rx with the horn antenna on a mechanical rotation unit, which can be rotated in both azimuth and elevation planes by step motors. An alternative implementation is to use the phased antenna array, whose principle is the phase-dependent superposition of signals. Hence, it can change the direction and beam shape of the radiated signals without the physical movement of the antenna. Compared with the rotating horn antenna, the phased antenna array can achieve fast switching and thus reduces the measurement time for DSS measurements~\cite{bas2019real}. However, to the best of our knowledge, apart from prototypes such as~\cite{Merkle2017} suitable phased arrays are not commercially available for the THz frequency range at the moment.

Another challenge of the rotating-horn approach is that, as the Tx/Rx distance increases, and the horn antenna gains increase, the alignment of Tx and Rx becomes harder, especially for double-directional measurements. Therefore, the measurement system requires tighter specifications for the positioning system, which is an area of ongoing study.

\subsubsection{Measurement Campaigns} \label{VNAMeasurementCampaigns}
VNA-based THz channel measurement systems and campaigns are summarized in Table~\ref{tab:channel_measurement_campaigns_VNA}, most of which were implemented focusing on frequencies from 140~GHz to 750~GHz, using directional horn antennas with antenna gain ranging from 15 to 26~dBi. The Tx/Rx distance typically varied from 0.1 to 14~m, and was extended to at most 100~m~\cite{abbasi2020double} with RFoF extension. The following describes channel measurement setups and measurements in sequence from smaller to larger covered distances. 

\begin{sidewaystable*}[ph!]
    \begin{center}
    \caption{VNA-based THz channel measurement systems and campaigns.}
    \includegraphics[width = \linewidth]{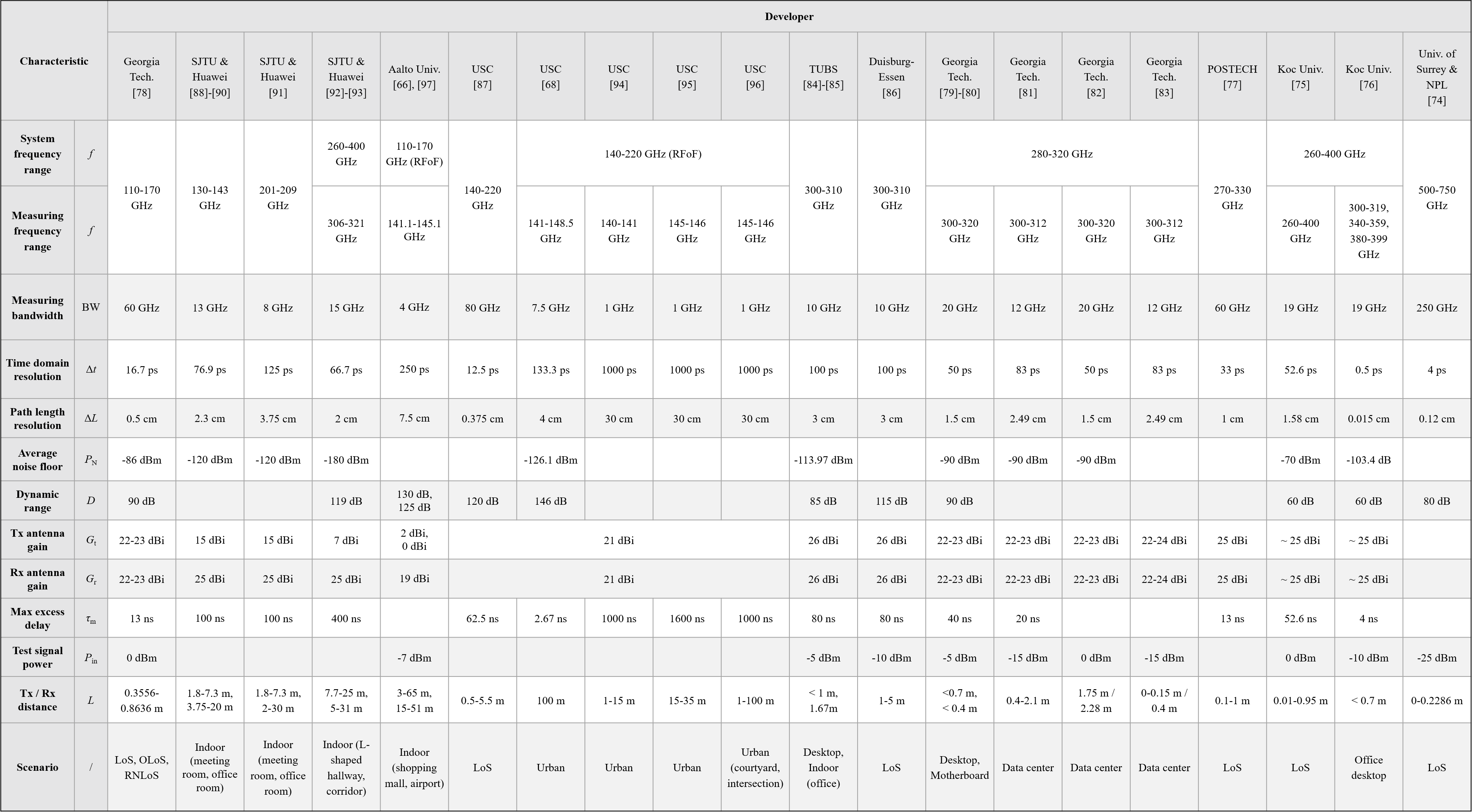}
    \label{tab:channel_measurement_campaigns_VNA}
    \end{center}
\end{sidewaystable*}

% 1. Directional
% 1.1 LoS scenarios (samller than 1~m)
% Univ. of Surrey & NPL 500-750 GHz LoS, directional
First, we mention measurements for distances $<1$ m focusing on line-of-sight (LoS) scenarios. The University of Surrey cooperated with National Physical Laboratory (NPL, Teddington, UK) in establishing a 500-750~GHz channel measurement system using Keysight PNA-X vector network analyzer, configured with VDI extender heads, and measured LoS scenarios within 0.23~m distance range in the 500-750~GHz frequency band~\cite{Serghiou2020Ultra}.
% Koc Univ. 260-400 GHz LoS, directional
Besides, LoS scenarios were measured by Koc University, varying Tx/Rx distance from 0.01~m to 0.95~m at frequencies from 260~GHz to 400~GHz~\cite{khalid2016wideband}, using a VNA combined with a sub-harmonic mixer~\cite{khalid2016wideband,khalid2019statistical}. They also investigated the effect of linear and angular displacement between the Tx and the Rx, and performed non-line-of-sight (NLoS) measurements with a reflective surface in~\cite{khalid2019statistical}.
% Pohang Univ. 270-330 GHz LoS, directional
Pohang University of Science and Technology (POSTECH, Korea) developed a 270-300~GHz VNA-based channel measurement system, and measured short-distance (0.1-1~m) LoS channels in this frequency band~\cite{Song2020LOS}.
% Georgia Tech 110-170 GHz LoS, OLoS, RNLoS, directional
Georgia Institute of Technology (Georgia Tech or GIT) developed a 110-170~GHz channel measurement system with the Agilent E8361C VNA and the N5260A mmWave controller and OML V06VNA2 mmWave test head modules, and conducted measurements in LoS, OLoS, and reflected NLoS environments with Tx/Rx distances of 0.3556-0.8636~m~\cite{kim2015d}.

% 1.2 Short-range & Indoor scenarios
%THz channel measurements were performed in short-range (< 1~m) and indoor scenarios.
% Georgia Tech 300 GHz indoor/motherboard/data center, directional
Various special environments were also investigated. For instance, GIT also developed a 280-320~GHz channel measurement system with a N5224A PNA vector network analyzer and VDI transceivers (Tx210/Rx148) and measured the short-range (up to 0.7~m) desktop~\cite{Kim2015} and computer motherboard~\cite{Kim2016Charaterization} scenarios at 300-312~GHz, as well as the data center~\cite{Cheng2020Characterization} scenario (0.4-2.1~m) at 300-320~GHz.
With the measurement system, \textit{Cheng et al.} measured the THz propagation in a real data center environment~\cite{Cheng2020THz} with the Tx/Rx distance of 1.75~m and 2.28~m at 300-312~GHz. The team also measured the multiple-input-multiple-output (MIMO) fading and Doppler effect in this scenario using a virtual $4\times4$ antenna array in~\cite{Cheng2020Terahertz}.
% TUBS 300 GHz LoS, directional (1/2)
Besides, Technische Universit\"at Braunschweig (TU Braunschweig, TUBS) developed a 300~GHz wave characteristic measurement system, including a 300~GHz transmission system and a VNA of Rohde and Schwartz~\cite{priebe2011channel}. Based on the new system, a short-range (up to 1~m) desktop scenario is measured by \textit{Priebe et al.} at 300~GHz~\cite{priebe2011channel}. Diffraction measurements are reported in~\cite{Jacob2012Diffraction} by \textit{Jacob et al.}.
% Duisburg-Essen 300GHz, LoS, directional
The University of Duisburg-Essen measured the LoS channel (1-5~m) across the 300-310~GHz spectrum with the R\&S ZVA67 VNA-based channel measurement system~\cite{zantah2019channel}.
% USC 140-220GHz indoor, directional
Moreover, the University of Southern California (USC) improved their existing 140~GHz VNA-based channel measurement system to support measurement campaigns at the frequency over 140~GHz up to 220~GHz, and \textit{Abbasi et al.} measured indoor LoS channels in an office environment with the measurement distance from 0.5~m to 5.5~m~\cite{abbasi2020channel}.

\begin{figure}
	    \centering
	    \includegraphics[width=0.9\linewidth,angle=270]{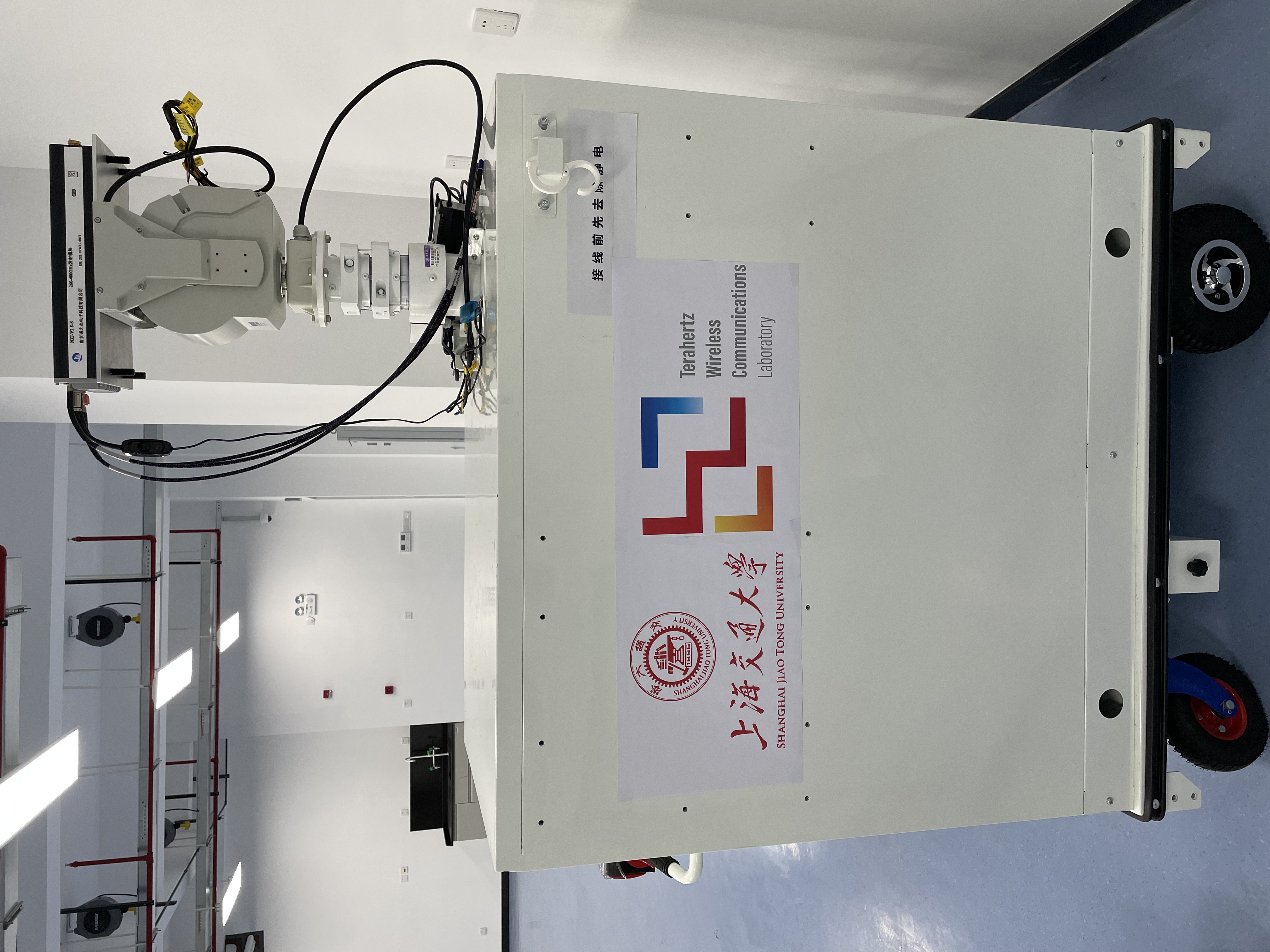}
	    \caption{The 260-400~GHz VNA-based channel measurement system.}
	    \label{fig:photo_measurement_system}
\end{figure}

% 2. Non-directional (rotating unit)
The aforementioned campaigns mainly focus on large-scale channel characterization, namely path loss and shadowing effects, when keeping the horn antennas in a fixed orientation, which is adequate for capturing the LoS path.
However, for many applications, and in particular as the Tx/Rx distance increases, multi-path components (MPCs) from various directions become non-negligible and must be characterized. Therefore, as introduced previously, rotation of Tx and/or Rx units is typically applied to resolve MPCs in angle.
% 2.1 Indoor scenarios
% TUBS 300 GHz indoor (non-directional)(2/2)
For instance, in~\cite{priebe2011channel}, a small indoor office ($2.43\times4.15$~m) scenario was scanned at multiple pairs of angles of arrival (AoA) and angles of departure (AoD) in the horizontal plane, with the Tx/Rx distance of about 1.67~m at 300~GHz.
% TWC 140 GHz meeting room & office room
Furthermore, Shanghai Jiao Tong University (SJTU) in collaboration with Huawei established a 140~GHz VNA-based channel measurement system~\cite{chen2021channel,Yu2020Wideband,chen2021140}. They carried out the directionally resolved channel measurements in a typical indoor meeting room with the Tx/Rx distance of 1.8-7.3~m~\cite{chen2021channel,Yu2020Wideband} and an office room with the Tx/Rx distance of 3.75-20~m~\cite{chen2021140}. The team analyzed the temporal and angular distribution of MPCs, and studied channel parameters and their correlation in detail.
% TWC 220 meeting room & GHz office room
They later extended the measurement frequency to 220~GHz~\cite{he2021channel} and conducted a wideband channel measurement campaign at 220~GHz in the same meeting room with the Tx/Rx distance of 1.8-7.3~m and the office room with the Tx/Rx distance of 2-30~m. The Rx was mounted on a rotation unit driven by step motors. In particular, the measurement campaign consisted of three cases, namely, (i) LoS office area, (ii) LoS hallway and (iii) NLoS. Based on the measurement results, path loss properties at 220~GHz were studied and compared to those in 140~GHz indoor scenarios.
% TWC 300 GHz indoor hallway scenarios
Besides, authors in~\cite{wang2022thz} and~\cite{li2022channel} established a 260-400~GHz VNA-based channel measurement system as shown in Fig.~\ref{fig:photo_measurement_system}, and carried out indoor channel measurements at 306-321~GHz in an L-shaped hallway (7.7-25~m) and a long corridor (5-31~m) in the SJTU campus.

% 2.2 long-range scenario with RFoF
The method of optical fiber extension allowed to increase the transmission distance and enable THz measurement campaigns that were no longer limited to short-range, indoor scenarios.
% USC 140 GHz (RFoF), non-directional
USC adopted this method for their 140~GHz RFoF-extended VNA-based remote channel measurement system, which greatly increased the original transmission distance of 8-10~m to 100~m~\cite{abbasi2020double}. Using this system, \textit{Abbasi et al.} carried out a series of medium-distance (1-35~m) and long-distance (up to 100~m), double-directional outdoor THz channel measurements in a larger variety of urban scenarios at THz frequencies~\cite{abbasi2020double,abbasi2021double,abbasi2021ultra,abbasi2021thz}.
% Aalto Univ. 140GHz (RFoF), non-directional
In a similar way, Aalto University used a 200-m optical fiber between the Tx and the VNA to extend the straight-line measuring distance of a 140~GHz channel measurement system with Rx on a rotor~\cite{Nguyen2018Comparing, Nguyen2016Dual}. With the distance-extended measurement system, the team measured and analyzed the 140~GHz indoor channels in a shopping mall with measurement distance from 3 to 65~m and an airport check-in hall with measurement distance from 15 to 51~m)~\cite{Nguyen2018Comparing,Nguyen2021Large}.

%% file: v1/measurement_Corr-based.tex
\subsection{Time-domain Sliding Correlation Method}

\subsubsection{Method}
The correlation-based sounding method employs signals whose autocorrelation is (similar to) a Dirac delta function. It then exploits the fact that - in a channel that is time-invariant for the duration of the signal - the concatenation of a transmit signal $x(t)$, channel, and matched Rx filter creates the same receive signal as a delta function passing through a channel~\cite{molisch2011wireless}, i.e., the channel impulse response. Since the matched filtering can also be described as correlation with the complex-conjugated transmit signal, this method is known as correlation-based sounding.
It has the advantage that signals with low PAPR can be used as input signals, enabling the use of higher transmit {\em energy} for the sounding even as the transmit {\em power}, which is limited by the capabilities of the transmit amplifier, stays low.
Popularly used transmit signals are pseudo-noise (PN) sequences, in particular maximum-length (M) sequences~\cite{pirkl2008optimal}. 
Compared to VNA-based measurements, correlation-based sounders offer the possibility of real-time measurements; however, PN-sequence-based sounders tend to have a power spectral density of the sounding signal, and thus an signal-to-noise ratio (SNR), that is not uniform over the band of interest. A major advantage of correlation sounders is the ease with which the sounding signal can be generated, as it is a binary sequence.

Since the correlation-based method requires an expensive wideband digitizer to support a very high sampling rate (equal to the Nyquist rate), most current correlation-based THz sounders use sliding correlation, which trades off measurement duration with sampling rate~\cite{alsaman2021survey}. In this approach, first introduced for cellular measurements by Cox~\cite{cox1973spatial}, the sequence generated at the Rx side for the correlation process has a chip rate $f^\prime_{chip}$ smaller than the chip rate of the transmit signal $f_{chip}$. The generated ``slower'' sequence is correlated with the received signal and then passes through a low-pass filter. The result is then proportional to $h\left(\frac{t}{\gamma}\right)$, where $\gamma=f_{chip}/(f_{chip}-f^\prime_{chip})$ is denoted as the slide factor, also known as slip rate~\cite{ferreira2015real}. In other words, the sliding operation reduces the required sampling rate by this factor $\gamma$, increases (compared to direct sampling) the measurement duration by $\gamma$, and also increases the SNR by the same factor. 

\begin{figure}
\centering
\subfigure[Correlation-based channel measurement system.]{
\includegraphics[width=0.95\linewidth]{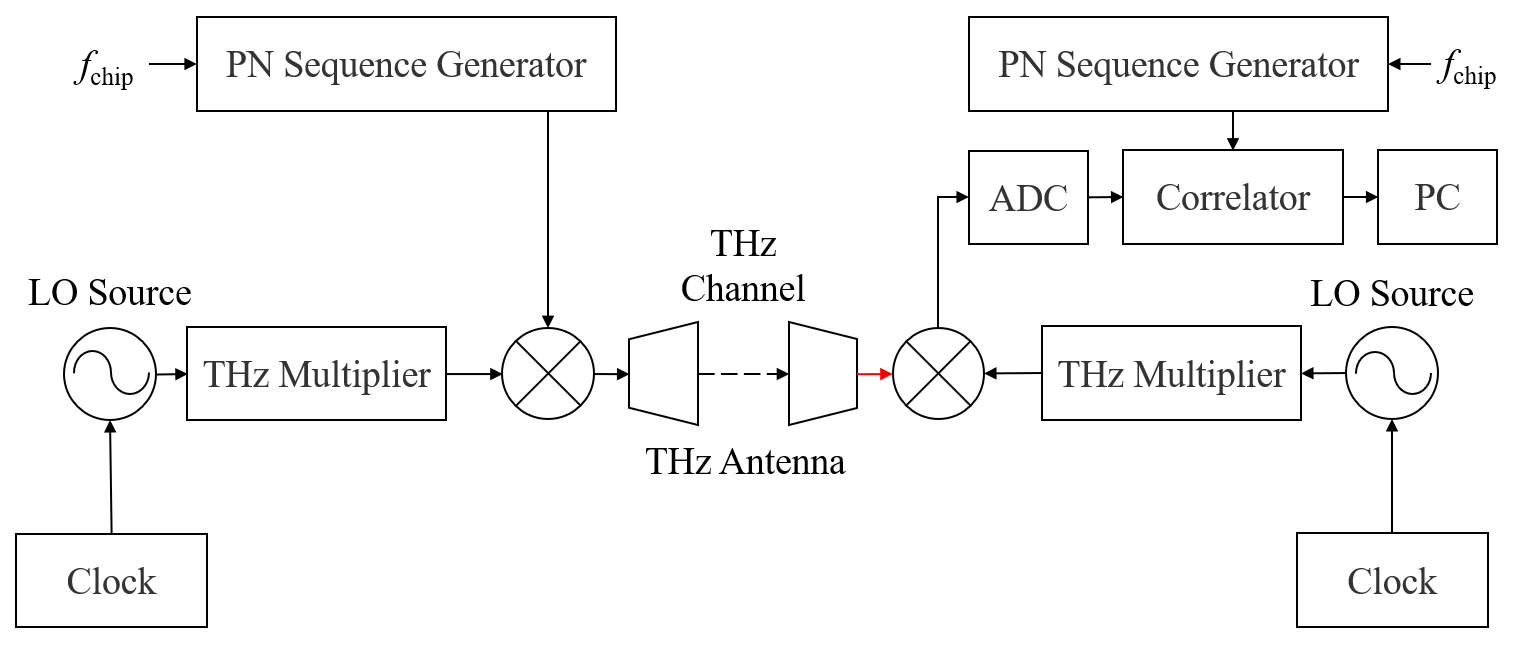}
}
\quad
\subfigure[Sliding-correlation-based channel measurement system.]{
\includegraphics[width=0.95\linewidth]{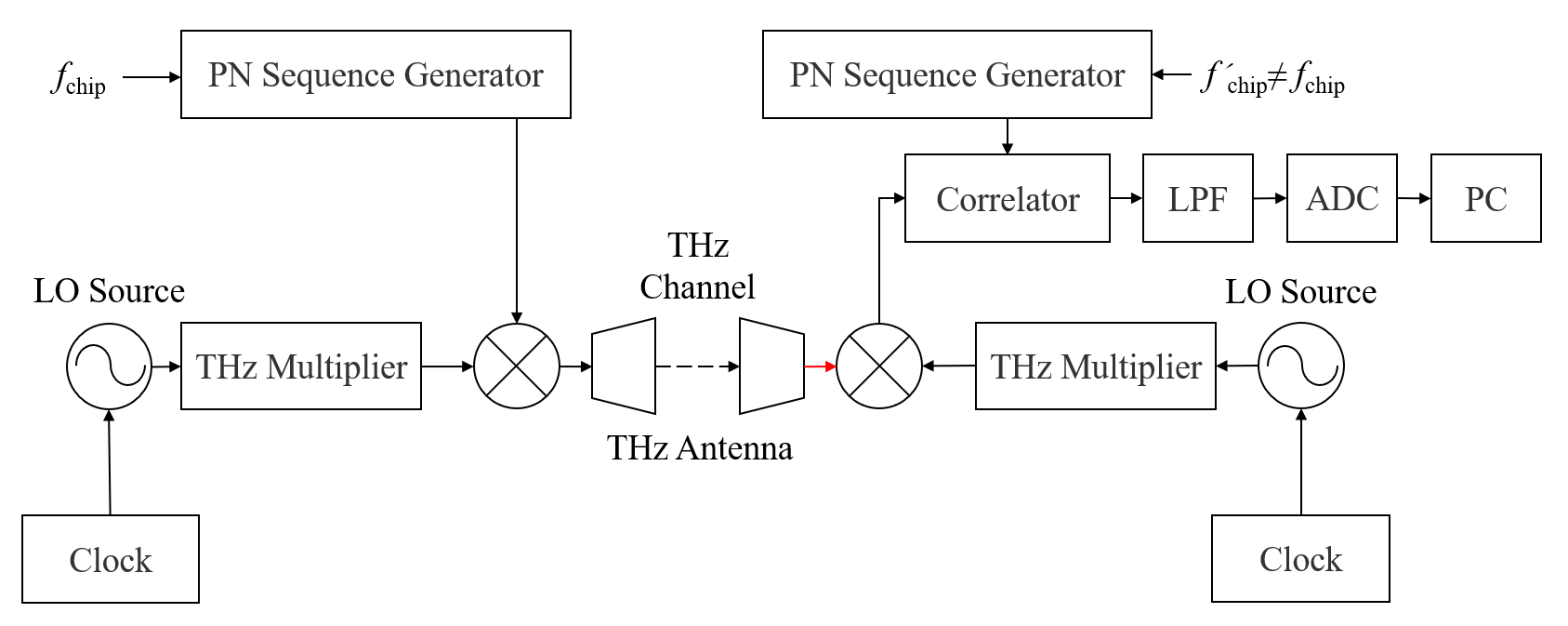}
}
\caption{The architecture of a time-domain correlation measurement system.}
\label{fig:Corr-based measurement}
\end{figure}

The time-domain correlation-based measurement module typically contains THz multipliers, THz mixers, and high-speed analog-to-digital (ADC) and digital-to-analog (DAC) converters, as shown in Fig.~\ref{fig:Corr-based measurement}(a).
At the Tx, the high-speed DAC transmits the wideband signal sequence pre-stored in the FPGA (Field-Programmable Gate Array). The signal is then up-converted to the THz band. Conversely, at the Rx, the received signal is down-converted to the baseband, collected by the high-speed ADC and output to the FPGA for analysis. By contrast, as shown in Fig.~\ref{fig:Corr-based measurement}(b), a sliding-correlation-based measurement module requires a slower copy of the transmitted signal at the Rx side from the PN sequence generator and a low-pass filter at the Rx side.

\subsubsection{Measurement Campaigns}
Channel measurement campaigns based on the sliding correlation method have been reported in the frequency band from 140~GHz to 300~GHz.

A measurement system with dual-mode switching between sliding correlation mode and real-time spread spectrum at 140 GHz was developed by NYU Wireless~\cite{maccartney2017flexible}.
Using the sliding correlation mode, the team measured and analyzed reflection and scattering characteristics of the wave propagation at 140~GHz\cite{Ju2019Scattering}. Furthermore, \textit{Xing et al.} measured rooftop surrogate satellite and backhaul (rooftop base station height $38.2$~m above the ground) to emulate ground-to-satellite and ground-to-unmanned aerial vehicle communications at 140~GHz~\cite{xing2021propagation}. Two clear LoS links and seven blocked LoS links are investigated, with Tx elevation angles ranging from 0 to $32^\circ$.
Combining this system with a rotating horn setup, NYU Wireless measured indoor channels at 140~GHz in scenarios including offices, conference rooms, classrooms, long hallways, open-plan cubicles, elevators, and the factory building~\cite{xing2018propagation,Xing2019Indoor,ju2021millimeter, ju2022sub}. The group also conducted directionally resolved outdoor wideband measurement campaigns in an urban microcell environment (base station height $4$~m)~\cite{xing2021propagation,kanhere2021outdoor,ju2021140,ju2021sub} at 140~GHz.

Technische Universit{\"a}t Ilmenau developed a THz measurement system by upgrading their mmWave M-sequence channel sounder in~\cite{muller2014ultrawideband} with up- and down-converters to reach the center frequency of about 190~GHz, \textit{Dupleich et al.} conducted directionally resolved measurements in a conference room at the campus of TU Ilmenau~\cite{dupleich2020characterization}. The measurements were verified by ray-tracing (RT) simulations results.

Furthermore, \textit{Undi et al.} conducted a channel measurement at 300~GHz in an urban microcell scenario, with a time-domain correlation-based channel sounder~\cite{undi2021angle}. The measurement was implemented on a courtyard (40~m $\times$ 18~m) in the Fraunhofer Institute for Telecommunications, Heinrich Hertz Institute, and was angle-resolved with respect to the azimuth direction~\cite{undi2021angle}.

Moreover, Technische Universit\"at Braunschweig has  a channel measurement platform at 300~GHz available, using the sliding correlation method based on M sequences with the order of 12~\cite{rey2017channel}. The clock frequency is 9.22~GHz, and the bandwidth is about 8~GHz, so that most of the sequence power is concentrated within 8~GHz in the frequency domain~\cite{rey2017channel}.
With this measurement system, first, Beijing Jiao Tong University (BJTU) and Technische Universit\"at Braunschweig carried out channel measurements and modeling for train-to-train (T2T)~\cite{guan2019channel_EuCAP}, infrastructure-to-infrastructure (I2I)~\cite{guan2019channel_EuCAP}, train-to-infrastructure (T2I) inside-station~\cite{guan2019measurement}, and intra-wagon~\cite{guan2020channel,guan2019channel_TVT} channels from 60~GHz to 0.3~THz, studying the propagation characteristics of mmWave and THz waves in railway scenarios~\cite{guan2021channel}. Among these, investigation of the intra-wagon channel is directionally resolved in the horizontal plane~\cite{guan2020channel,guan2019channel_TVT}. Besides, these measurement campaigns were carried out in static scenarios and further dynamic measurement of the scenario is still an open issue~\cite{guan2017millimeter}.
Second, Technische Universit\"at Braunschweig also cooperated with the University of Tampere, Finland~\cite{Petrov2020Measurements,Eckhardt2021channel}. \textit{Petrov et al.} measured the signal propagation in vehicular environments at 300~GHz~\cite{Petrov2020Measurements}.
\textit{Eckhardt et al.} conducted a comprehensive measurement study of signal propagation at 300~GHz in single-lane and multi-lane vehicular environments~\cite{Eckhardt2021channel}. The study revealed that vehicular communications at 300~GHz are characterized by complex multi-path propagation, among which several important components are identified.
These contributed results provided theoretical foundations for the design of future vehicle-to-vehicle communications.
Third, \textit{Eckhardt et al.} conducted a measurement campaign in a real data center at 300~GHz, which divided the scenario into inter- and intra-rack parts. The trial evaluated the path attenuation, the power delay profile (PDP) and the power angular spectrum (PAS), and demonstrated the feasibility of wireless communication at 300~GHz in a data center~\cite{eckhardt2019measurements}.
Finally, with the same channel measurement platform, \textit{Eckhardt et al.} also measured the indoor-to-outdoor scenario in a Boeing 737 aircraft to analyze the attenuation caused by the window and the fuselage~\cite{Eckhardt2020Indoor}.

%% file: v1/measurement_Pulse-based.tex
\subsection{Direct Pulse Method / THz-TDS}

\subsubsection{Method}

The conceptually simplest way of measuring impulse responses is the transmission of a train of very narrow pulses, a principle known as THz-TDS. Each pulse is very narrow (short duration), and the period of the train is greater than the maximum excess delay of the channel. The amplitude of a sampling instance can be considered as the amplitude of the channel impulse response at the time of the exciting pulse, at a delay that is equal to the difference between excitation pulse transmission and sampling time of the observation. Therefore, the channel impulse response can be directly derived after the receiver samples the received signal at a high speed in the time domain. Depending on the speed of the sampler, either direct sampling or swept-delay sampling is possible. 

\begin{figure}
    \centering
    \includegraphics[width=0.9\linewidth]{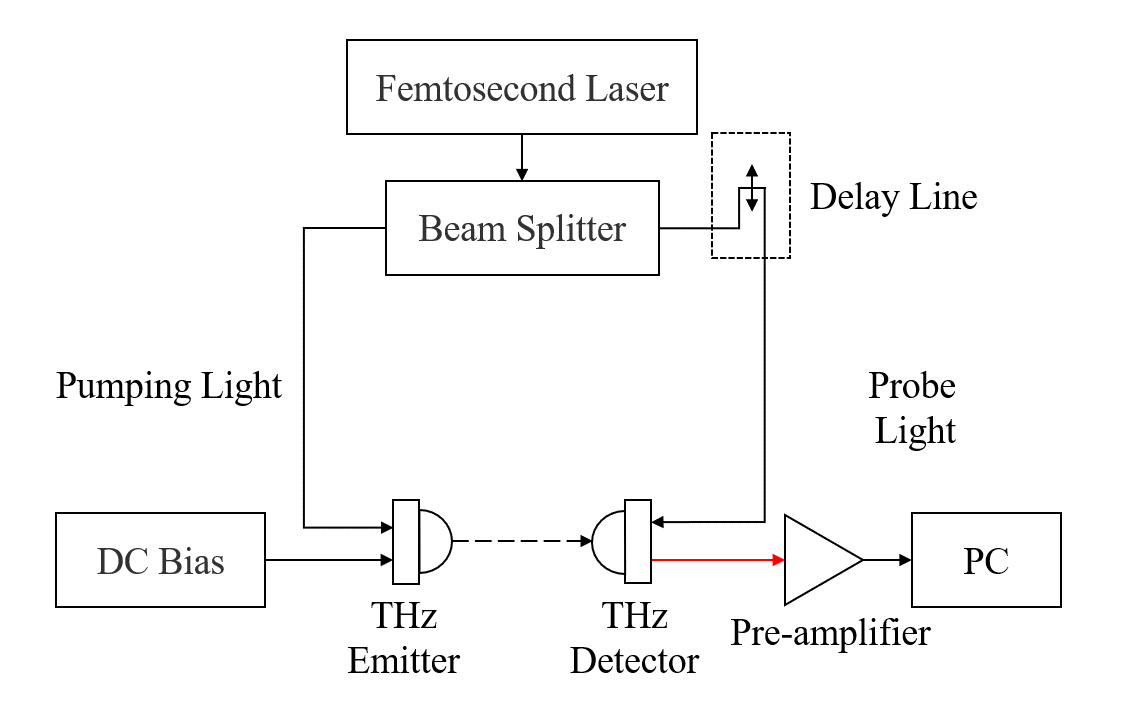}
    \caption{The architecture of a pulse-based measurement system.}
    \label{fig:Pulse-based measurement}
\end{figure}
As shown in Fig.~\ref{fig:Pulse-based measurement}, a typical THz-TDS contains a femtosecond laser, a beam splitter, a THz emitter, a mechanical delay line and a THz detector~\cite{neu2018tutorial}.
The output of a femtosecond laser pulse is split by the beam splitter into two beams. One beam that is known as pumping light is directed to the THz emitter and used to generate THz radiation, where a DC bias can separate the optical carrier and the THz pulse. The other beam known as probe light is routed through a delay line to the THz detector and used to detect the THz radiation. The detection is activated when the beam arrives simultaneously with the THz signal, for which the temporal delay should be found by sweeping the path length of the delay line~\cite{neu2018tutorial}. The detected signal is then sent to a pre-amplifier and a computer.

The THz-TDS features huge and scalable bandwidth in the THz band~\cite{Harde1997terahertz}. However, it suffers from the large size of the setup and low output power. To compensate for the low power, lenses are usually required at both ends of the transceiver to improve the intensity of the pulse signal. Still, the THz-TDS is usually used for channel measurements within a short distance, i.e., less than a few meters. Besides, the transmitting beam is very narrow, which makes it mainly suited for measuring material characteristics, such as reflection, scattering and diffraction characteristics, in the THz band.

\subsubsection{Measurement Campaigns}

Most existing THz-TDS-based channel measurement campaigns use the Picometrix T-Ray 2000 THz-TDS, which can emit terahertz pulses with a bandwidth of 0.1-3~THz~\cite{Hossain2019Stochastic,Ma2015Comparison,Federici2016Review}.
For instance, \textit{Hossain et al.} used the THz-TDS to measure the interference between THz devices in the 300~GHz frequency band, and used stochastic geometrical approaches to model and analyze the interference~\cite{Hossain2019Stochastic}.
Furthermore, aiming at the outdoor scenario, \textit{Federici et al.} measured the attenuation of THz signal caused by the weather with the THz-TDS, implemented theoretical analysis, and summarized the influence of different weather factors on THz links~\cite{Ma2015Comparison,Federici2016Review}.

Moreover, since 2007, the Terahertz Communication Laboratory from TU-Braunschweig has been focusing on channel measurement, simulation, and antenna technology for 300~GHz and below~\cite{Piesiewicz2007short,priebe2011channel,Koch2007Terahertz, kurner2013towards}.
The team, in collaboration with Brown University, used a THz-TDS to measure reflection coefficients~\cite{piesiewicz2007scattering,Jansen2008impact}, and scattering coefficients~\cite{Jansen2011diffuse} of various indoor materials from 60~GHz to 1~THz.

Besides, based on a THz-TDS, Sichuan University and China Academy of Engineering Physics measured the reflection characteristics of common indoor materials in the 320-360~GHz frequency band~\cite{Wang2014Thz}.

%% file: v1/measurement_discussion.tex
\subsection{Comparison and Discussion}

During the measurement, systematic errors mainly come from two sources, namely, the Tx/Rx module and the cable, which need to be further calibrated through post-processing. The first kind of error can be removed by directly connecting the Tx and Rx modules, and dividing this response $S_{\rm calib}$ into the original response $S_{\rm measure}$. The response of the channel can be derived as $S_{channel}=S_{\rm measure}/S_{\rm calib}$. The second kind of error, happening when high-frequency signals are transmitted through the cable, can be removed with the help of electronic calibration module. Nevertheless, the extra noise or phase jitter can not be eliminated by the calibration procedure.

So far, many terahertz channel measurement systems have been built in the 140~GHz, 220~GHz and 300~GHz frequency bands, based on VNA, sliding correlation method and THz-TDS.
Most of the current THz channel measurements focus on the below-300-GHz band, while extensive channel measurements above 300~GHz are still missing from the literature.

\begin{table*}[htbp]
    \centering
    \caption{Methodologies for THz channel measurement.}
    \includegraphics[width = 0.8\linewidth]{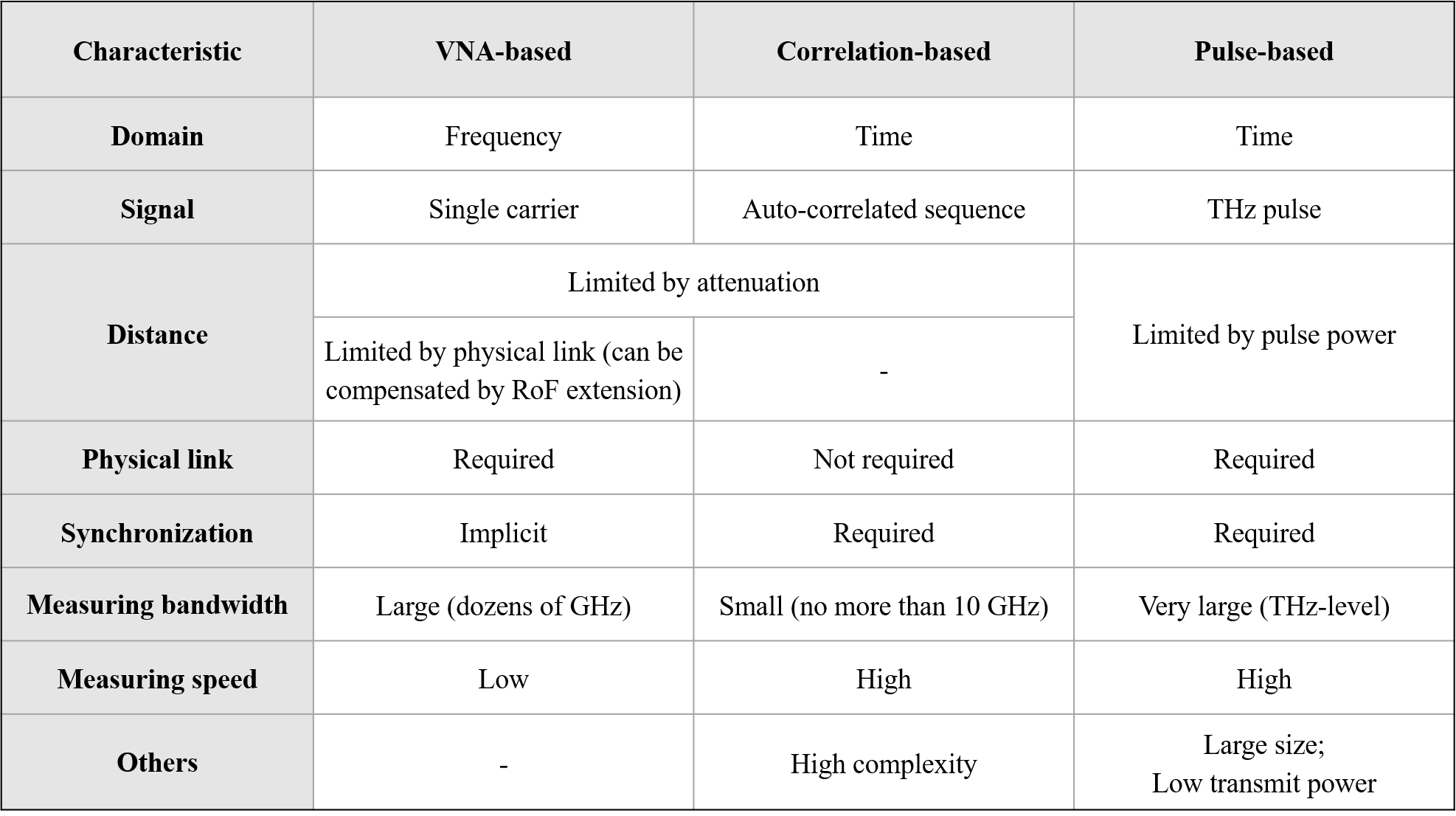}
    \label{tab:channel_measurement_methods}
\end{table*}
\begin{figure}
    \centering
    \includegraphics[width=\linewidth]{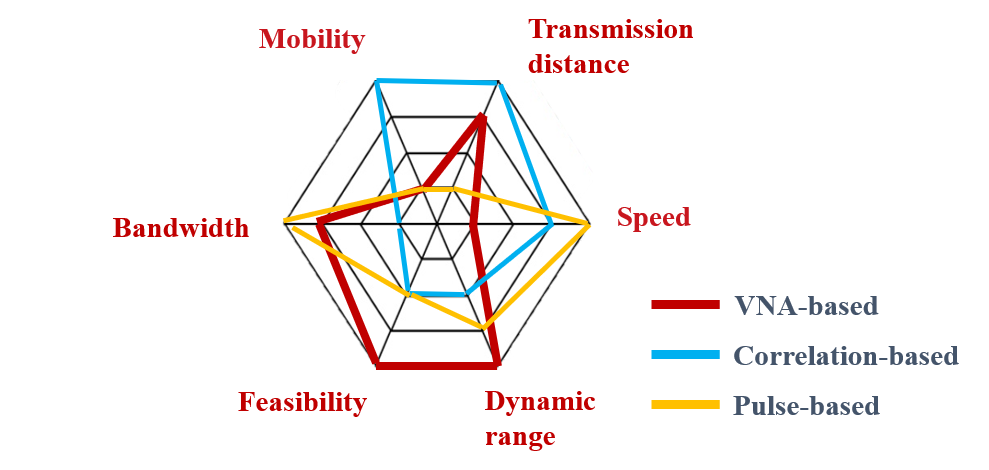}
    \caption{Comparison of THz channel measurement methodologies.}
    \label{fig:radar}
\end{figure}

Table~\ref{tab:channel_measurement_methods} and Fig.~\ref{fig:radar} compare the characteristics, strengths and weaknesses of the three measurement methods.
Among them, VNA-based measurement systems implement frequency stepping. The S parameters of the channel within a certain frequency band are directly measured and recorded, and the amplitude-frequency response of the channel is obtained after system calibration. Advantages of the method include high time-domain resolution, inherently synchronized transceivers, and low complexity of measurement system\footnote{From the standpoint of a user. A VNA is of course in itself a highly complex piece of equipment.}. However, frequency sweeping is time-consuming, and in the most basic setup, the transceivers need to be connected to the VNA by transmission lines, which limits the measuring distance and thus application scenarios, making it not applicable to outdoor scenarios. However, extension of the setup with radio-over-fiber connections eliminates this bottleneck, and long-distance outdoor measurements have been performed.

In channel measurement systems based on the sliding correlation method, a PN sequence is transmitted, and the Rx conducts a cross-correlation operation on the received signal to obtain the channel impulse response in the time domain. The advantages include instantaneous broadband measurement, fast measurement and direct access to the time-domain information of the channel. Separate synchronization between Tx and Rx needs to be performed, which can be done either through (optical) cables or GPS-disciplining of precision clocks. Unlike the other two kinds of sounders which require connection through cables, the correlation-based channel sounder can be synchronized by two separate rubidium clocks and relieve itself of cable connection, and thus avoids tripping hazard for long-range measurement. Sliding correlator systems enable greater flexibility for the trade-off between measurement speed, sampling rate, and SNR.

The third channel measurement system is based on THz-TDS, which emits THz pulses, and the Rx can directly obtain the time-domain THz channel response after detection and sampling. THz-TDS-based measurement systems have huge and scalable system bandwidth. However, the THz wave emitted by the THz-TDS is down-converted from the optical band, and the signal power is much lower than that of the previous two channel measurement systems, which makes it suitable for channel measurements only over short distances of few meters. Besides, the THz-TDS has large size and narrow beam width, which makes it not suitable for directional scanning. Furthermore, standardization of TDS-based measurement, calibration, and data analysis for the THz band should also be established~\cite{naftaly2017metrology}.

Recently, channel measurement systems integrating more than one channel sounding mode have been developed for trade-off between the high dynamic range and fast sounding speed, for example, the sliding correlation mode and real-time spread spectrum mode in the NYU wireless sounder~\cite{xing2018propagation}. The real-time spread spectrum is faster than the correlation mode at the cost of the reduction of dynamic range. Furthermore, to deal with the variety of the communication scenarios with different requirements on the sounding speed, dynamic range, Tx-Rx distance, etc. for THz channel measurement, new channel measurement systems integrating both temporal and frequency domain methods are worthy of exploration. For example, stimulated by the exploration of the THz band through frequency extension from both electrical and photonic directions, the method of electro-optic sampling (EOS), which is similar to TDS as measuring techniques and is analogous to VNA at the application level, is mentioned as a heuristic measurement approach~\cite{naftaly2017metrology}.

%% file: v1/modeling.tex
\section{Channel modeling in the THz Band} \label{sec:modeling}

\begin{figure*}
    \centering
    \includegraphics[width=\linewidth]{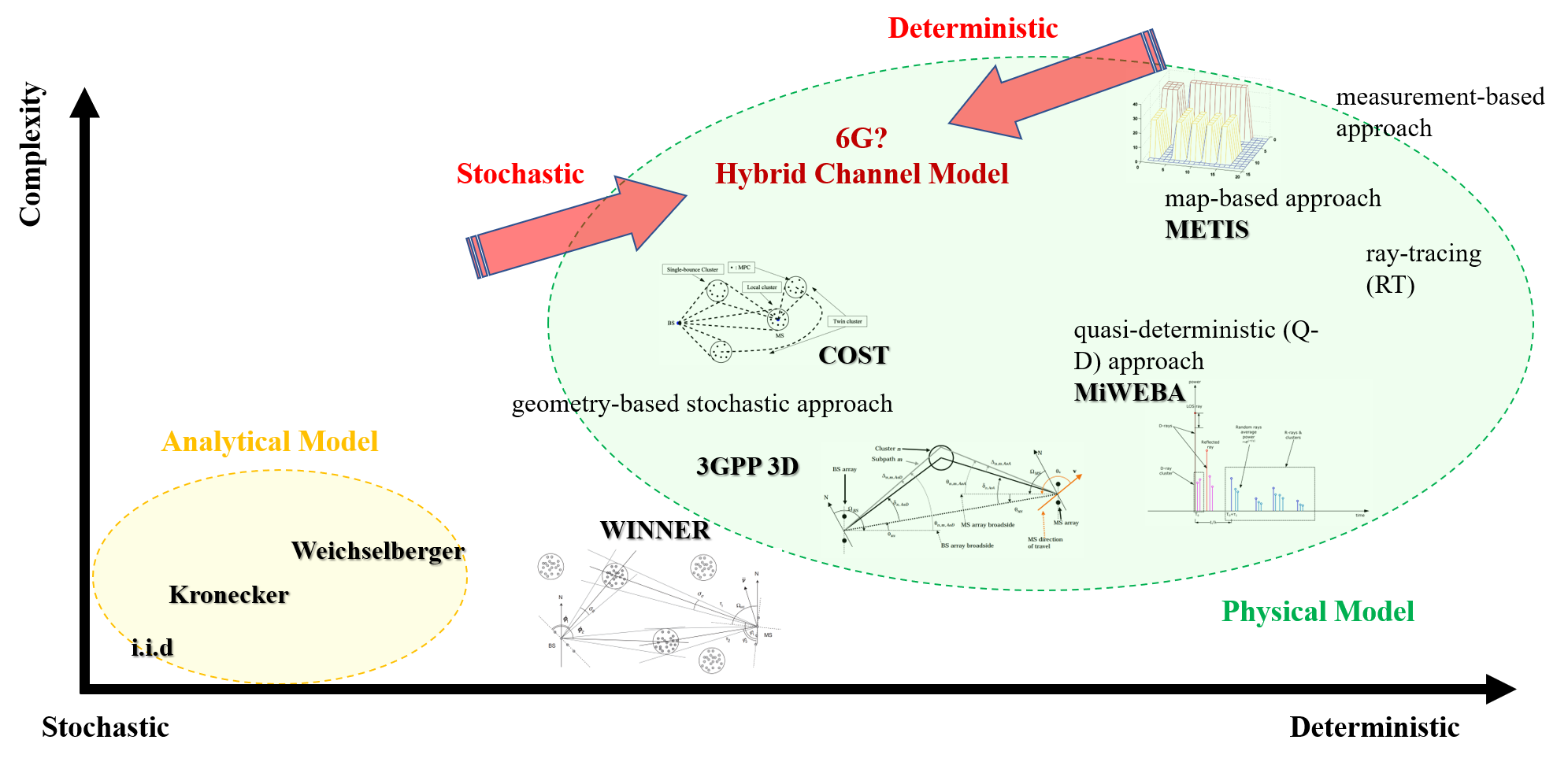}
    \caption{Trend of channel modeling methodologies (i.i.d.: identically distributed Rayleigh fading model; Kronecker: Kronecker-based stochastic model~\cite{ozcelik2003deficiencies}; Weichselberger: Weichselberger model~\cite{weichselberger2006stochastic}; WINNER: WINNER II spatial channel models~\cite{Kyosti2008WINNER}; 3GPP: 3GPP spatial channel models~\cite{3GPP-SCM}; COST: COST259/273/2100 models~\cite{liu2012cost}; MiWEBA: MmWave Evolution for Backhaul and Access channel models~\cite{weiler2016quasi}; METIS: Mobile and wireless communications Enablers for the Twenty-twenty Information Society channel models)~\cite{nurmela2015deliverable}.}
    \label{fig:axis_channel_modeling_methods}
\end{figure*}

Channel characteristics in the THz band are quite different from those of lower (mmWave and cmWave) frequency bands.
THz wave peculiarities mainly include i) the high isotropic free-space path loss, ii) the strong molecular absorption phenomena that makes different spectral windows suitable for communications over different bands~\cite{jornet2013fundamentals}, severe frequency selectivity and the resulting temporal dispersion effects~\cite{han2015multi}, iii) the significant Doppler effect, and iv) strong penetration loss and rough-surface scattering, among others.
Thus, although channel models for lower frequencies have been well explored~\cite{andersen1995propagation,bertoni1999radio,saunders2007antennas,molisch2011wireless,molisch2014propagation,molisch2021millimeter} and can be used as a starting point, they cannot be uncritically applied to THz channels and need to be modified and tailored to accurately characterize the aforementioned THz wave peculiarities.
This section provides an overview of the challenges and requirements of channel modeling in the THz band and describe fundamental THz channel modeling methodologies as well as some existing models. Fig.~\ref{fig:axis_channel_modeling_methods} provides an overview of common channel modeling methods and models and trends towards 6G. Existing channel models are emphasized in bold black, while modeling methods are labeled in light black. The x-axis denotes the amount of stochastic/deterministic degree in the modeling methods, while the y-axis represents the complexity of the methods.

\subsection{Overview}

We start by listing the challenges and requirements to be addressed for the analysis and design of THz band channels. First, the THz wave propagation for LoS, reflected, scattered and diffracted paths, respectively, needs to be modeled. In addition to the various propagation paths, the different channel scenarios including static and time-varying environments need to be accounted for. Second, effects from THz antennas or antenna arrays may need to be considered. Most situations can make use of double-directional channel models~\cite{steinbauer2001double}, which are independent of the considered antenna structure, and which can be combined with arbitrary antennas (or compound antenna-plus-casing, and possibly including effects of user hand or body~\cite{harrysson2011evaluation}).
However, for (i) the case of body-area networks (where antenna surroundings and channels cannot be separated easily) and (ii) for antenna arrays with dimension larger than the stationarity region of the channel, a separate analysis is required. The latter case of large dimension of arrays can be caused either by sparse antenna arrays where multiple array panels are separated physically, or by ultra-massive multiple-input-multiple-output (UM-MIMO) that are equipped with hundreds or thousands of densely packed antennas~\cite{akyildiz2016realizing}. UM-MIMO systems are of special interest as they can effectively increase communication range and further enhance capacity in THz wireless communications. By contrast, spare arrays have benefits of reduced cost and complexity that are attractive for THz imaging, compared to UM-MIMO with dense arrays.
Third, the channel parameters of the THz spectrum such as the path gain, delay spread, temporal broadening effects and wideband channel capacity need to be accurately described.

\begin{figure*}
    \centering
    \includegraphics[width=\linewidth]{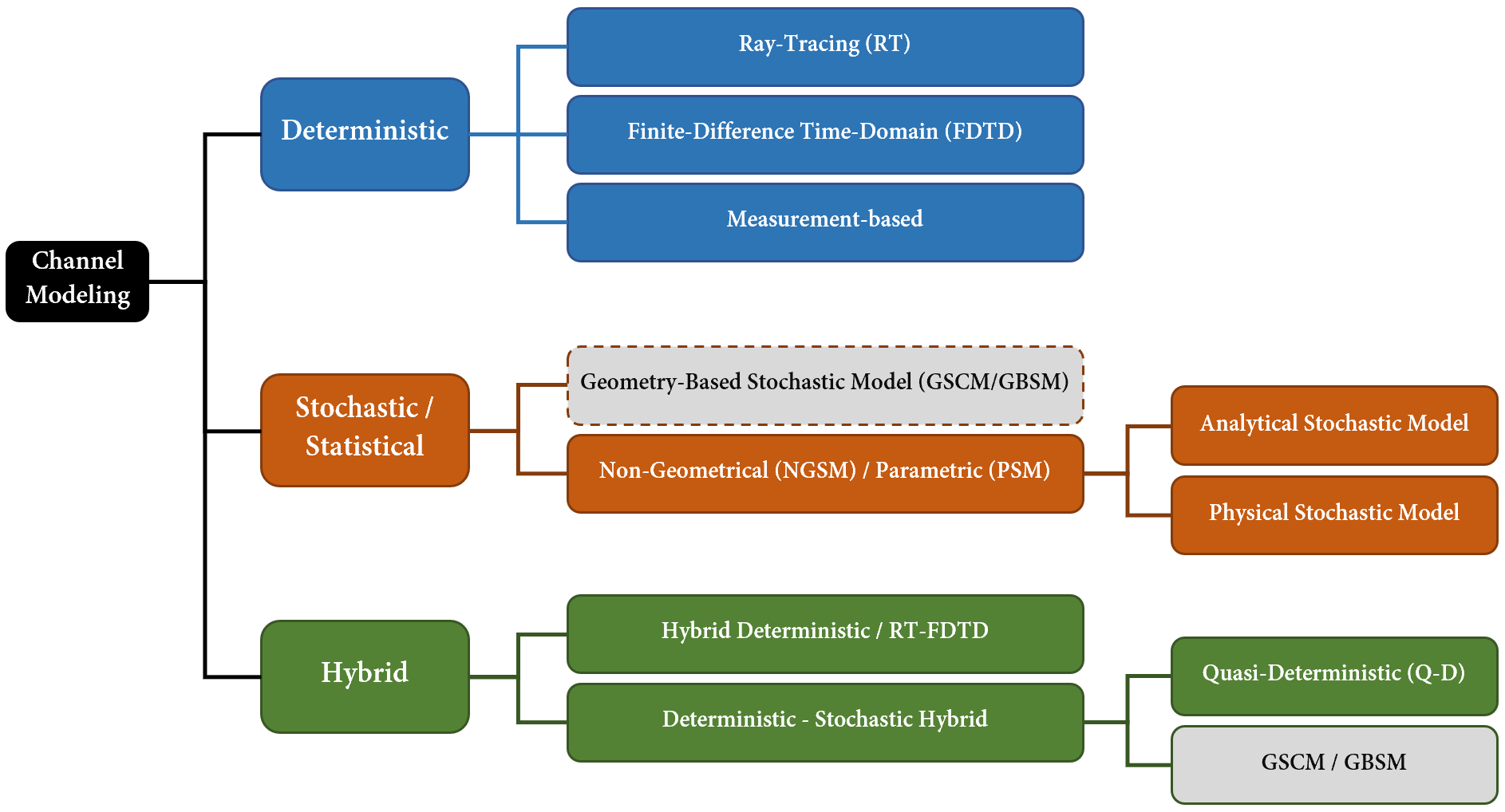}
    \caption{Channel modeling methodologies.}
    \label{fig:summary_channel_modeling_methods}
\end{figure*}

As summarized in Fig.~\ref{fig:summary_channel_modeling_methods}, the methodologies for physical wireless channel modeling can be broadly categorized as deterministic, stochastic (or statistical), and hybrid approaches.
Deterministic methods solve (approximately) Maxwell's equations in a given environment, and can achieve high accuracy, but require detailed information about the geometry and electromagnetic properties (dielectric constants, loss factors) of the environment and have high computational complexity. Statistical methods use random distributions and processes to model the channel parameters, including the path gain, delay, number of paths, and coupling, among others. Statistical channel models are obtained mainly based upon measurements and describe types of environments (e.g., indoor office environments) rather than specific locations. While the parameterization of such models from measurements involves a large effort, creating model realizations from the models has much lower complexity than deterministic models. Therefore, deterministic channel models are suitable for deployment planning, while statistical ones are suitable for the development and testing of wireless systems.
Hybrid channel models, which combine deterministic and statistical methods provide a balance between accuracy and efficiency. For example, the dominant paths in the THz band, such as the LoS and reflected rays, are individually captured based on the deterministic method, while other paths, such as scattered and diffracted paths, could be statistically generated. We note that as a fourth approach, generative neural networks have recently emerged. However, they have as of yet not been investigated in the context of THz channels, so that they are beyond the scope of this survey.

\input{v1/modeling_Deterministic}

\input{v1/modeling_Statistical}

\input{v1/modeling_Hybrid}

\subsection{Remarks and Discussions}
\begin{table*}[htbp]
    \centering
    \caption{Methodologies for THz channel modeling.}
    \includegraphics[width = 0.8\linewidth]{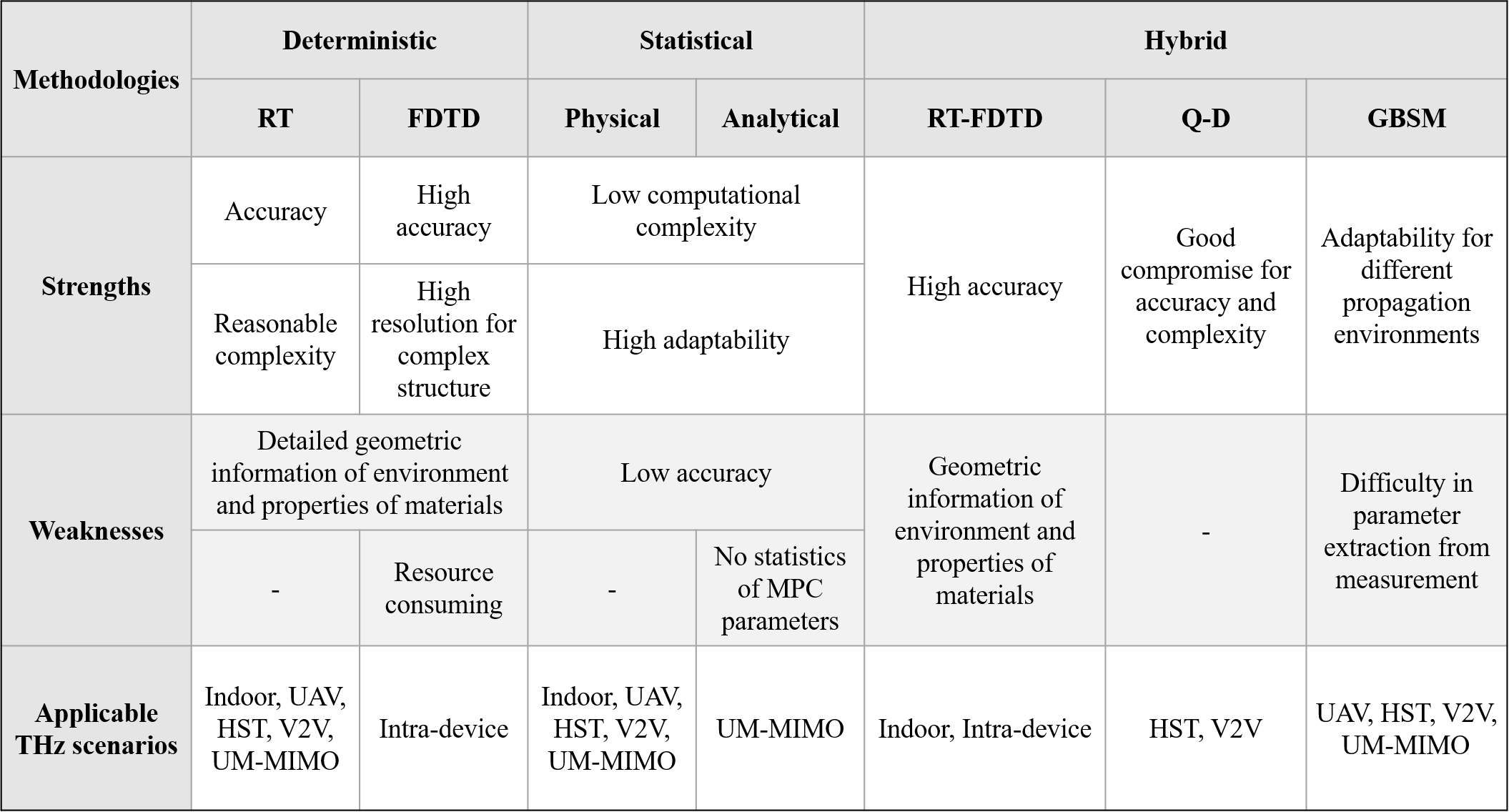}
    \label{tab:channel_modeling_methods_SISO}
\end{table*}

In 6G wireless communication, use of the THz band will enable technologies like UM-MIMO, and THz communication systems will be used in various scenarios such us indoor, intra-device (e.g. on-chip), high-speed train (HST), vehicle-to-vehicle (V2V), unmanned aerial vehicle (UAV), and inter-satellite~\cite{nie2021channel} communications. Different channel modeling methods need to be suited for various scenarios. A comparison of the strengths and weaknesses of the aforementioned channel modeling methodologies for THz communication scenarios are summarized in Table~\ref{tab:channel_modeling_methods_SISO} and described as follows.

First, RT has the ability to model the EM wave propagation with reasonable resource consumption. In the THz band, EM waves behave quasi-optically due to the small wavelength, and thus a geometrical optics method such as RT is able to accurately capture the wave propagation. Modeling by RT is applicable for most channels in THz scenarios, though further studies of material properties in the THz band are required.
In contrast, the purley numerical method, FDTD, provides better performance in complex small-scale structures and in wideband. Though FDTD has advantages in accuracy, it is limited to small-scale areas due to the lack of efficiency. Thus it is applicable for intra-device channels in the THz band, for example, ultra-high-speed on-chip communications~\cite{akyildiz2014terahertz,fricke2014time}.

Second, the computational complexity of stochastic/statistical modeling methods is much lower than deterministic counterparts. Though the result may deviate from the actual measurements, stochastic/statistical models are widely applied due to their simplicity for generation and ability to describe the statistics of channel properties. Resulting from the flexibility, statistical models are adaptable to most propagation scenarios in the THz band.
In particular, compared with physical statistical models which characterize statistics of {\em propagation} channel characteristics (e.g., arrival time and angle distribution of the MPCs), analytical statistical models focus more on statistics of the {\em radio channel}, which subsume the antenna characteristics and array geometry with the propagation channel into an ``effective'' channel whose characteristics are described, e.g., by correlation matrices.  

Third, hybrid approaches are developed to achieve acceptable accuracy and low complexity simultaneously. The deterministic hybrid modeling method improves the accuracy of the geometrical optical method like RT when dealing with complex structures. Taking the computational effort into account, the hybrid deterministic approach is suitable for indoor and intra-device channel modeling in the THz band.
Moreover, Q-D approaches can be adapted to  different scenarios and show high accuracy with low complexity because critical components that dominate in the channel are traced deterministically.
Furthermore, GSCMs naturally support multiple links and time evolution of a channel, and give insight into physical propagation by the statistics of spatial distributions of scatterers. They are widely used in THz channel modeling due to their flexibility by adjusting environment geometries and channel characteristics for different scenarios. They can provide spatial consistency in particular when introducing the concept of visibility regions and the birth-death process of clusters.
Besides, for UM-MIMO channels with properties such as the spherical wavefront, non-stationarity, and cluster visibility, GSCMs are applicable since they support accurate propagation distance calculation for each individual antenna element.

%% file: v1/modeling_Deterministic.tex
\subsection{Deterministic}
Deterministic channel models accurately model the wave propagation based on the theory of electromagnetic (EM) wave propagation~\cite{sarkar2003survey}. The approach is site-specific, and requires detailed geometric information of the propagation environment, dielectric properties of materials and spatial positions of the Tx and the Rx~\cite{wcomm_book}. Therefore, a deterministic approach provides a good agreement between the simulation results and the measurements in general~\cite{sarkar2003survey}, though the accuracy varies based on the specific method, the accuracy of the environmental information, and the analyzed frequency band. The results from the deterministic modeling can be useful by themselves (e.g., for deployment planning), to provide statistical channel information by applying Monte Carlo analysis on many random transmit/receive locations, and/or as input for statistical channel modeling. In particular, RT~\cite{yang1998ray,son1999deterministic, wcomm_book,molisch2011wireless,han2015multi} and finite-difference time-domain (FDTD)~\cite{yee1966numerical, Taflove2005computational,zhao2007FDTD} are two representative methods of deterministic channel modeling, while the use of measured, stored impulse responses (or equivalent) is another possible deterministic approach.

\input{v1/modeling_RT}
\input{v1/modeling_FDTD}

\subsubsection{Measurement-based}
The measurement-based approach relies on channel measurement along with data storage. The concept of ``stored measurements'' has been used at least since the 1990s, when channel sounder measurements started to be stored digitally. Various projects, such as the Metamorp project, attempted to standardize formats for data storage both in the time domain (as impulse response) and frequency domain (transfer function). Major challenges revolve in particular around unified formats of metadata such as calibration data of the channel sounders, and descriptions of the measurement parameters and environments. 
More recently, the principle of ``open source'' data has motivated many researchers to place measurement results online for download. Various standardization groups, including the NextG Channel alliance~\cite{nist2021} aim to facilitate data exchange. The challenges in the context of THz channels revolve around the size of the measured data, both due to the large bandwidth, and large antenna arrays.

%% file: v1/modeling_RT.tex
\subsubsection{Ray-Tracing}

RT has emerged as a popular technique for the analysis of site-specific scenarios, due to its ability to analyze very large structures with reasonable computational resources~\cite{yun2002ray}.
The ray-tracing algorithm models the propagation of electromagnetic waves based on the high-frequency approximation of Maxwell's equations, geometrical optics. The locations of the Tx and the Rx are first specified, followed by determining all possible routes between the transceivers, based on high-frequency-approximation rules like geometric optic (GO), geometric theory of diffraction (GTD), uniform theory of diffraction (UTD), and Kirchhoff theory~\cite{ragheb2007modified}. The technique is especially suitable for THz channels due to the fact that these approximations become more accurate due to the stronger corpuscular property in the THz band, which is associated with the wave-particle (wave corpuscle) duality of light~\cite{han2018propagation}.

\begin{figure}
\centering
\subfigure[Propagation scenario.]{
\includegraphics[width=0.3\linewidth]{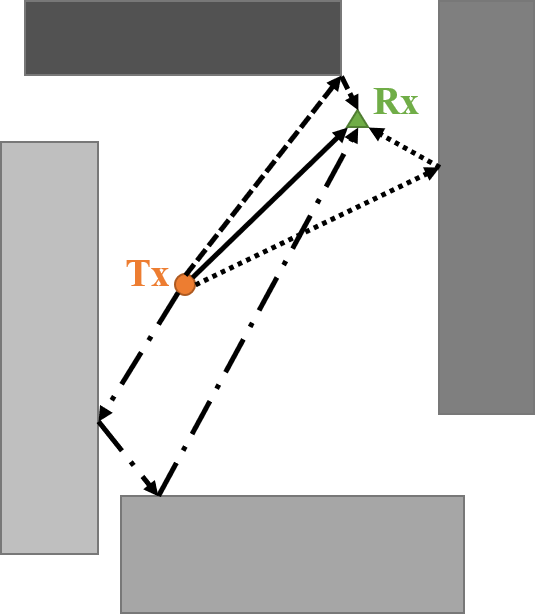}
}
\quad
\subfigure[An instructional and incomplete visible tree.]{
\includegraphics[width=0.5\linewidth]{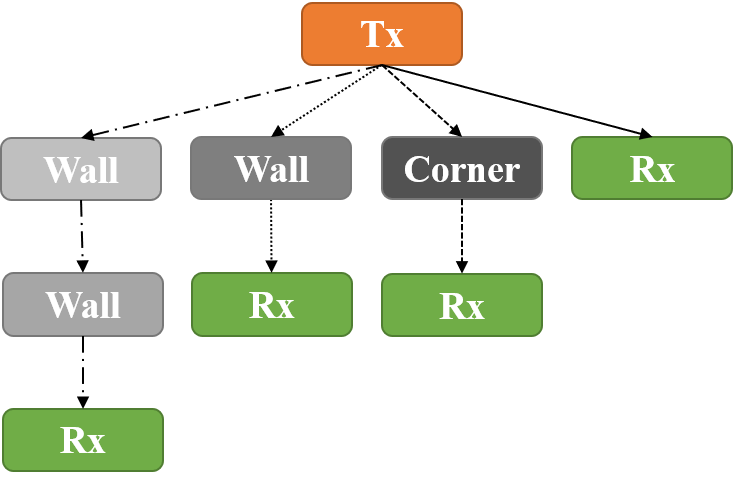}
}
\caption{Illustration of the visible tree in a simple propagation scenario.}
\label{fig:RT_tree}
\end{figure}

One strategy for efficiently capturing the individual propagation paths in the tracing process is the so-called visibility tree~\cite{rachid2012UWB}. 
The visibility tree has a layered structure with nodes and branches. Each node represents an object of the scenario, and each branch represents a LoS connection between two nodes. The root node denotes the antenna of the Tx. An instructional propagation scenario and the corresponding visible tree are shown in Fig.~\ref{fig:RT_tree}. Like a typical tree, the construction of the visible tree follows a recursive approach. Starting from the first layer of the tree, every two nodes in adjacent layers are connected by branches representing the LoS path between the nodes. The branch is terminated with a leaf where the Rx is reached. Therefore, the total number of leaves in the visible tree equals the number of paths identified by the RT procedure. The process is repeated until reaching the highest layer with the pre-set prediction order.
After the visibility tree is built, a backtracking procedure determines the path of each ray by traversing the tree upwards from leaves to the root node, and applying geometrical optics rules at each traversed node.

An alternative approach is ray launching~\cite{kreuzgruber1993ray,durgin1997advanced,rose2014analytical,lu2018discrete}, where rays are transmitted into a grid of directions (either into a set of azimuth directions when 2D simulations are done, or azimuth/elevation combinations for 3D simulations). The simulation follows the path of each ray, taking into account changing of direction due to reflections, diffraction, etc., until the ray either leaves the area of interest, or its strength falls below a pre-determined threshold. Ray launching requires longer simulation times when simulating a single transmit/receive location pair, but can simulate many Rxs (for the same transmit location) without significantly increasing effort, and is thus well suited for simulating channel characteristics throughout a cell. 

\begin{figure}
    \centering
    \includegraphics[width = 0.9\linewidth]{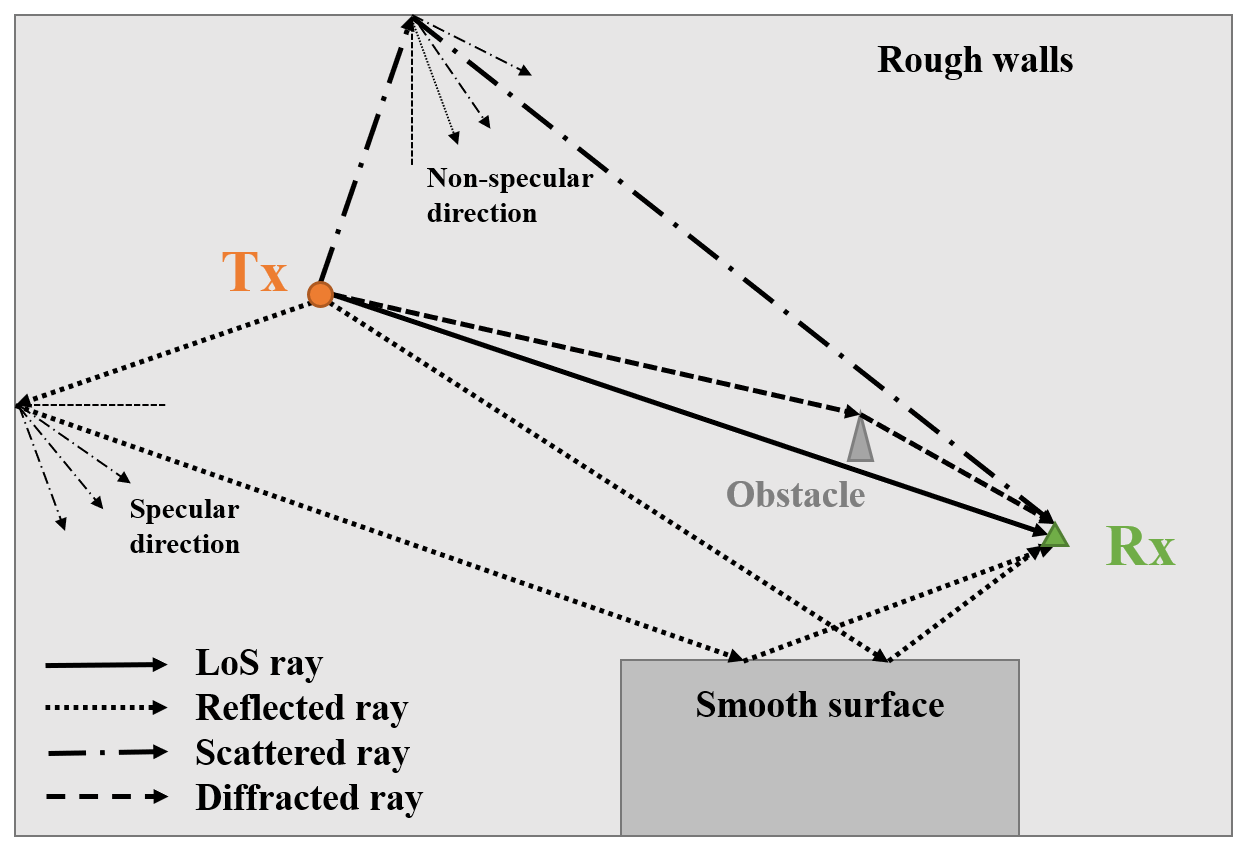}
    \caption{The ray-tracing modeling method for propagation in the THz band.}
    \label{fig:RT}
\end{figure}

For wave propagation in the THz band, the free space propagation needs to consider the spreading loss and the molecular absorption loss~\cite{han2018propagation}. The spreading loss can be characterized by Friis' law. The molecular absorption results from the fact that part of the wave energy is converted into the internal kinetic energy of the molecules in the propagation medium~\cite{jornet2011channel}.
In addition, the radio signal transmitted from a source will encounter multiple objects in the environment, forming reflected, diffracted, or scattered rays to the Rx, as illustrated in Fig.~\ref{fig:RT}, which need to be traced as well.
Surface roughness is typically measured by the surface height standard deviation, which is then compared with the wavelength to distinguish ``smooth'' surfaces from ``rough'' surfaces in a specific frequency band~\cite{ragheb2007modified}. When the wavelength is comparable to or smaller than the surface roughness, the surface is considered as ``rough''. Therefore, surfaces that are considered smooth at lower frequencies become rough in the THz band~\cite{han2015multi}, for example the plastic or the wallpaper in~\cite{piesiewicz2007scattering}. When the incident wave encounters a rough surface, the reflected ray will radiate in the specular direction and the scattered rays will radiate in all other directions. Among these rays, specularly reflected rays are dominant in the THz band~\cite{Jansen2011diffuse}, while other scattering components are conjectured to have a strong impact on NLoS channels since the surface roughness becomes comparable with the wavelength, and thus the diffusely scattered power increases with increasing frequency. Ref.~\cite{Jansen2011diffuse} suggests to determine scattered rays by dividing the scattering surface into smaller square tiles around specular reflection points and each tile contributes a scattered ray. Moreover, the correlation length of a rough surface is typically much greater than the wavelength in the THz band, and thus sharp irregularities are not present. Diffraction effects are assumed to be mostly negligible in ray-tracing due to the extremely high diffraction loss in the THz band~\cite{Jacob2012Diffraction,priebe2013ultra, han2015multi}.

The results are heavily dependent on the geometrical-optic models for reflection, scattering and diffraction, since in the THz band, the multi-ray propagation phenomenon becomes severely polarization-, angle- and frequency-dependent~\cite{piesiewicz2007scattering}. Since the Kirchhoff theory is widely used to capture multi-path loss, the most frequently used models for reflection and scattering are based on modifying the classical Kirchhoff theory model to agree with the experimental data in the THz band~\cite{ragheb2007modified, Jansen2011diffuse, piesiewicz2007scattering}. Similarly, for diffraction, UTD and the Fresnel Knife Edge Diffraction (KED) theory are good approximations~\cite{Jacob2012}.

In face of the sparsity of accurate environmental data and empirical measurement datasets, it is common to develop ray-tracers and then calibrate ray-tracing results from measurements~\cite{hur2016proposal}.
So far in the THz band, ray-tracing has been calibrated by measurements for indoor~\cite{priebe2013ultra} and the T2I inside-station~\cite{guan2019measurement} scenarios at 300 GHz. \textit{Sheikh et al.} developed a ray-tracing algorithm (RTA) to model the diffuse scattering from rough surfaces at THz frequencies based on Beckmann-Kirchhoff model~\cite{sheikh2018novel}, and conducted a simulation in a 7~m$\times$7~m$\times$3~m office room covering both LoS and NLoS scenarios and three types of plasters with different degrees of roughness~\cite{sheikh2020study}. Complete and generic multipath channel models based on the RT method for the entire THz band have been built in~\cite{han2015multi}, evaluating the capacity and analyzing key channel parameters of low THz band (0.1-1~THz). Besides, the elevation plane is involved to develop a 3D end-to-end channel model for the THz spectrum in~\cite{han2017three}. However, calibration/validation for 1-10~THz is still missing due to the lack of material parameters.

\begin{figure}
\centering  
\subfigure[Virtual point for MIMO channel.] {
\includegraphics[width = 0.8\linewidth]{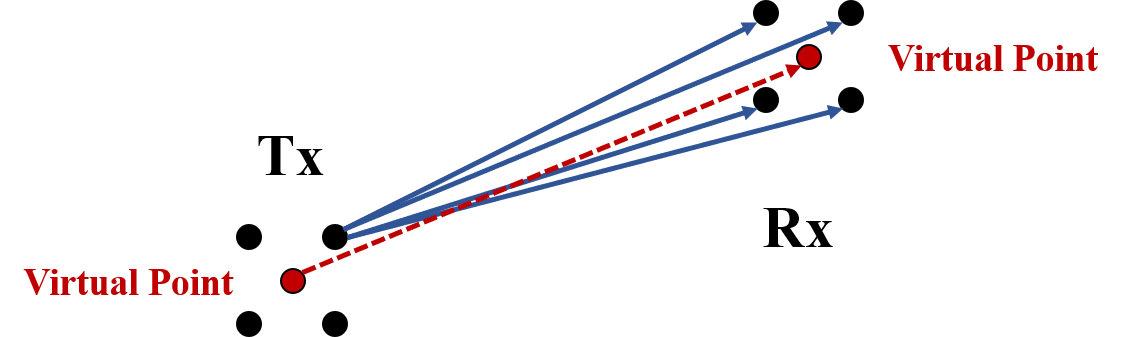} 
}     
\subfigure[Virtual point for UM-MIMO channel.] {
\includegraphics[width = 0.8\linewidth]{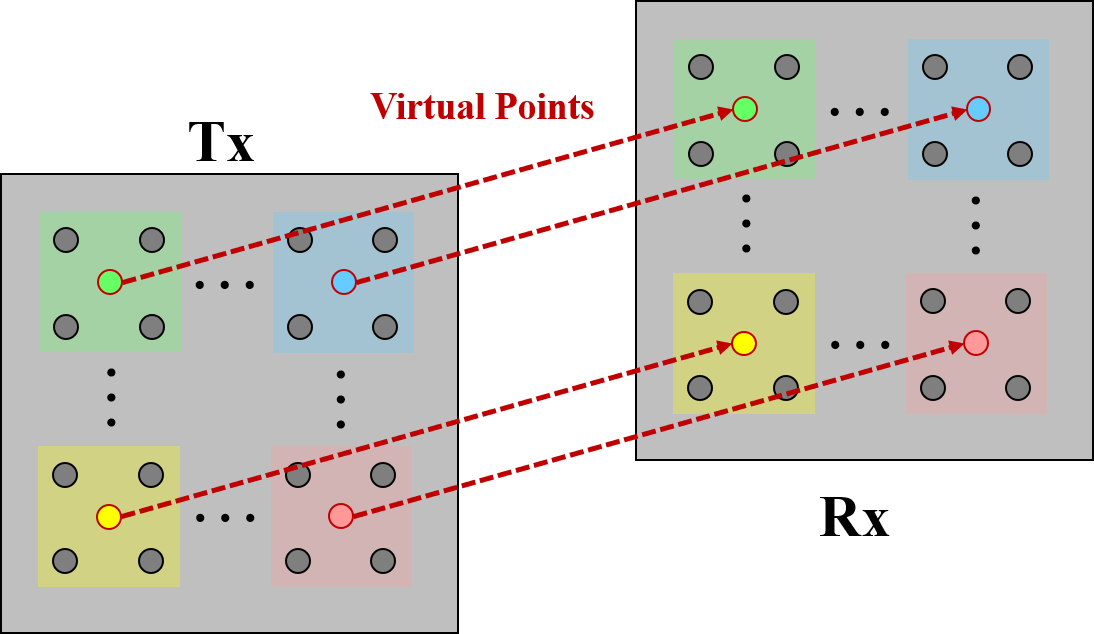}   
}
\caption{Virtual point approximation approach.}     
\label{fig:RT_virtual_point}
\end{figure}

Conventional RT models for a single antenna system perform a point-to-point analysis between the transceiver, while for multiple antenna systems, the operation can be costly if RT is performed for every Tx-Rx link~\cite{ng2005modelling}.
The computational burden can be reduced by performing a single ray-tracing simulation from which not only the amplitudes and delays, but also the directions of the paths are extracted. This information can be combined with the array characteristics to provide the transfer function between each transmit and each receive antenna pair, and is thus independent of the antenna array size~\cite{molisch2002modeling}. This approach was later called ``virtual point approximation'' in~\cite{ng2005modelling}. As shown in Fig.~\ref{fig:RT_virtual_point}, the MIMO channel matrix is synthesized from a point-to-point traced channel between two virtual points instead of those between every pair of antennas. The virtual point approximation is based on the assumption that all sub-channels share a very similar set of rays by properly selecting the location of the virtual points. Post-mapping and checking procedures are implemented based on actual positions of antenna elements after the ray-tracing operation on virtual points. Especially for UM-MIMO channels where the antenna array is larger than the stationary region, the whole antenna array can be divided into smaller sub-arrays each of which corresponds to a virtual point. Therefore, a UM-MIMO channel matrix is obtained from a double-directional channel among virtual points obtained by ray-tracing.

Another way to relieve the computational burden is to apply simplified RT models, for instance, map-based models. Note that some researchers refer to the term ``map-based'' as any site-specific propagation model~\cite{lim2017map} and categorize them into map-based deterministic, map-based stochastic, and map-based hybrid channel models. However, in this paper, we apply the term ``map-based'' to denote the deterministic modeling approach, which is based on ray-tracing and uses a simplified 3-D geometrical description of the environment~\cite{nurmela2015deliverable,carton2016validation}. Note that a number of researchers categorize map-based models as ``hybrid'' models. 

One kind of map-based channel model was proposed by METIS~\cite{nurmela2015deliverable}. It inherently accounts for significant propagation mechanisms including line-of-sight propagation, diffraction, specular reflection, diffuse scattering and blocking. It is suitable for evaluating massive MIMO, beamforming, and realistic path loss modeling in the case of device-to-device (D2D) and vehicle-to-vehicle (V2V) communications. Since conventional ray-tracing approaches are strictly site-specific and subject to high complexity, the METIS map-based model addresses these issues by assigning random objects to represent cars or humans after the map is defined. Point sources for diffuse scattering, transceiver locations, and pathways (path length and arrival/departure angles) are then defined 
additionally. The complexity of the map-based modeling approach can be adjusted by the selection of the number of rays and types of relevant propagation mechanisms~\cite{carton2016validation}.

Conventional environmental databases provide resolution on the order of $5 \times 5$ m, which is orders of magnitude larger than the wavelength at THz frequencies. Much better accuracy can be obtained from methods like laser scanning of the environment, which yields point-cloud data of the environment~\cite{jarvelainen2016indoor,jarvelainen2016evaluation, pascual2016importance, pascual2018wireless, virk2018site}. Ray-tracing can be done based on those point clouds, with increased accuracy, though also typically significantly increased runtime of the simulation. Point-cloud ray-tracing has been applied to THz channels in \cite{gougeon2019ray,Nguyen2021Large,papasotiriou2021experimentally}.

%% file: v1/modeling_FDTD.tex
\subsubsection{Finite-Domain Time-Domain}

FDTD is also known as Yee's method named after the Chinese American applied mathematician Kane~S.~Yee~\cite{zhao2007FDTD}. It is a numerical analysis technique that directly solves Maxwell's equations. FDTD can resolve the impact of small and complex scatterers, and rough surfaces in the THz band, but suffers from very high computational complexity when applied to an  environment that has large dimensions in units of wavelength, as often occurs for THz channels. Furthermore, a database of the environment with sufficient resolution, e.g., a point cloud from laser scanning (see above) is required. 

\begin{figure}
    \centering
    \includegraphics[width=0.7\linewidth]{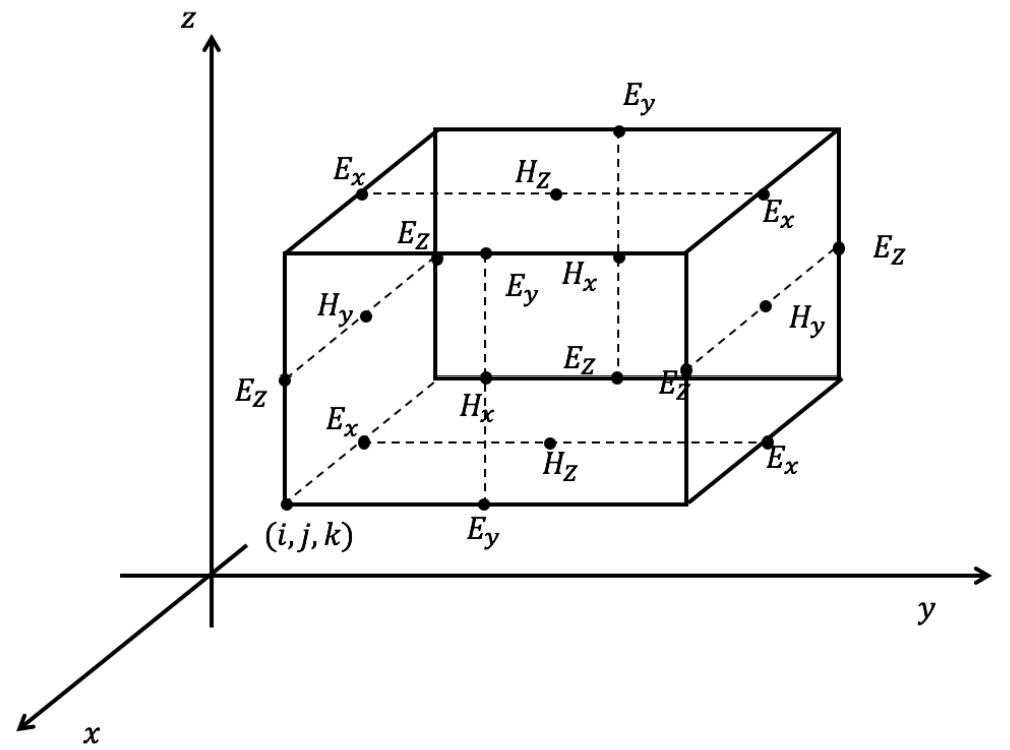}
    \caption{FDTD.}
    \label{fig:FDTD}
\end{figure}

To implement FDTD, the space is partitioned into grids called Yee cells in the first step. Then the magnetic and electric fields are sampled alternately in both temporal domain and spatial domain so that each sampled point for magnetic field (H-field) is surrounded by six points for the electric field (E-field) and vice versa, as shown in Fig.~\ref{fig:FDTD}. We assume the strength of electric field or magnetic field $F(x,y,z,t)$ (i.e., $F$ can be replaced by $E$ or $H$), is discretized in space and time domain as $F^n(i,j,k) = F(i\Delta x, j\Delta y, k\Delta z, n\Delta t)$, where $\Delta x$, $\Delta y$, $\Delta z$ and $\Delta t$ are the steps of three-dimensional space and time. By the central difference method, Maxwell's equations are discretized for further computation.

As an approximate modeling technique to solve Maxwell's equations, FDTD can achieve high accuracy. Moreover, this method is especially adaptable for small and complex scattering scenarios in the THz band, where the surface material has a higher roughness level relative to the small wavelength. While RT generally requires a modified or calibrated scattering model, FDTD retains - in principle - accuracy under arbitrary conditions. However, since the spatial discretization of the grid must be sufficiently small to resolve the smallest electromagnetic wavelength and the smallest geometrical feature in the model, FDTD requires large amounts of memory to keep track of the solution at all locations, as well as substantial time and computational resource to update the solution at successive instants of time~\cite{wang2000hybrid}. Since the required resolution increases with frequency, computations at THz frequencies can be beyond the capabilities of even advanced computers. A comparison for ray-tracing with FDTD in a very small intra-device channel has been presented in~\cite{fricke2014time}.

%% file: v1/modeling_Statistical.tex
\subsection{Statistical}
Although deterministic modeling methods provide accurate channel modeling results, they require detailed geometric knowledge of the propagation environment and suffer from high computational complexity. Alternatively, statistical modeling methods could be invoked to describe THz propagation characteristics.
Statistical approaches  capture statistical behaviors of wireless channels for different scenarios. They thus describe channels in a particular type of environment, not for a particular location. 
A main strength of statistical channel modeling is the low computational complexity which allows fast channel model construction based on key channel statistics, and thus fast system simulations. For this reason, they are popular for system design and testing, and have been used in the majority of standardized channel models. 

Stochastic models are generally categorized into geometry-based stochastic channel models (these have been originally abbreviated GSCMs, though the acronym GBSMs is also in use) and nongeometrical stochastic channel models (NGSMs). 
GSCMs have a geometrical component, while NGSMs follow a completely stochastic manner.
The idea of GSCM, first suggested in the 1990s by several groups independently~\cite{blanz1998flexibly,petrus2002geometrical,fuhl1998unified,norklit1998diffuse}, has some similarities to deterministic models, which are also based on geometry. The difference is that deterministic models prescribe scatterer locations based on an environmental database, while in GSCM, scatterer locations are chosen in a stochastic fashion according to a certain probability distribution~\cite{molisch2003geometry}. The similarity is that after the placement of scatterers, wave propagation is captured by applying the fundamental laws of specular reflection, diffraction, and scattering as in RT, though in practice the interaction processes are significantly simplified, with only a subset (such as first- and second-order specular or point scattering) taken into account.

In contrast, NGSMs, also named parametric stochastic models (PSMs), are purely stochastic models. They describe and determine parameters such as direction of departure (DoD), direction of arrival (DoA), and delay, by prescribing underlying probability distribution functions, without taking into account the underlying propagation environment. NGSMs only define paths between transceivers, and aim to get the statistical properties of the parameters in a given channel response by using measurements or ray-tracing. NGSMs are popular in channel modeling due to their simple structure, and thus low computational complexity. However, they have difficulty describing complex relationships between parameters, in particular spatial consistency, and the relationship between the temporal changes of DoDs, DoAs, and delays as a device is moving over larger distances. 

The wideband characteristics of the channel can be described by tapped delay line formulas. To support MIMO systems as well as novel antenna concepts in THz communications, not only amplitude, phase and temporal, but also spatial/directional channel information need to be considered~\cite{priebe2013stochastic}. Therefore, the spatial-extended approach specifies statistical distributions on the multi-path parameters including DoD, DoA, time of arrival (ToA), and complex amplitudes\cite{han2018propagation}.

Due to the large feasible bandwidth at THz frequencies, statistical models need to take into account the fact that MPCs may be resolvable, and thus need to be described in a different manner than the fading models like Rayleigh or Rician commonly used at lower bandwidth; this aspect bears some similarity to ultra-wideband channels that have been explored in the 3-10 GHz range \cite{molisch2005ultrawideband}, though it must be noted that the {\em relative} bandwidth in the THz band is typically small due to the high carrier frequency. 

Stochastic modeling methods can also be divided into physical and analytical models. Physical channel models characterize the statistics of the double-directional channel characteristics, such as power delay profile, arrival time and angle distribution, which are all independent of the antenna characteristics. In contrast, analytical models directly characterize the impulse response of the channel and antenna characteristics in a mathematical way without explicitly accounting for wave propagation~\cite{almers2007survey, imoize2020standard }.

\input{v1/modeling_NGSM}

\input{v1/modeling_Analytical}

%% file: v1/modeling_NGSM.tex
\subsubsection{Physical Model}

The Saleh-Valenzuela (S-V) model is based on the observation that multipath components arrive in clusters. Specifically,  the classical S-V model assumes that ToA of clusters and of MPCs within a cluster, follow Poisson processes with different densities, while the power of the clusters, and of the MPCs within a cluster, are exponential functions of the delay (again, with different decay time constants)~\cite{saleh1987statistical}. Early works on statistical channel modeling for the mmWave or THz band concentrate on calibrating and extending the S-V model~\cite{gustafson2014on, park1998analysis, kunisch1999median}. For instance, to incorporate the spatial (angular) domain, the DoA and DoD must be specified. In the spatially-extended S-V model, a zero-mean second-order Gaussian Mixture model (GMM) has been identified as a good approximation for the DoA/DoD distributions in the THz band~\cite{choi2013geometric, priebe2011AoA}.
Besides, other research works retain the idea of cluster while adopting different distributions rather than the Poisson process to characterize ToA for better agreement with measurement results~\cite{priebe2013stochastic, samimi20153D, smulders2009statistical, akdeniz2014millimeter, azzaoui2010statistical}. A statistical channel model at 0.3~THz is developed in~\cite{priebe2013stochastic}.

Another NGSM is the Zwick model~\cite{zwick2000stochastic,zwick2002stochastic}, which is described by MPCs rather than clusters, and omits amplitude fading. Designed for indoor scenarios, the model characterizes each MPC by its loss, delay, and DoA/DoD, and models the appearance and disappearance of MPCs over time as a birth and death process, a marked Poisson process~\cite{zwick2000stochastic}. Taking the direct connection from Tx to Rx as a reference for the angles, the mean angular power distributions become independent of the actual geometry~\cite{zwick2000stochastic}. Therefore, by using different local spherical coordinate systems, a fully 3-D non-geometrical modeling approach can be achieved. The original Zwick model is improved to include the application of the model to MIMO systems in~\cite{zwick2002stochastic}.

%% file: v1/modeling_Analytical.tex
\subsubsection{Analytical Model}
Compared with physical models, analytical models subsume the channel and antenna characteristics and thus describe the impulse responses from the antenna connector of a Tx antenna element to the antenna connector at an Rx antenna element (henceforth called a ``sub-channel''. It then arranges these individual impulse responses in a matrix, and then describes the statistical properties, including correlations, of those matrix elements.

Unlike the independent and identically distributed (i.i.d.) Rayleigh fading channel model, which was widely used in the early days of MIMO research, the Kronecker-based stochastic model (KBSM) is based on correlation properties between sub-channels. As simplification, the Kronecker-based model assumes that the correlation at the transmit and receive arrays are separable, which, however, becomes less valid with increasing number of antennas and dominance of single-reflection propagation in the THz band~\cite{svantesson2003tests}. Due to the expansion of the antenna array, the physical antenna becomes too large to be a point source. Therefore, the far-field assumption fails and the assumption of separate correlation is no longer valid~\cite{oestges2006validity}.

Other models consider massive MIMO channels in beam- or eigen- spaces.
For instance, the virtual channel representation (VCR) characterizes the physical propagation by sampling rays in a beam space~\cite{sayeed2002deconstructing,you2017BDMA}.
However, the model is valid only for uniform linear arrays and requires modification for other array configurations~\cite{han2018propagation}.
To improve the modeling accuracy of MIMO channels, the Weichselberger model was proposed in 2006~\cite{weichselberger2006stochastic} as a generalization of the Kronecker and VCR models. It decorrelates the channel coefficients in the eigen space (rather than the beam space) and establishes a more generic framework of UM-MIMO channels~\cite{han2018propagation}.

The models mentioned above (Kronecker, VCR, Weichselberger) can be generally categorized into correlation-based stochastic models (CBSMs). CBSMs can be used to evaluate the performance of massive MIMO systems due to their low complexity at the cost of spatial determinism capability~\cite{wang2018survey}.
However, the aforementioned conventional CBSMs did not consider the near-field effect and non-stationarity, making them not suitable for massive MIMO channel modeling~\cite{imoize2020standard} in the THz band. To utilize these models in UM-MIMO scenarios in the THz band, \cite{sun2015beam} considered dropping the far-field assumption and proposed to use a beam-domain channel model (BDCM). Furthermore, \cite{imoize2020standard} describes a BDCM that includes the near-field effect, spherical wavefront and space-time non-stationarity.

%% file: v1/modeling_Hybrid.tex
\subsection{Hybrid}
\label{sec:hybrid}
As described above, deterministic channel modeling shows high accuracy with high time and resource consumption, while statistical channel modeling benefits from low computational complexity at the cost of accuracy. Therefore, an interesting and promising trend, is to develop hybrid methods by combining the benefits from two or more individual approaches, as shown in Fig.~\ref{fig:axis_channel_modeling_methods}.
Following this path, several attempts on hybrid channel modeling methods have already been presented~\cite{priebe2011AoA, wang2000hybrid, reynaud2006hybrid, thiel2008hybrid, molisch2002virtual} and channel models based on hybrid modeling methods are also proposed~\cite{nurmela2015deliverable, lecci2020quasi, weiler2016quasi, maltsev2014Quasi, liu2012cost, oestges2012pervasive, zhu20193gpp}.
Next, we introduce several hybrid channel modeling methods. Some of them combine two deterministic approaches (i.e., RT and FDTD) and are categorized by hybrid deterministic (RT-FDTD) approach, while others combine deterministic and stochastic approaches.

\subsubsection{Hybrid deterministic / RT-FDTD Approach}

\begin{figure}
    \centering
    \includegraphics[width=0.9\linewidth]{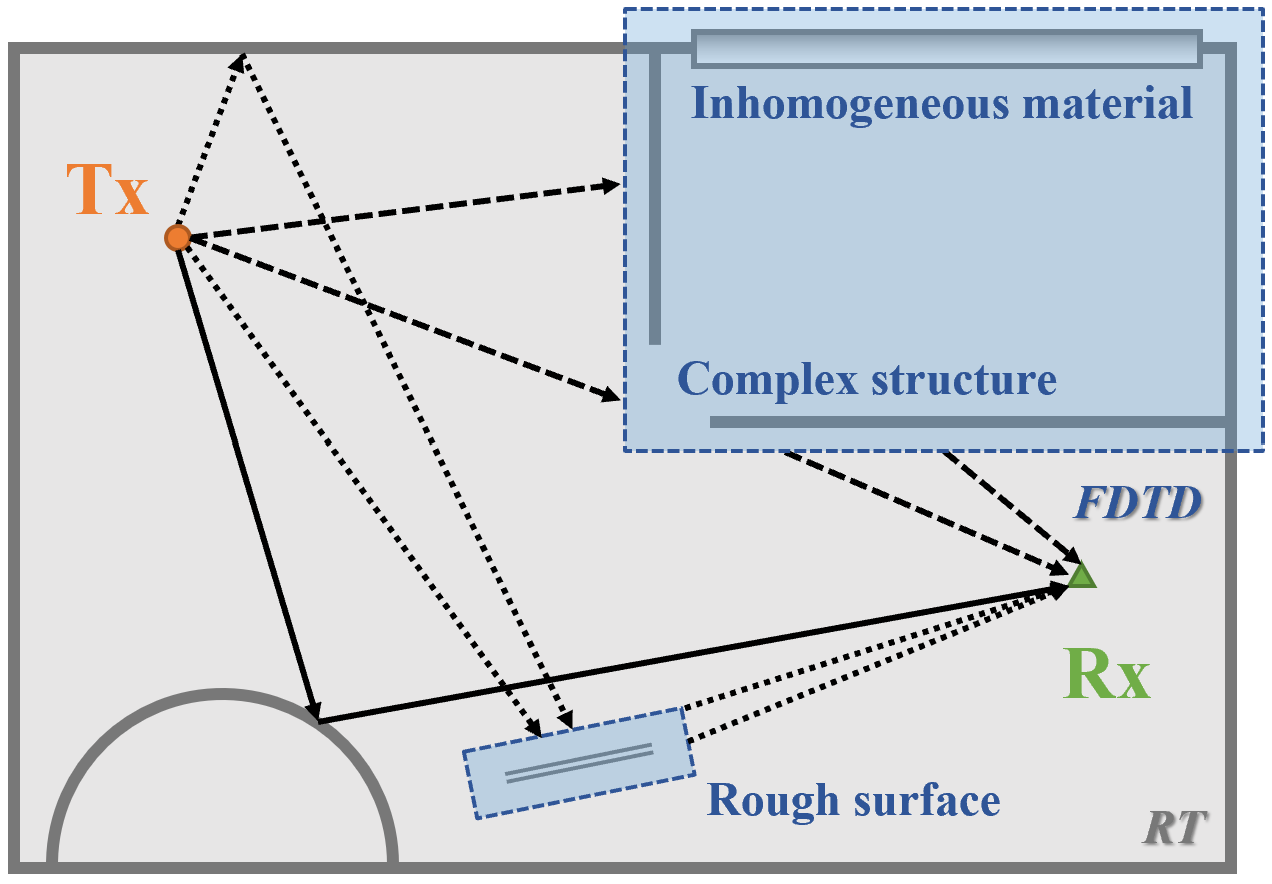}
    \caption{Ray-tracing and FDTD hybrid approach.}
    \label{fig:RT-FDTD}
\end{figure}

In the RT-FDTD hybrid modeling, FDTD is used to study regions close to complex discontinuities where ray-based solutions are not sufficiently accurate and RT is used to trace the rays outside of the FDTD regions, as shown in Fig.~\ref{fig:RT-FDTD}.
At higher frequencies, the surface of the objects becomes relatively rough and thus it becomes harder for geometrical optics, as RT, to approximate the properties of reflection and scattering. As introduced before, FDTD can resolve small and complex scatterers and rough surfaces in the THz band, whereas it suffers from high computational costs. The RT-FDTD hybrid technique can retain accuracy while improving computation time by applying FDTD only in a small portion of the entire modeling environment, namely the near field of the scatterers, and leaving the rest for RT~\cite{wang2000hybrid}. This hybrid method is suitable for mmWave and THz channels, and is also implemented in some commercial EM solvers such as High Frequency Structure Simulator (HFSS).
However, a critical challenge for this method is to smoothly transit between RT and FDTD methods and computation of the boundary results.
Ref.~\cite{wang2000hybrid} studies single interactions between ray-tracing and FDTD where the location of the Rx is restricted in the FDTD region, for example the corner of the room. The proposed hybrid approach is then extended for time-efficiency and multiple interactions between ray-tracing and FDTD~\cite{reynaud2006hybrid, thiel2008hybrid}.

\subsubsection{Deterministic-Stochastic Hybrid Approach}

While statistical channel models offer high efficiency, they cannot easily reproduce spatial consistency and the temporal evolution of cluster correlations. A variety of models thus exist that combine statistical and geometrical modeling approaches. This allows to provide important features of the channel model that cannot be obtained from a purely stochastic model. First, statistical impulse response channel models fail to naturally capture the correlation between links in a multi-user communication system since the random generation for each link cannot guarantee all generated links in the same physical environment. Moreover, the rigid structure of the approach does not naturally support continuous channel descriptions over intervals larger than the stationarity distance, hence hampering the simulation of a large movement of the mobile station. Similarly, the inherent link between changes in angle and directions that follows from geometrical considerations is not easy to reproduce in statistical channel models. For all these reasons, hybrids between geometric and stochastic approaches are useful. In the following, we discuss a variety of such hybrid models, in descending order of their geometric contribution.

A quasi-deterministic (Q-D) channel model determines dominant MPCs from a strongly simplified, environmental map, and then adds clusters of stochastically modeled MPCs. These clusters are associated with the dominant MPCs\footnote{Alternatively, the ray-tracing can provide the mathematical center of the cluster, without necessarily putting a dominant MPC there.}, and thus have similar DoAs, DoDs, and delays, though with some small spread around them. Further clusters of MPCs corresponding to small and possibly mobile scatterers can be added as well. The Q-D approach was first suggested in the early 2000s \cite{kunisch2003ultra,molisch2002virtual} for cmWave channels. 
In the mmWave band, the 801.11.ad channel model has adopted the hybrid modeling concept for indoor channels at 60~GHz~\cite{maltsev2010statistical}. The Q-D approach is also applied by the MmWave Evolution for Backhaul and Access (MiWEBA) model~\cite{maltsev2014Quasi, weiler2016quasi} and the IEEE 802.11ay channel model~\cite{maltsev2016channel}, which is an extension of IEEE 802.11ad channel model.

In~\cite{choi2013geometric, priebe2011AoA}, the authors implement this hybrid approach in an indoor channel at 300 GHz with GMM distribution for generating DoAs of intra-cluster rays.
In~\cite{chen2021channel}, \textit{Chen et al.} develop a semi-deterministic channel model for indoor THz channels at 140~GHz through the hybrid SRH modeling method, with Von Mises distributions for generating the azimuth AoA of both inter-cluster MPCs and intra-cluster subpaths. Besides, the arrivals of inter-cluster MPCs and intra-cluster MPCs are modeled by a Poisson process.

GSCMs can also be seen as a hybrid of stochastic and deterministic approaches: while the placement of the scatterers is stochastic, the simplified ray-tracing is deterministic. Designing details of GSCMs include the PDFs of the scatterer placement that give information about the tested propagation scenario, and parameters such as the number of scatterers and cluster assignment. 
On the one hand, GSCMs can be classified into regular-shaped GSCMs and irregular-shaped GSCMs based on the shape of scatterer placement. Regular-shaped GSCMs assume effective scatterers to be located on regular shapes (one-ring, two-ring, ellipses, cylinders, etc.), while irregular-shaped GSCMs assume effective scatterers to be located on irregular shapes. The interaction processes can be restricted to single scattering, but more general approaches that involve multiple reflections and waveguiding are required for analyzing MIMO systems \cite{molisch2004generic}. 
Recently, \cite{bian2021general} derived the correlation functions of a general 3D space-time-frequency (STF) non-stationary GSCM called beyond 5G channel model (B5GCM). Regular-shaped GSCMs are mainly used for theoretical analysis of, e.g., correlation functions, while irregular-shaped GSCMs can better reproduce measured results. Especially notable in this category are the COST259/273/2100 models~\cite{molisch2006cost259,liu2012cost}, which places clusters of scatterers, resulting in clusters of MPCs with similar delays and directions. A set of clusters with consistent stochastic parameters are first generated throughout the simulation environment based on the location of the base station (BS). With this given geometrical cluster distribution, Large-scale parameters (LSPs) of a channel are actually controlled by the clusters that are visible to the MS (i.e., clusters that contribute to the channel). Therefore, after defining the mobile station (MS) location, the scattering from visible clusters is determined and LSPs can be synthesized at each channel instance. The coupling between angles and delays of the clusters, as well as the LSP evolution when the MS moves over large distances, and the correlation between channels of different users are all inherently provided by the cluster location and visibility information. 

Another widely used approach is the 3GPP Spatial Channel Model (SCM)~\cite{3GPP-SCM} and WINNER II model~\cite{meinila2009WINNER, Kyosti2008WINNER}. While these are called ``geometry-based stochastic models'' in the standardization documents, the contribution of the geometry is small: namely, a ``drop'' (placement) of the MS in a cell-centered on a BS determines the run length and angle of the LoS component, relative to which all other angles and delays are defined. However, the model then proceeds to prescribe the deviations of MPC parameters (delays, DoA, DoD, complex amplitudes) in a purely stochastic manner.
LSPs are first generated from their stochastic distribution. AoAs, AoDs, delays, and amplitudes of the MPCs in the various clusters are then generated according to these LSPs. Note that in most versions of the 3GPP and Winner models, different ``drops'' have uncorrelated statistical realizations of their LSPs. Thus, correlations between different drops are not taken into account; at the same time, the LSPs, once they are chosen, do not change with the movement of the MS (though a recent addition to the 3GPP model provides a partial enabling of such spatial consistency). 
After determining LSPs, small-scale parameters (SSPs) are drawn randomly based on tabulated distribution functions and LSPs. In other words, LSPs are used as control parameters when generating SSPs. In the last step, initial phases are randomly assigned and channel coefficients are obtained. 

Even though the 3GPP model claims validity up to 100~GHz~\cite{3gpp.38.901}, it is based only on a small number of measurements $>6$~GHz, and in the interest of backward compatibility with the $<6$~GHz models omits a number of effects, that are important for mmWave and THz channels.
For example, 3GPP and WINNER SCMs consider that one temporal cluster contains only one spatial cluster, i.e., two separated spatial clusters cannot occur at the same delay. However, measurements in the mmWave and THz bands indicate that a temporal cluster may contain several spatial lobes at the Rx, i.e., clusters may come from diverse angles almost simultaneously~\cite{samimi20153D,rappaport2013millimeter,Yu2020Wideband}, which is an important characteristic for multiple antenna systems applied at higher frequencies.
Therefore, a number of other works propose models that draw on the philosophy of 3GPP models, with some possible modifications. Refs.~\cite{hur2016proposal,sun2017novel} provide 3GPP-type models that are parameterized from extensive ray-tracing results. \textit{Samimi et al.} use temporal clusters and spatial lobes to deal with the temporal and spatial components respectively, where a spatial lobe contains more than one traveling cluster with different arrival time~\cite{samimi2014ultra}, and created a simulator parameterized from measurements at New York University (NYU) (see Section~\ref{sec:simulator}). 
Studies of the family include the models suggested in 3GPP TR 38.901~\cite{3gpp.38.901}, mmMAGIC~\cite{haneda2017measurement}, METIS\cite{nurmela2015deliverable}, 5GCMSIG~\cite{haneda20165g,haneda2016indoor}, and QuaDRiGa~\cite{jaeckel2014QuaDRiGa,QuaDriGa}. 
All of these models are derived for mmWave frequencies, with the exception of \cite{ju2021millimeter}, which parameterizes an NYUWireless model structure based on measurements at 140 GHz.

%% file: v1/simulator.tex
\section{THz Channel Simulator} \label{sec:simulator}

\begin{table*}[htbp]
    \centering
    \caption{State-of-the-art THz channel simulators (non-exhaustive).}
    \includegraphics[width = 0.8\linewidth]{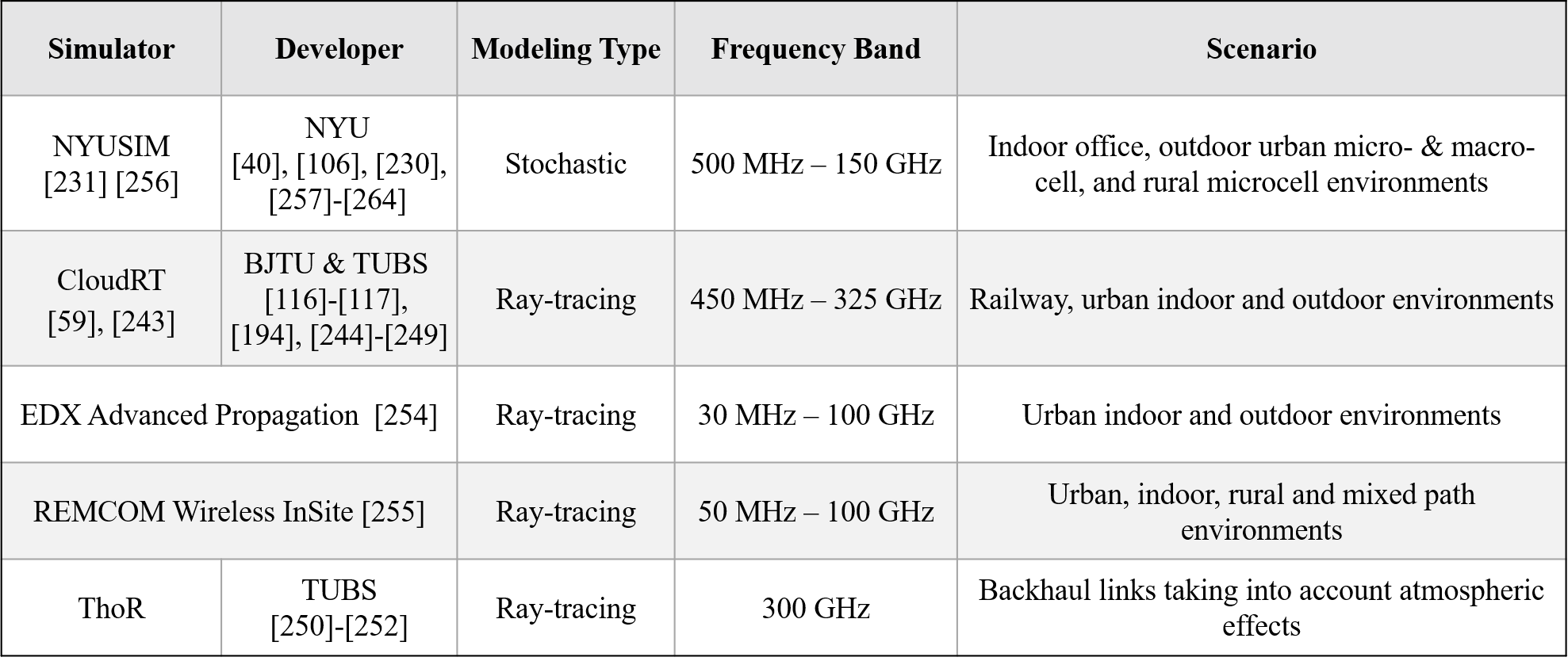}
    \label{tab:channel_simulator}
\end{table*}
	
THz channel simulators are developed based on channel models and measurements described in Section~\ref{sec:measurement}, Section~\ref{sec:modeling} in order to efficiently reproduce channel characteristics without repeating channel measurements~\cite{sun2018propagation}.
To be specific, numerous different scenarios can be generated for further system evaluations based on validated channel models without carrying out additional expensive and time-consuming measurements~\cite{he2018channel,priebe2013ultra,loredo2008indoor}. For this purpose also stochastic channel models are frequently applied~\cite{priebe2013stochastic,ieee802153d_CMD}.  Similar to the channel models, also channel simulators can be classified as deterministic, geometry-based stochastic, and purely stochastic. 

In the deterministic category, an RT model with a defined environmental database, antennas, and trajectory of Tx, Rx, and scatterers can be seen as a channel simulator. A widely used one based on this approach, called CloudRT~\cite{wang2016accelerated,he2019design}, is jointly developed by BJTU and TU Braunschweig~\cite{guan2020channel,guan2019measurement,ma2020characterization}. It integrates a V2V RT simulator~\cite{guan2013deterministic,nuckelt2013comparison} validated by measurements in~\cite{guan2013deterministic, abbas2015simulation} and a ultra-wideband THz RT simulator~\cite{priebe2013stochastic} validated by measurement results in~\cite{priebe2013stochastic,he2017stochastic}. In particular, CloudRT is validated to support railway, urban indoor and outdoor environments at frequencies from 450~MHz to 325~GHz~\cite{he2019design,guan2018towards,abbas2015simulation,priebe2013stochastic,guan2013deterministic,he2017stochastic,guan2019measurement}. More information is available at \textit{http://www.raytracer.cloud/}.

Recently, the Horizon 2020 ThoR project (\textit{www.thorproject.eu}) has proposed an automatic planning algorithm for 300~GHz backhaul links. This has been applied and analyzed in the Hannover scenario including BSs with wireless and fiber backhaul links, taking atmospheric effects into account~\cite{jung2019simulation,jung2021link,JungEuCAP2021}.
Besides, \textit{Peng et al.} have developed and used a new broadband ray-launching-based channel simulator~\cite{peng2020electromagnetic}. It is based on the deterministic modeling by ray-launching as introduced in~\cite{rose2014analytical}, and can generate deterministic channel models according to realistic scenario descriptions.

In addition to simulators developed by certain universities, off-the-shelf commercial softwares are also available. EDX Advanced Propagation~\cite{EDX}, developed by EDX Wireless, supports urban indoor and outdoor environments at frequencies from 30~MHz to 100~GHz, while Wireless InSite~\cite{InSite}, developed by Remcom supports urban, indoor, rural and mixed path environments at frequencies from 50~MHz to 100~GHz.

For geometry-based stochastic and purely stochastic models, the most widely used academic simulator is Quadriga \cite{QuaDriGa}, created by the Heinrich Hertz institute in Berlin. It contains a variety of modules that allow the creation of purely stochastic models such as 3GPP channel models (compliant with the specification TS 38.901), as well as map-based models and geometry-based stochastic models. It has been parameterized based on measurements of up to 80 GHz. More information can be found at 
\textit {quadriga-channel-model.de}.

Another widely used model for mmWave and THz channels, created by NYU based on their measurements is called NYUSIM~\cite{sun2017novel,NYUSIM150}. The simulator is first developed based on extensive real-world measurements at multiple mmWave frequencies from 28 to 73~GHz. Extensions based on measurements up to 150~GHz are currently underway~\cite{ju2021millimeter,xing2021millimeter}. It now supports indoor office, outdoor scenarios in urban microcell (UMi), urban macrocell (UMa), and rural macrocell (RMa) environments~\cite{sun2017novel,rappaport2013millimeter,rappaport2015wideband,samimi20163D,samimi2015local,sun2016investigation,sun2015synthesizing,maccartney2015indoor,maccartney2016millimeter,maccartney2017study}. More information is available at \textit{https://wireless.engineering.nyu.edu/nyusim/}. Extensions based on measurements at 140~GHz are currently underway; at the time of this writing, an indoor environment has been parameterized~\cite{ju2021millimeter}.

A non-exhaustive collection of state-of-the-art THz channel simulators, both academic and commercial, are summarized in Table~\ref{tab:channel_simulator}.

%% file: v1/characterization.tex
\section{Channel Characterization}  \label{sec:characterization}
\par We now turn to the numerical values of the key channel characteristics that are required for system evaluations. 
For example, the knowledge of large-scale fading such as path loss and shadowing helps to calculate the link budget in order to derive the required SNR. In addition, to evaluate the effect of the inter-symbol interference (ISI), the root mean square (RMS) delay spread and/or coherence bandwidth are essential. Furthermore, non-stationary features in the temporal, frequency, and spatial domains shed light on possible design choices for physical-layer communications, including waveform and antenna array design, among others. Besides being of value by themselves, all these channel characteristics can also help in the parameterization of the more detailed channel models.
This section elaborates the behavior and numerical values of channel properties in the THz band mainly based on measurement results reported in the literature. Simulation results based on deterministic simulators like ray-tracing are only included when measurement data is not sufficient to draw convincing observations.
\subsection{Large-scale and Small-scale Channel Characteristics} \label{channelchar}
\begin{table}[tbp]
    \centering
    \caption{Set-ups of measurement campaigns in the THz band in indoor scenarios.}
    \includegraphics[width =\linewidth]{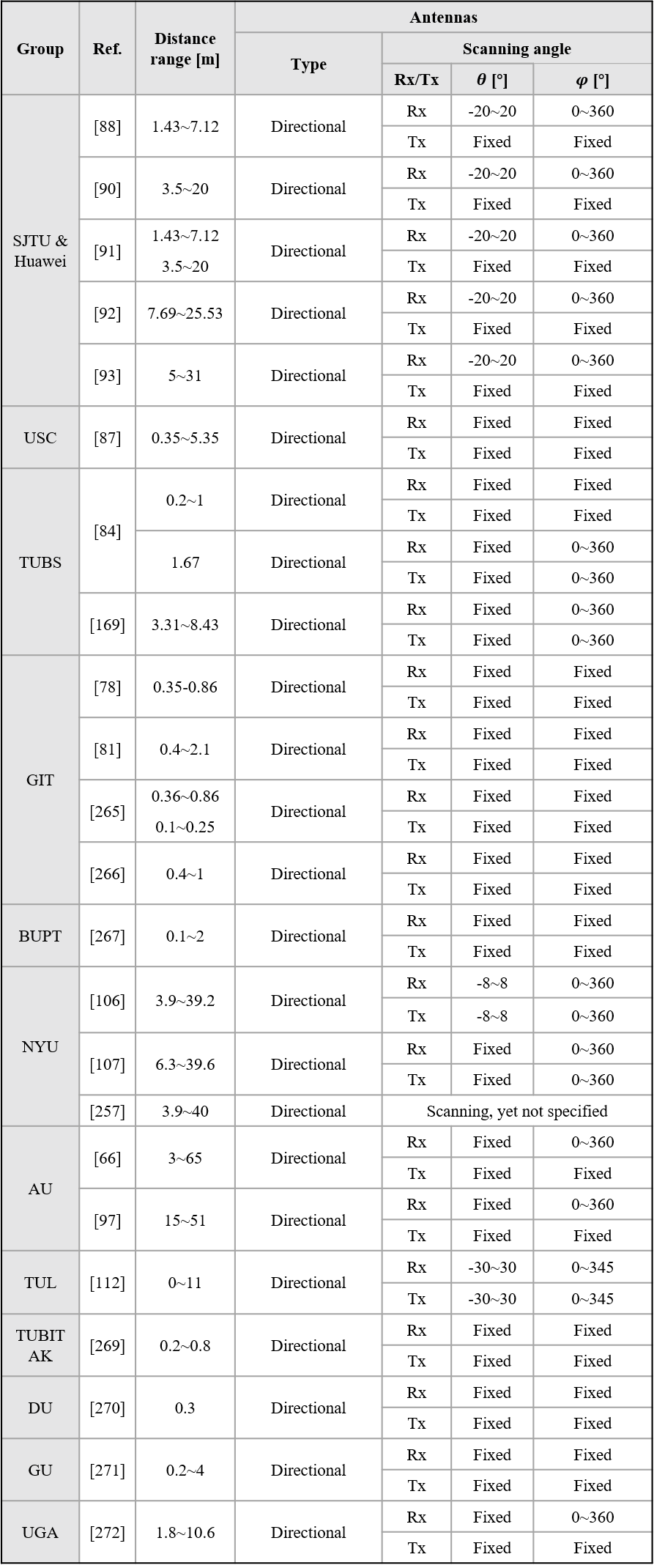}
    \label{tab:indoor_results}
\end{table}
\begin{table}[tbp]
    \centering
    \caption{Set-ups of measurement campaigns in the THz band in outdoor scenarios.}
    \includegraphics[width =\linewidth]{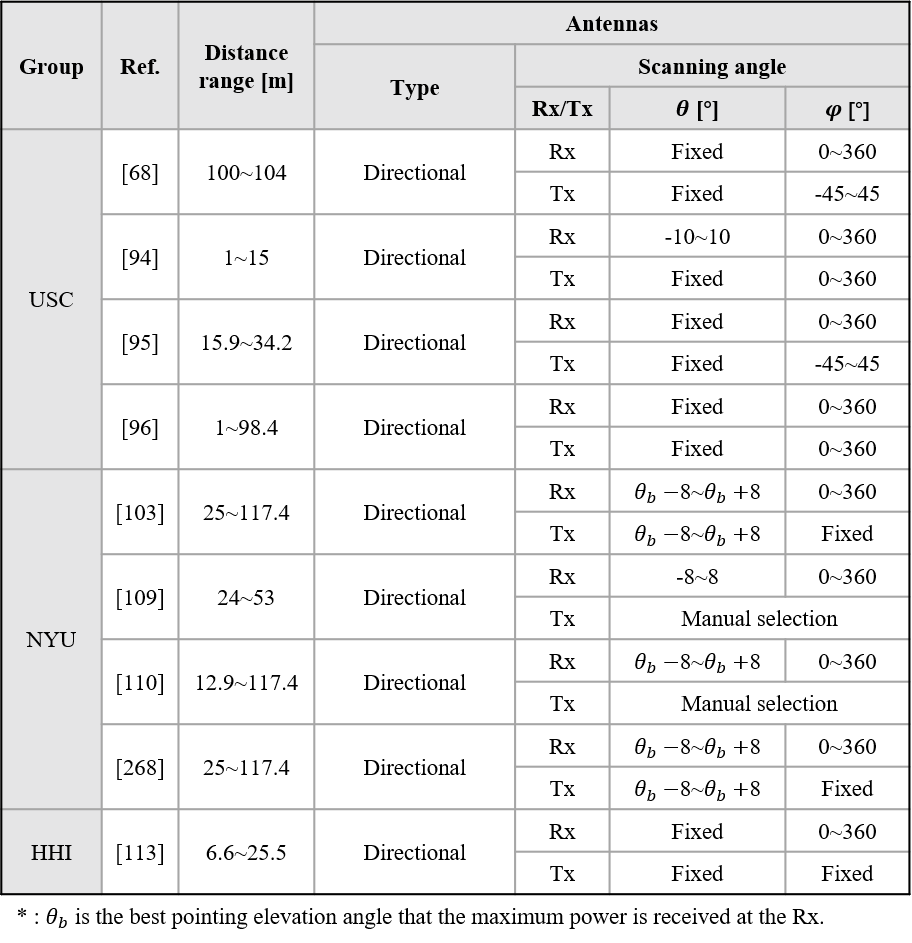}
    \label{tab:outdoor_results}
\end{table}
We start with the analysis of the key channel characteristics, including the large-scale fading due to path loss and shadow fading as well as the small-scale fading due to multi-path effects, such as delay spread (DS), angular spread (AS), etc., for indoor and outdoor communications systems. The measurement setups, including the measurement distance range and the antenna configurations, of existing studies in both indoor and outdoor scenarios are summarized in Table~\ref{tab:indoor_results} and~\ref{tab:outdoor_results}, from different groups including Shanghai Jiao Tong University collabrated with Huawei (SJTU $\&$ Huawei)~\cite{chen2021channel,li2022channel,wang2022thz,he2021channel,chen2021140}, University of Southern California (USC)~\cite{abbasi2020channel,abbasi2021double,abbasi2021thz,abbasi2021ultra,abbasi2020double}, Technische Universität Braunschweig (TUBS)~\cite{priebe2011channel,priebe2013ultra}, Georgia Institute of Techonology (GIT)~\cite{kim2015d,Cheng2020Characterization,cheng2017comparison,cheng2019thz}, Beijing University of Post and Telecommunications (BUPT)~\cite{tang2021channel}, New York University (NYU)~\cite{ju2021millimeter,xing2021millimeter,ju2021140,xing2021propagation,xing2021millimeter_letter,ju2021sub,ju2022sub}, Aalto University (AU)~\cite{Nguyen2018Comparing,Nguyen2021Large}, Technische Universität Ilmenau (TUL)~\cite{dupleich2020characterization}, Türkiye Bilimsel ve Teknolojik Araştırma Kurumu (TÜBITAK)~\cite{ekti2017statistical}, Durham University (DU)~\cite{raimundo2018channel}, Ghent University (GU)~\cite{de2021directional}, University Grenoble-Alpes 
 (UGA)~\cite{pometcu2018large}, Heinrich Hertz Institute (HHI)~\cite{undi2021angle}. The channel characteristics established in these studies are discussed in the following parts.
\subsubsection{Path Loss and Shadow Fading} 
The large-scale characteristics, including path loss and shadow fading, characterize the variation of the received power as the position of the Tx, the Rx, or both, change. Specifically, path loss refers to the power ratio between the transmitted signal and the received signal averaged over an area that contains a large number of realizations of both the small-scale and large-scale fading. It is often modeled as a function of only the (Euclidean) distance between Tx and Rx, though this need not be the best, or most physically reasonable, model, as discussed below. 
The shadow fading, which is often due to blockage of objects, describes the variations of the received power (when averaged over the small-scale fading) around the path loss. 

The path loss can be caused by one, or multiple, physical effects that affect the different multipath components propagating from the Tx to the Rx. These effects, such as free-space path loss, reflection loss, diffuse scattering and diffraction, are all frequency-dependent, and might lead to higher path loss at THz. Besides these, from the wave propagation perspective, the atmospheric attenuation and weather effects might also be different in the THz band due to the increase of frequency.

 \begin{figure}
    \centering
    \includegraphics[width=0.8\columnwidth]{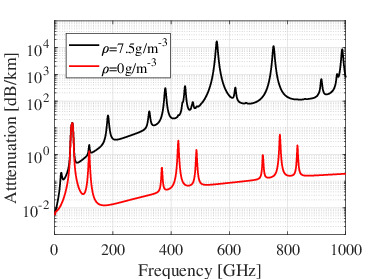}
    \caption{The specific attenuation due to atmospheric gases.}
    \label{fig:molecular}
\end{figure}
\par The atmospheric attenuation in the THz band can be calculated using the HITRAN (HIgh resolution TRANsmission molecular absorption database)~\cite{jornet2011channel} or using the recommendation ITU-R P. 676~\cite{ITU-RP676} from International Telecommunication Union (ITU). Since the ITU-R P.676 is easier to implement and understand compared to HITRAN database, we choose ITU-R P.676 here to discuss the influence of atmospheric attenuation. According to  ITU-R P. 676~\cite{ITU-RP676}, the atmospheric attenuation for frequencies up to 1000~GHz can be approximately calculated as a summation of individual spectral lines from oxygen and water vapor, together with other additional factors. Using the formula and spectral lines provided by ITU-R P. 676, the attenuation can be evaluated at any value of pressure, temperature, and humidity. For instance, the atmospheric attenuation values, for a pressure of 1013.25~hPa, temperature of 15~$^\circ$C, water vapor density of 7.5~g/$\text{m}^3$ for standard case and 0~g/$\text{m}^3$ for dry air, are shown in Fig.~\ref{fig:molecular}, where we draw the observations as follows. First, atmospheric attenuation in the THz band is more noticeable than those in the microwave or mmWave band. Second, there are several attenuation peaks, e.g., $f=60$~GHz, 120~GHz, 183~GHz, 325~GHz, 380~GHz, etc., which should be avoided when designing communication systems for medium or longer distances\footnote{Those bands can be advantageous for extremely short-range communications as interference from farther-away emitters is effectively suppressed}. Instead, the formed spectral windows between attenuation peaks are more advantageous for this purpose. Furthermore, for frequencies up to 300~GHz, the atmospheric attenuation is lower than 10~dB per kilometer, which justifies that current THz channel measurements (within 200~m) omit consideration of the atmospheric attenuation. However, for higher frequencies, especially for those above 600~GHz, this effect is much more significant and thus needs to be appropriately considered even for lower distances.
\par Apart from the standard atmospheric attenuation, THz waves may suffer from attenuation related to specific weather conditions, such as rain, fog, snow, sand, etc~\cite{siles2015atmospheric,moon2015long}. Many efforts have been taken to investigate the propagation characteristics of THz waves in rain and fog, either from theoretical analysis~\cite{ishii2016rain,norouzian2019rain,Marzuki2019char} or experiments~\cite{ishii2011measurement,norouzian2019rain,golovachev2019propagation,yang2015broadband}. These studies indicate that theoretical calculations based on Mie scattering~\cite{mie1908beitrage} and empirical formulas provided by ITU~\cite{ITU-RP838,ITU-RP840} are effective for calculating the loss values caused by fog and rain. To be specific, attenuation in rain remains nearly constant in the THz band, around \SI{20}{dB/km} at a rainfall rate of \SI{50}{mm/hr}~\cite{ishii2016rain}, since the size of rain drops is much larger than the THz wavelength. In contrast, attenuation in fog gradually increases as the frequency increases in the THz band. For example, for dense fog with visible range of \SI{10}{m}, a \SI{100}{dB/km} attenuation factor is measured at \SI{330}{GHz} in~\cite{golovachev2019propagation}, while the attenuation factor at \SI{60}{GHz} is calculated as \SI{9.4}{dB/km} by using existing empirical formulas~\cite{ITU-RP840,altshuler1984simple}. Furthermore, rare studies have been conducted for the sand attenuation in the THz band~\cite{du2017characterisation}. Moreover, several measurement campaigns have been conducted to investigate the propagation loss of THz waves during snowfall~\cite{Norouziari2018low,norouzian2019experimental}. Unlike studies under other weather conditions, no theoretical basis is provided, since the shape and size distributions of the snowflake are difficult to characterize. Nevertheless, measurement results have shown that the snow attenuation at \SI{300}{GHz} is lower than \SI{20}{dB/km} for snowfall rate lower than \SI{20}{mm/hr} at \SI{300}{GHz}~\cite{norouzian2019experimental}.
\par Generally, the atmospheric attenuation and weather effects are very important for long-range application scenarios, such as remote sensing, or satellite-to-ground communications. However, for existing studies of channel characterization in the THz band, which mostly focus on short-range terrestrial communications (normally shorter than \SI{200}{m}), these effects are typically not a major factor. As a result, the summarized path loss results in this article do not take into account atmospheric attenuation and weather effects.
\begin{figure} 
\centering  
\subfigure[Indoor scenarios] {
 \label{ple_indoor}     
\includegraphics[width=0.8\columnwidth]{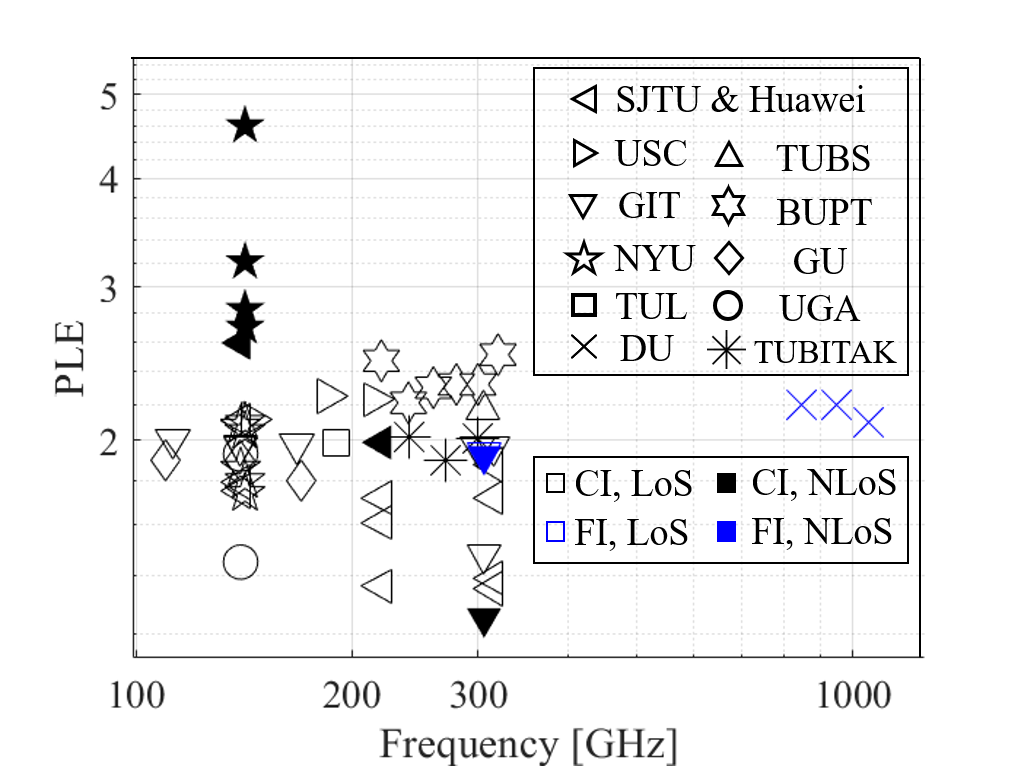}  
}     
\subfigure[Outdoor scenarios] { 
\label{ple_outdoor}     
\includegraphics[width=0.8\columnwidth]{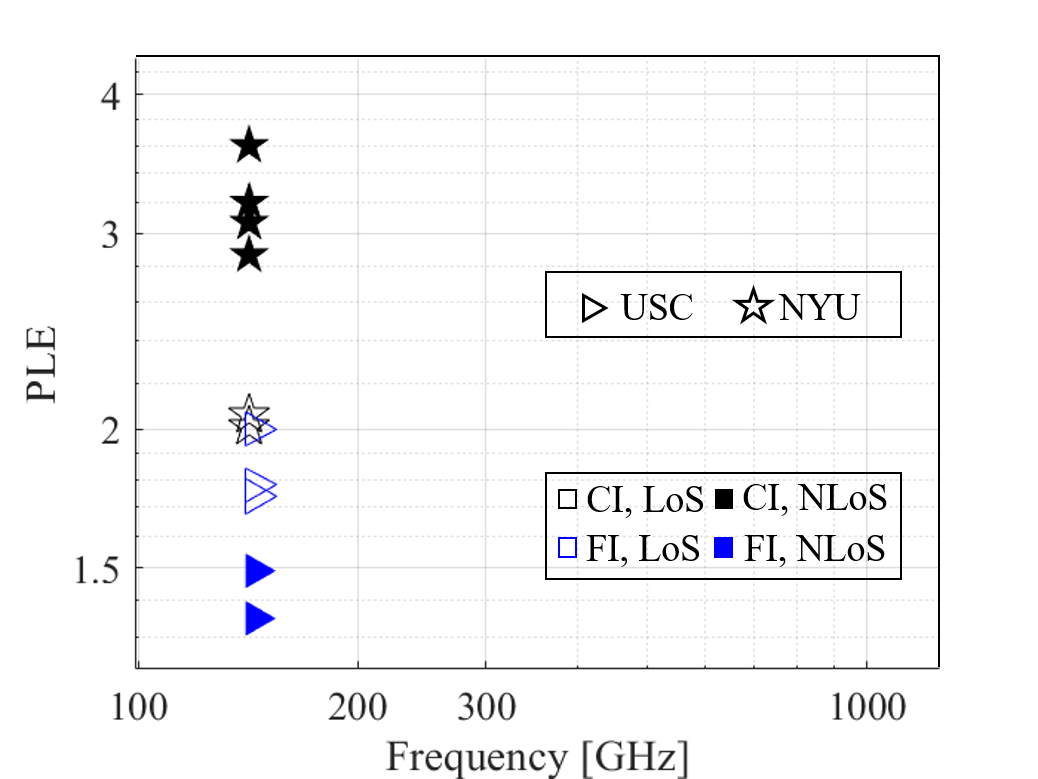}     
}
\caption{Representative results of path loss exponents in the THz band. Different types of markers represent different research organizations. Empty markers are results in the LoS case and filled markers are those in the NLoS case, respectively. Furthermore, different colors of the markers are for different linear models that are used to obtain the PLE.}
\label{fig:ple}
\end{figure}
\par Representative results for the PLEs in the THz band are shown in Fig.~\ref{fig:ple}, where the indoor scenarios include office, data center, etc.~\cite{cheng2017comparison,Cheng2020Characterization,priebe2011channel,abbasi2020channel,xing2021millimeter,ekti2017statistical,cheng2019thz,raimundo2018channel,chen2021channel,de2021directional,pometcu2018large,dupleich2020characterization,ju2021millimeter,li2022channel,wang2022thz,chen2021140,ju2022sub,tang2021channel,he2021channel,kim2015d}, and the outdoor scenarios are mainly in urban scenario~\cite{ju2021140,xing2021propagation,abbasi2021thz,abbasi2021double}, respectively.

It must be noted that two different types of such linear models are commonly used. 
The most widely used path loss models include \emph{close-in (CI)} free space reference distance model and \emph{float-intercept (FI)} or $\alpha-\beta$  model, as summarized in~\cite{rappaport2017overview,rappaport2021radio}.
The FI model has two free parameters, slope $\alpha$ and intercept $\beta$ of the linear fit, while the CI model fixes the intercept to be equal to the free-space path loss at a reference distance (often $1$~m), and thus fits only one parameter. Due to the different degrees of freedom in the fitting, the PLEs obtained from those two methods might differ, and those differences can be significant particularly for NLoS scenarios. As shown in Fig.~\ref{fig:ple}, the results using FI and CI model exhibits similar PLEs in LoS case, while the PLEs of studies based on FI model are clearly smaller than those using CI model.
It must also be noted that any model parameterized from measurements is only applicable in the range in which the measurements were taken. 

Care must be taken in the interpretation of the path loss results. Experimentalists tend to place Tx and Rx at such locations that measurable Rx power can be anticipated. Situations where, in an indoor environment, the two link ends are separated by steel-concrete walls tend to be avoided; similarly, outdoor setups generally avoid locations that are deep in a shadow of a building (without reflecting paths that could carry energy). Thus, the measured path loss is impacted by this pre-selection. Locations with very high path loss (i.e., being in an outage) may occur in THz channels with appreciable probability. Ray-tracing and other deterministic methods tend to avoid the selection bias when simulations occur for a regular grid of Tx and/or Rx locations. 

\begin{figure} 
\centering  
\subfigure[Indoor scenarios] {
 \label{sf_indoor}     
\includegraphics[width=0.8\columnwidth]{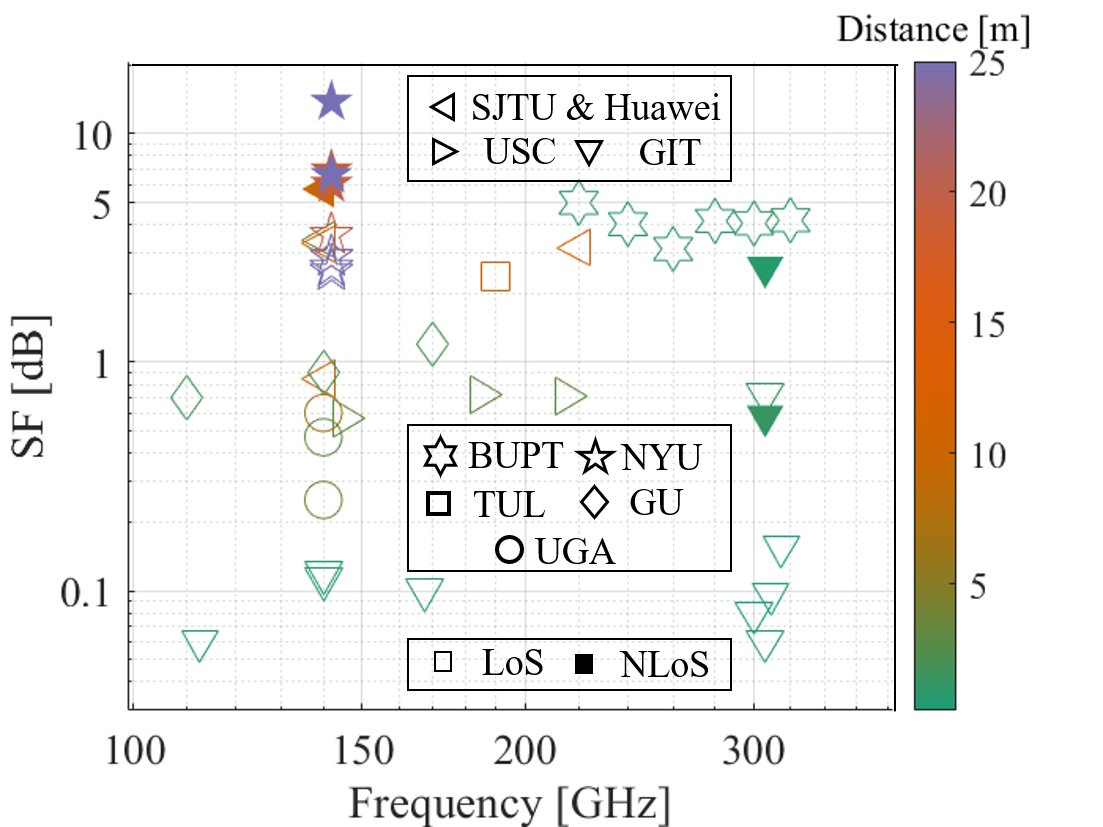}  
}     
\subfigure[Outdoor scenarios] { 
\label{sf_outdoor}     
\includegraphics[width=0.8\columnwidth]{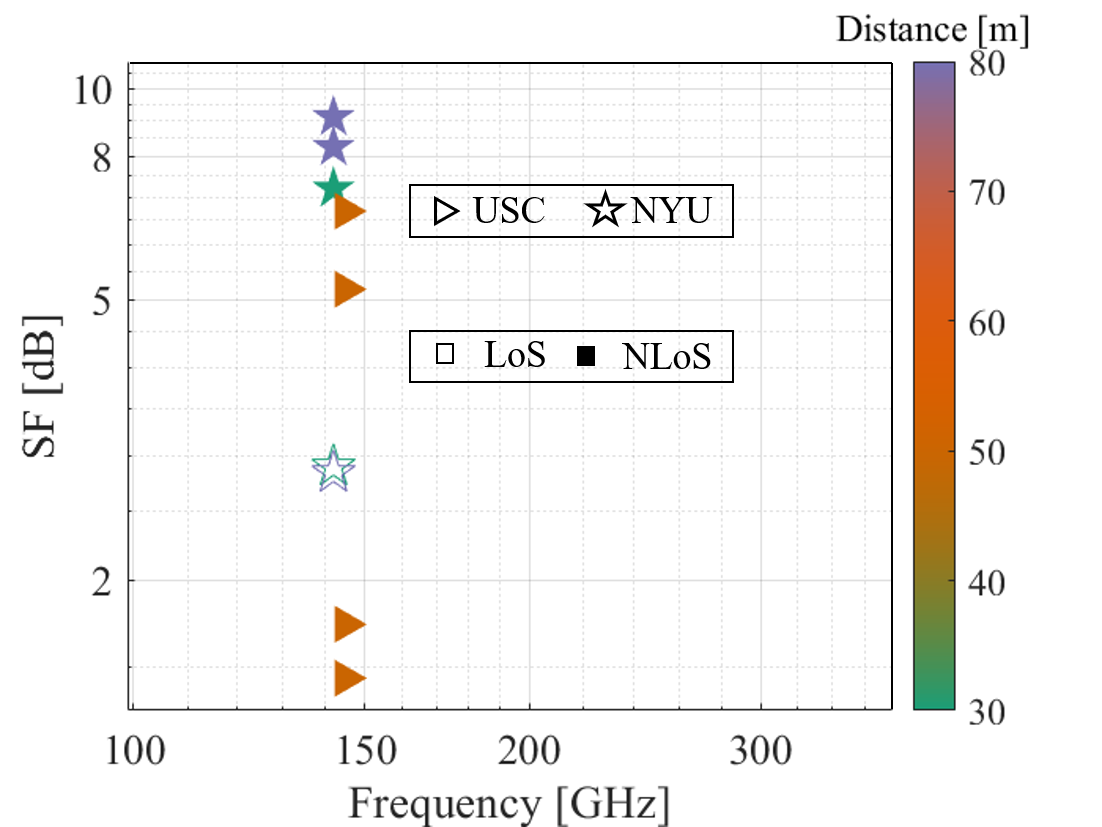}     
}
\caption{Representative results of shadow fading standard deviation in the THz band. Different types of markers represent different research organizations. Empty markers are results in the LoS case and filled markers are those in the NLoS case, respectively. Furthermore, different colors of the markers are for different linear models that are used to obtain the PLE.}
\label{fig:sfsd}
\end{figure}
\par We can draw a number of observations from Fig.~\ref{fig:ple}. Firstly, there is no significant difference in terms of the PLE values between indoor and outdoor scenarios. Second, the number of measurements in the LoS case is much larger than that in the NLoS case in indoor scenarios. In contrast, the amounts of measurements for LoS case and NLoS case in the outdoor scenarios are similar. In fact, as reported in~\cite{cheng2017comparison,Cheng2020Characterization,priebe2011channel,abbasi2020channel,xing2021millimeter,ekti2017statistical,cheng2019thz,raimundo2018channel,chen2021channel,de2021directional,pometcu2018large,dupleich2020characterization,ju2021millimeter,li2022channel,wang2022thz, ju2021140,xing2021propagation,abbasi2021thz,abbasi2021double,chen2021140,ju2022sub,tang2021channel,he2021channel}, in indoor scenarios, some of the studies only consider LoS cases, while most studies in outdoor scenarios consider both LoS and NLoS cases. This is reasonable since in outdoor scenarios, there is a larger probability that the LoS path could be blocked by objects like buildings, trees, etc.; at the same time, in outdoor scenarios there is a higher probability that reflected paths can provide significant energy to the Rx, while in indoor, NLoS often means blockage of the LoS as well as other multipath components by a wall. 
Third, PLEs in LoS cases are close to 2, which equals to the PLE for free space path loss. By contrast, larger PLE values are observed in the reported NLoS cases. Both this effect, and the observed values for PLE in LoS and NLoS are consistent with other frequency bands. 
Since diffraction can be neglected at THz frequencies~\cite{Jacob2012Diffraction}, NLoS propagation is mainly attributed to specular reflection and diffuse scattering~\cite{Ma2019}, while the contribution of the latter grows as the surface roughness intensifies.

\par The shadow fading effect is usually characterized by a zero-mean Gaussian distributed random variable, with standard deviation $\sigma_{SF}$. Representative results for $\sigma_{SF}$ are plotted in Fig.~\ref{fig:sfsd}. 
First, many $\sigma_{SF}$ values are below 1 dB in indoor scenarios, which are in general less than the ones in outdoor scenarios. This implies the increasing probability of blockage in outdoor cases. Besides, the $\sigma_{SF}$ values are clearly larger for NLoS cases compared to those in LoS cases, which is reasonable as the power in NLoS would fluctuate more due to blockage of objects without a dominating LoS path. Furthermore, the $\sigma_{SF}$ increases as distance increases in both indoor and outdoor scenarios, which is consistent with the conjecture that shadowing is more severe for larger communication distance. A physical interpretation of this effect and sample results in the mmWave range is given in \cite{karttunen2017spatially}. Moreover, $\sigma_{SF}$ varies substantially for both indoor and outdoor scenarios, ranging from nearly zero to over ten~dB. This can be explained by the fact that the shadow fading effect is strongly dependent on the detailed geometry, resulting in different $\sigma_{SF}$ values. We also refer to our previous discussion of the selection bias in measurements and its impact on the measurement of deep fades.

\subsubsection{K-factor}
Caused by the constructive or destructive superposition of MPCs, the channel fluctuates rapidly due to the multi-path effect. To evaluate how strong the multi-path effect is, the K-factor, defined as the ratio between the power of the strongest path and the sum of powers of other paths, is evaluated. To be specific, a larger K-factor indicates that the channel behavior is dominated by a single path (e.g., often, but not always, the LoS path), indicating a weaker multi-path effect.
\begin{figure}
    \centering
    \includegraphics[width=0.8\columnwidth]{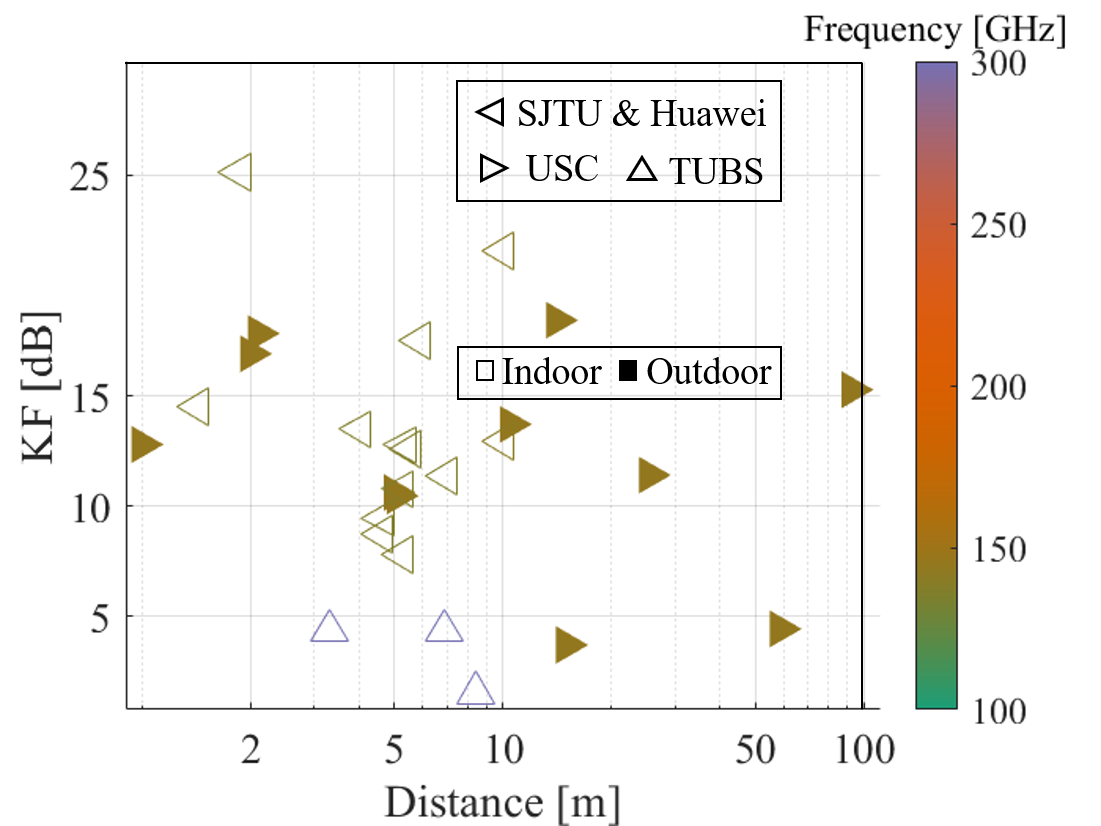}
    \caption{Representative results of K-factor in the THz band. Different types of markers represent different research organizations. Empty markers are results in indoor scenarios and filled markers are those in outdoor scenarios, respectively.}
    \label{fig:kfch}
\end{figure}
\par Representative results of K-factor in the THz band are summarized in Fig.~\ref{fig:kfch}, based on the recent studies in both indoor scenarios~\cite{chen2021channel,chen2021140,priebe2013ultra} and outdoor scenarios~\cite{abbasi2021double,abbasi2021thz}.
Note that all these measurements are conducted either with omni-directional antennas, or directional antennas while scanning over the 3D spatial domain to incorporate all MPCs. The existing measurements investigating K-factor are mostly conducted at 140~GHz and 300~GHz, while extensive measurements are still needed for other, especially higher, carrier frequencies. Particularly for indoor scenarios, it is clear that the K-factor values at 300~GHz are smaller than those at 140~GHz. However, with limited results, it might be insufficient to make any conclusion on the trend of K-factor varying with frequencies. Furthermore, the K-factor values at 300~GHz for outdoor scenarios are slightly larger than those in indoor scenarios, which might be caused by the reflections in enclosed indoor spaces being normally stronger than those in open outdoor environments.
\par Intuitively,  the THz band exhibits larger K-factors compared with lower frequency bands, since the reflection loss and diffraction loss in the THz band are larger than those in lower frequency bands.
According to the specification 3GPP 38.901~\cite{3gpp.38.901}, typical K-factor values in channels below 100~GHz are around 10 dB. The K-factor values in Fig.~\ref{fig:kfch} at 140~GHz are slightly larger, while the results at 300~GHz are close to what has been observed at low frequencies. Therefore, further measurements in the same environment with the same dynamic range are needed to carefully verify the trend of K-factor as the carrier frequency increases.

\subsubsection{Delay Spread and Angular Spread}
Different MPCs may have either, or both, different delay and different AoD and AoA, creating delay and angular dispersion of the channel, respectively.
Due to the delay dispersion, an Rx receives several copies of the same signal with different arrival times, which may result in inter-symbol interference. A compact description of delay dispersion is given by the RMS DS, defined as the second central moment of the power delay profile. While other measures for the delay dispersion, such as the interference quotient (percentage of impulse response energy contained within a certain window) are more relevant for the design of equalizers, the RMS delay spread is the most widely used measure in the literature \cite{molisch2011wireless}. Since antenna patterns weigh the importance of different MPCs, the DS depends on the specific antenna patterns with which the measurements are done. Two quantities are commonly used: (i) the omnidirectional delay spread, which assumes omnidirectional (or even isotropic) antennas at both link ends, and (ii) the directional delay spread, which is the delay spread encountered when directional (e.g., horn) antennas are used at both link ends; typically, the horn orientation pairing that results in the highest receive power is chosen as a basis. Since with directional antennas, some of the MPCs fall outside the antenna main lobe, there are fewer significant multi-path components and consequently usually smaller DS. Obviously, the results depend on the specific beamwidth of the horns used in the measurements. 
\par Besides delay dispersion, the THz channel also exhibits angular dispersion. To be specific, the MPCs depart from the Tx to different directions, interact with objects in the environment, and then arrive at the Rx from different directions.
Several RMS angular spreads describe a link, namely the azimuth spread of departure (ASD), elevation spread of departure (ESD), azimuth spread of arrival (ASA), and elevation spread of arrival (ESA). Different definitions of the RMS angular spread exist. A commonly used one refers to the second central moment of the angular power spectrum, in analogy to the RMS delay spread. However, due to the $2 \pi$ periodicity of the angular power spectrum, this can lead to ambiguity and non-intuitive results (e.g., two MPCs at $15$ and $25$ degrees having a different RMS angular spread from the case of two components at $355$ and $5$ degrees). The definition of \cite{fleury2000first} avoids these problems. 
\begin{figure}
    \centering
    \includegraphics[width=1.0\columnwidth]{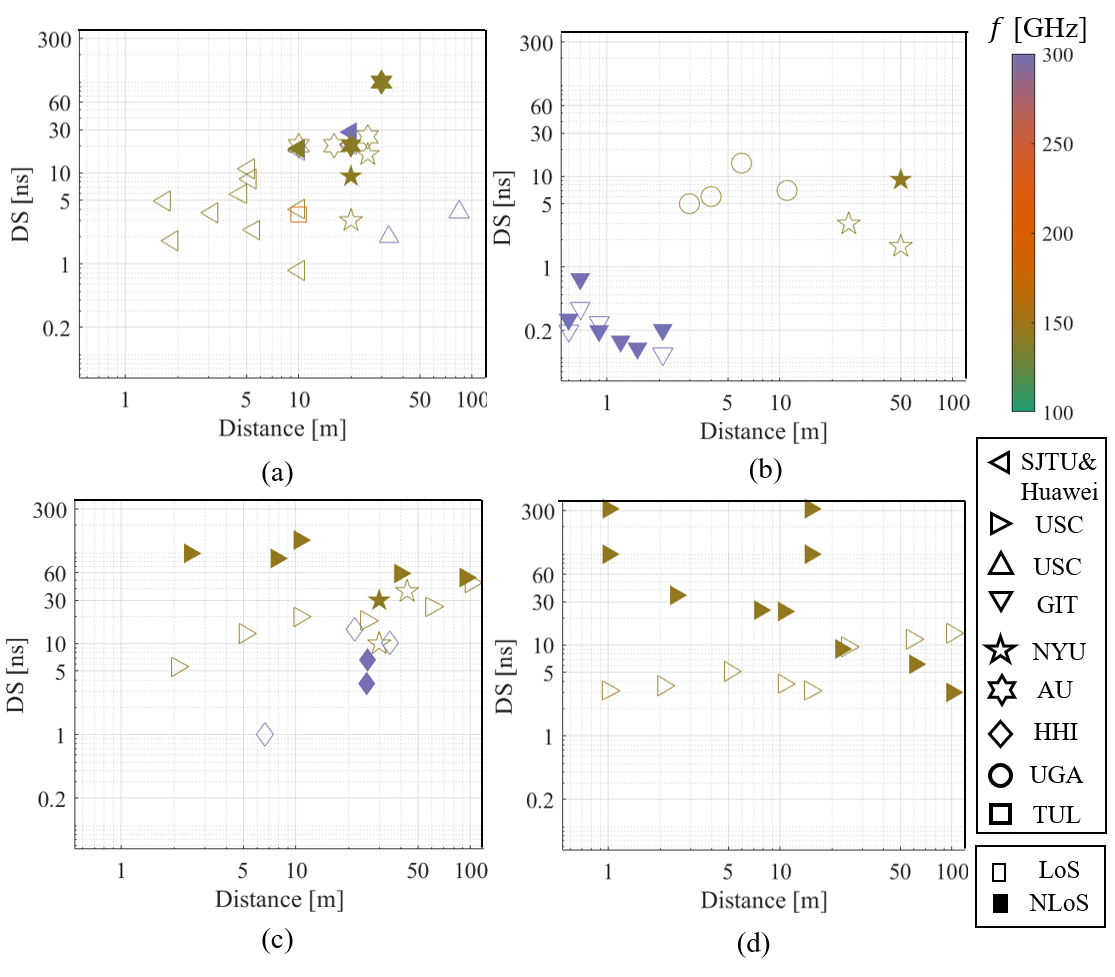}
    \caption{Representative results of DS in the THz band. Different types of markers represent different research organizations. Empty markers are results in the LoS case and filled markers are those in the NLoS case, respectively. (a) Indoor, omnidirectional results; (b) Indoor, directional results; (c) Outdoor, omnidirectional results; (d) Outdoor, directional results.}
    \label{fig:dschar}
\end{figure}
\par Representative values of DS are summarized in Fig.~\ref{fig:dschar}, based on research results in both indoor scenarios~\cite{chen2021140,ju2022sub,priebe2013ultra,dupleich2020characterization,Cheng2020Characterization,xing2021millimeter,cheng2019thz,chen2021channel,pometcu2018large,Nguyen2021Large,Nguyen2018Comparing,li2022channel,wang2022thz} and outdoor scenarios~\cite{ju2021140,undi2021angle,xing2021propagation,abbasi2021thz,abbasi2021double,xing2021millimeter_letter}.
Note that the results in Fig.~\ref{fig:dschar} are separated into four parts, considering whether the measurement campaigns are conducted in indoor or outdoor scenarios and whether the effects of directional antennas are eliminated through scanning over the 3D spatial domain. The omnidirectional results are either directly obtained by using omnidirectional antennas or synthesized using directional antennas while scanning over the 3D spatial domain.
\par Based on the results in Fig.~\ref{fig:dschar}, we draw several observations as follows. Firstly, up to now date, researchers have implemented extensive investigations for DS in the frequency bands at 140~GHz and 300~GHz, while those in 200-300~GHz and above 400~GHz are still open. 
Besides, the typical DS values for outdoor THz communication are profoundly larger than those for indoor communication. This is reasonable since the reflected paths in outdoor scenarios normally have a larger excess delay compared to reflected paths in indoor scenarios due to a much longer propagation distance. Moreover, as the frequency increases, the delay spread values decrease, which is reasonable as fewer multi-path components can be received in higher frequencies due to the larger free space path loss and reflection loss. Furthermore, as the distance increases, delay spread values generally increase, especially for directional results, as the environment becomes more complex for longer propagation distance and more MPCs occur. Last but not least, no significant difference between delay spread values in the LoS case and NLoS case is observed, which may need more measurement campaigns to further investigate.
\par
\begin{figure}
    \centering
    \includegraphics[width=1.0\columnwidth]{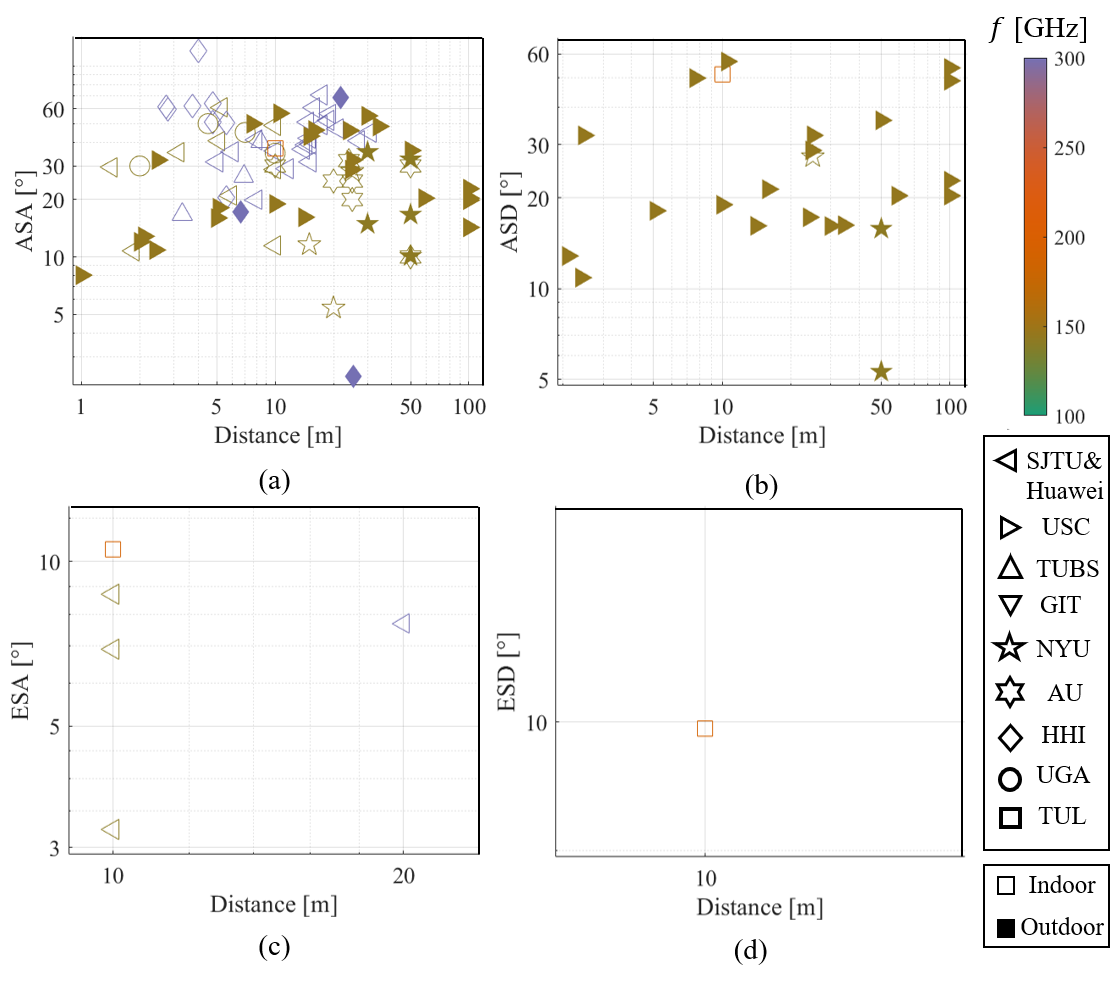}
    \caption{Representative results of AS in the THz band. Different types of markers represent different research organizations. Empty markers are results in the LoS case and filled markers are those in the NLoS case, respectively. (a) Indoor, omnidirectional results; (b) Indoor, directional results; (c) Outdoor, omnidirectional results; (d) Outdoor, directional results.(a) ASA; (b) ASD; (c) ESA; (d) ESD.}  
    \label{fig:asch}
\end{figure}

 As for the angular spread, representative results for ASA, ASD, ESA and ESD are shown in Fig.~\ref{fig:asch}, based on research results reported in indoor scenarios~\cite{chen2021140,ju2022sub,pometcu2018large,chen2021channel,priebe2013ultra,dupleich2020characterization,Nguyen2021Large,li2022channel,wang2022thz} and outdoor scenarios~\cite{abbasi2020double,ju2021140,undi2021angle,abbasi2021ultra,abbasi2021double,abbasi2021thz,xing2021millimeter_letter,ju2021sub}. As different angular spread definitions are used in the literature, those results based on the definition of \cite{fleury2000first} are translated into equivalent degree values. Note that all these results are conducted either with omni-directional antennas(in the case of ray tracing, where directions of the MPCs can be obtained directly from the simulator), or directional antennas while scanning over the 3D spatial domain to incorporate all MPCs. 
Note that the AS values in LoS and NLoS cases are not separately denoted. 
Since heights of the Tx and Rx are comparable in most investigated scenarios, MPCs show similar behaviors in the elevation plane, resulting in smaller elevation spreads than azimuth spreads, so that most measurements focus on angular spreads, while elevation spreads are rarely investigated. 
Moreover, the angular spreads range from several degrees to tens of degrees, which are larger than the beamwidth in the THz band (i.e., typically several degrees). This indicates that some multi-path components may fall outside the antenna beam, so that the use of directional antennas in a communication system further alleviates the multi-path effects. Last but not least, to compare indoor results and outdoor results, the azimuth spreads are very close, while the elevation spreads for indoor cases are observably larger than those for outdoor cases, due to contributions of the reflections from floors and ceilings in indoor scenarios. 

\subsubsection{Cross-polarization Ratio}
Cross-polarization ratio (XPR) is an essential parameter to examine the feasibility of polarization diversity, which is defined as the ratio between the electric field strength of the received co-polarization signal and the received cross-polarization signal, where co-polarization refers to identical polarization and cross-polarization represents orthogonal polarization with respect to the polarization of the transmitted signal. For example, if the transmitted signal is vertically polarized, the received signal with the vertical polarization is the co-polarized signal and that with the horizontal polarization is the cross-polarized one. A large XPR indicates the feasibility of multiplexing with orthogonal polarization without further signal processing to separate the data streams, i.e., two individual links with orthogonal polarization from Tx to Rx can be achieved with little interference. In cases with small XPR, polarization diversity may be utilized to enhance the link performance.
\begin{figure}
    \centering
    \includegraphics[width=0.8\columnwidth]{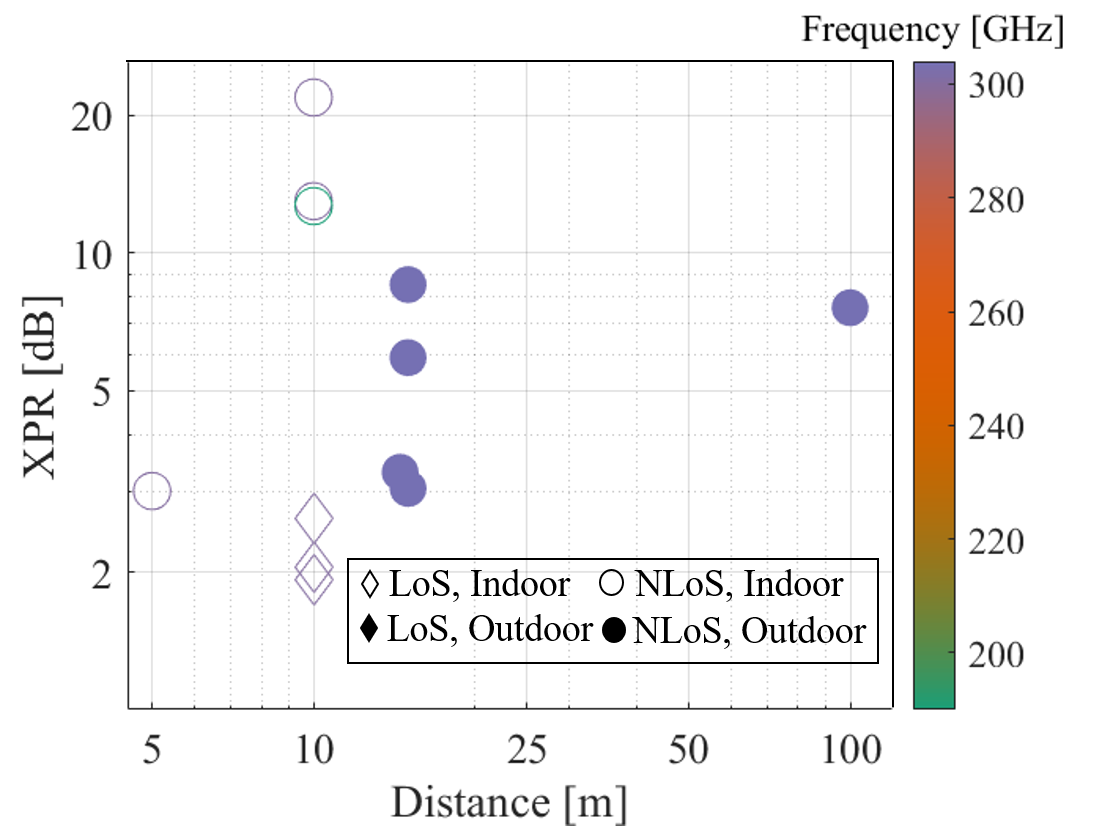}
    \caption{Representative results of XPR in the THz band. Diamond markers and circle markers represent results obtained in the LoS case and NLoS case, respectively. Empty markers are results in indoor scenarios and filled markers are those in outdoor scenarios, respectively. Furthermore, different colors of the markers are for different linear models that are used to obtain the PLE.}
    \label{fig:xprchar}
\end{figure}
\par Due to the hardware limitation, very limited measurement campaigns has conducted to study the cross-polarization ratio in the THz band~\cite{dupleich2020characterization}, which is not enough to analyze the polarization characteristics. As a result, ray tracing simulation results from groups in Beijing Jiatong University (BJTU) are included here to obtain some insights~\cite{guan2019channel_TVT,guan2020channel,guan2021channel,guan2019measurement}. The results of XPR values are depicted in Fig.~\ref{fig:xprchar}. It can be seen that the XPR in the LoS case is larger than those in NLoS cases, which is due to the existence of the purely co-polarized direct path in the LoS case. However, there is no clear trend on how the XPR varies in different scenarios or frequencies. Most existing results are geometry-dependent, while more extensive measurements are needed to provide polarization statistics in the THz band.

\subsubsection{Cross-correlation among  Channel Characteristics}
The aforementioned channel characteristics, including the path loss, shadow fading, K factor, delay spread, angular spread, and XPR, are cross-correlated and hence, exhibit spatial-temporal correlation, which should be taken into consideration, as done in COST 259/273/2100, 3GPP TR. 38.901 and QuaDriGa~\cite{liu2012cost,3gpp.38.901,QuaDriGa} for the microwave and mmWave bands. In the THz band, the cross-correlation between channel characteristics is still under-investigated while only several papers, limited to a subset of the scenarios of interest, have studied this phenomenon~\cite{chen2021channel,yi2019characterization,guan2021channel,guan2019measurement}, mostly using ray-tracing simulations. The value of the cross-correlation varies between -1 and 1, where -1 or 1 represents negatively or positively linearly correlated, while 0 denotes no correlation, respectively. To show the cross correlation in a quantitative way, three levels of cross correlation are noted, namely weakly-correlated (from -0.3 to 0.3), medium-correlated (from 0.3 to 0.6 or from -0.6 to -0.3) and strongly-correlated (larger than 0.6 or smaller than -0.6), as summarized in Table~\ref{tab:cc}.
\begin{table}
    \centering
    \caption{The cross correlation between channel characteristics. The notation ``W'', ``M'', and ``S'' stand for weakly-correlated, medium-correlated and strongly-correlated, respectively.}
    \resizebox{\linewidth}{!}{
    \begin{tabular}{c|c|c|c|c|c|c|c|c}
         &SF&KF&DS&ASA&ASD&ESA&ESD&XPR  \\
         \hline
         SF&1&-&-&-&-&-&-&-\\
         \hline
         KF&W&1&-&-&-&-&-&-\\
         \hline
         DS&W&W&1&-&-&-&-&-\\
         \hline
         ASA&W&+M&+M&1&-&-&-&-\\
         \hline
         ASD&W&-M&+M&+M/+S&1&-&-&-\\
         \hline
         ESA&W&W&W/+M&+M/+S&+M&1&-&-\\
         \hline
         ESD&W&W&W/+M&+M&+M/+S&+M&1&-\\
         \hline
         XPR&W&W&W&W&W&W&W&1\\
         \hline
    \end{tabular}}
    \label{tab:cc}
\end{table}
\par We draw the observations from Table~\ref{tab:cc} as follows. First, the shadow fading and the XPR show a weak correlation to other parameters. This is in contradiction to the well-established results at lower frequencies that shadowing is correlated with the RMS delay spread \cite{greenstein1997new,asplund2006cost,3gpp.38.901}. 
Second, the K factor is weakly correlated with the DS and elevation angle spreads, while being slightly correlated with the azimuth angle spreads. The reason for this interesting phenomenon needs further investigation. Third, DS and AS are positively correlated, since these spreads are both dependent on the power of MPCs. Besides, the azimuth spreads are more correlated with the DS compared to elevation spreads. This can be explained that for terrestrial scenarios in~\cite{guan2019measurement,guan2019channel_TVT,yi2019characterization}, elevation spreads are not significant since the heights of Tx and Rx are close, which results in weaker correlation for the elevation spreads and DS. Fourth, ASA and ASD are strongly positively correlated, which is reasonable since the MPCs with large azimuth of departure would have large azimuth of arrival. However, the correlation between ESA and ESD is weaker, as elevation angles show limited dispersion in terrestrial channels.
Fifth, the correlation between ASA and ESA and the one between ASD and ESD are positively strong, while the correlation between ASA and ESD and that between ESA and ASD is weak. Note that only a qualitative analysis is given in Table~\ref{tab:cc}, while quantitative values for cross correlation are dependent on the communication scenarios and need to be characterized based on extensive measurements. Generally we can state that due to the small number of available measurements, the observed correlations need further investigation. 
\subsubsection{Coherence region}
The coherence region in temporal/frequency/spatial domain can be defined as the temporal/frequency/space interval within which the channel auto-correlation function is larger than a certain threshold. The coherence region is usually calculated using empirical formulas, e.g., the coherence time as a proportion to the reciprocal of the Doppler spread and the coherence bandwidth as a proportion to the reciprocal of the delay spread.
\par In~\cite{chen2019time}, the coherence time is evaluated as a proportion to the reciprocal of the maximum Doppler frequency, where a 0.2~ms value at 110~GHz is obtained. Furthermore, as a proportion to the reciprocal of the delay spread, the coherence bandwidth is calculated in~\cite{Cheng2020Characterization}, ranging from 0.5~GHz to 2~GHz at 300~GHz carrier frequency. Regarding the coherence distance in antenna arrays for UM-MIMO communications, Bian~{\em et al.} have calculated the array coherence distance~\cite{bian2021general}, where the coherence distance is even lower than the distance between two antenna elements, indicating the strong non-stationarity in the array spatial domain. More studies are needed to further investigate the coherence region in the THz band.
\subsubsection{Comparison with lower frequency bands}
\begin{table}[!tbp]
    \centering
    \caption{Comparison of channel characteristics in indoor office scenarios.}
    \includegraphics[width =0.99\linewidth]{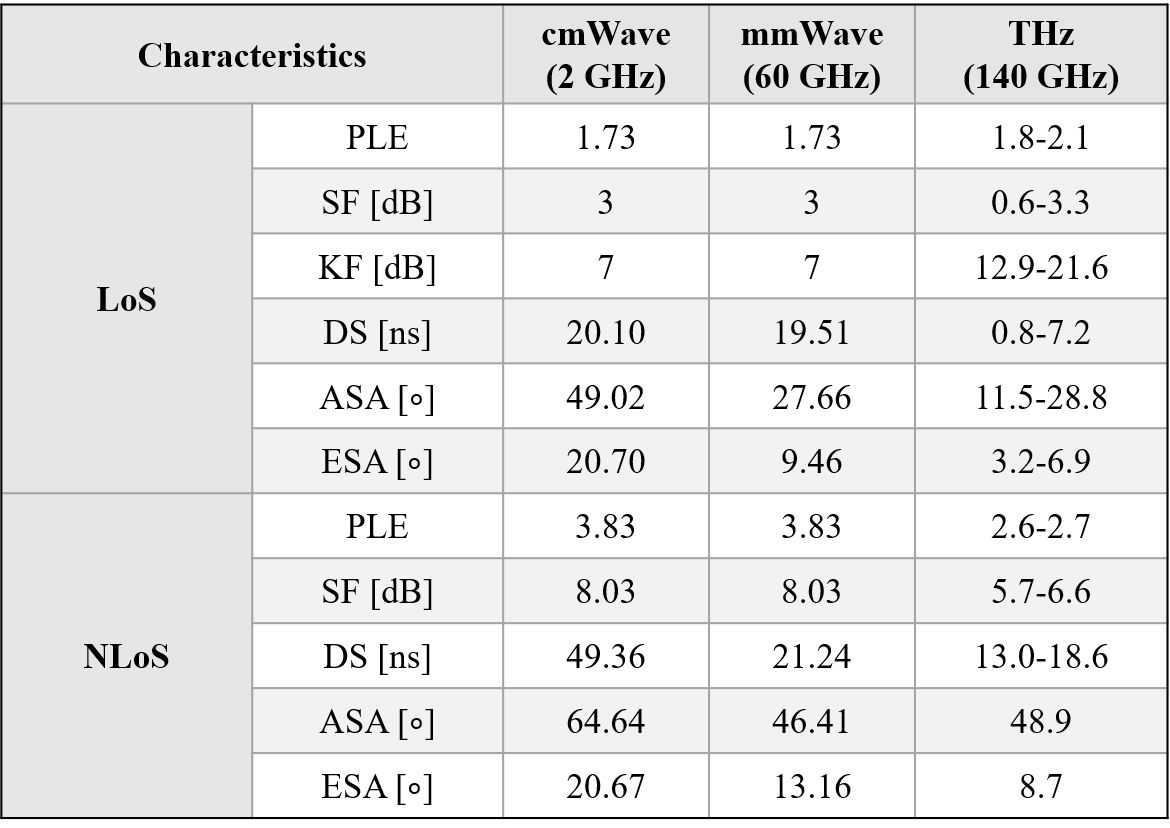}
    \label{tab:com_indoor}
    \vspace{0.5cm}
\end{table}
\par To reveal the uniqueness of the THz band, we compare the aforementioned large-scale and small-scale channel characteristics in the THz band with those in cmWave and mmWave bands in indoor office scenarios. Representative values of channel characteristic in different frequency bands are summarized in Table~\ref{tab:com_indoor}. In particular, the results in the cmWave band (at \SI{2}{GHz}) and mmWave band (at \SI{60}{GHz}) are from the 3GPP standardized channel model~\cite{3gpp.38.901}, while the results in the THz band are based on the measurement campaigns at \SI{140}{GHz} in the same indoor office scenarios~\cite{chen2021140,ju2021millimeter,abbasi2020channel}.
\par From Table~\ref{tab:com_indoor}, we draw the following observations. First, the PLE values in the THz band are closer to 2 (i.e., the PLE value in the free space) compared to those in cmWave and mmWave bands, especially in the LoS case. The reason behind this is that the reflection loss, diffraction loss, penetration loss, and other NLoS losses, are very large in the THz band. Therefore, the strong MPCs including the LoS path and once-reflected path, dominate in the THz band, whose power is mainly affected by the free space path loss with a PLE of 2. Second, the shadow fading values in the THz band are comparative to those in cmWave and mmWave bands. Third, the K-factor values in the THz band appears larger than those in lower frequency bands, since the LoS path is more dominant in the THz band. Fourth, the measured delay spread values in the THz band are smaller than those in lower frequency bands, especially in the LoS case. This is explained that due to the larger reflection loss and diffraction loss, the dominant paths, such as LoS path and once-reflected paths, have more significant power than other MPCs, resulting in smaller delay spreads. Fifth, pertaining to the angular spreads, the ASA and ESA values in the THz band are generally smaller or at least comparative than those in lower frequency bands. Admittedly, the measured results are insufficient to reach final judgement on angular spreads of departure and cross-polarization ratio, which are thus omitted in Table~\ref{tab:com_indoor}.
\par Though the above-mentioned comparison among the three different frequency bands suggests some useful insights, we should clarify that even at the same frequency and in the same campaign, variation of Tx/Rx positions setup of channel sounders, as well as methods of post-data processing would influence the channel results. We discovery that our results are aligned with the ones in multi-band measurement campaigns~\cite{xing2021millimeter,xing2021millimeter_letter,Nguyen2018Comparing}, which provide similar discussions and observations about the unique propagation characteristics in the THz band. Therefore, to summarize the trends, the increase of frequency incurs the decrease of multipath effects, resulting in PLEs closer to 2, smaller delay and angular spread values, etc. Moreover, the larger reflection and diffraction losses in the THz band cause NLoS paths weaker, whereas the LoS path, if existed, is more dominant with a larger K-factor.
\subsection{Channel Characteristics in Chip-scale and Nano-scale Networks}
\par THz communications are not only promising for the aforementioned indoor and outdoor scenarios, but also exhibiting great potential to be utilized in much smaller environments, such as chip-scale and nano-scale communications. Molecular absorption that might impede communications at certain frequencies over larger distances has less impact, so that ultra-short-range communications are easier to achieve, and the advantages of THz band including ultra-large bandwidth and abundant spectrum resources can enable the realization of ultra-high data rates. Channel characteristics in such micro- and nano-scale scenarios are different from that in macro-scale scenarios like indoor and outdoor environments. In this subsection, reported results from the literature for chip-scale and nano-scale channel characterization are elaborated.
\subsubsection{Channel Characteristics in Chip-scale Channels}
Existing chip-scale communications rely on metal wires or printed circuits to exchange data between chips or components on chips. However, the wired structure exhibits several problems, such as cable management problem, large difficulty for system design, high power consumption and high latency for on-chip links, etc~\cite{Kim2016Charaterization,chen2019channel}.
To address such problems, a wireless alternative is envisioned as a promising solution, either using optic or electromagnetic wave. Optical solutions may suffer from the drawback that different manufacturing processes and materials are required for realizing optical signal generation and reception. In contrast, the THz band shows good promise to support low-cost, high data rate connections for the inter-chip and intra-chip data exchange. To realize THz chip-scale communications, it is essential to understand the channel first. As the THz chip-scale communication is still a new topic, limited studies focus on the channel characterization in chip-scale THz channels, mainly involving path loss and multi-path effect.
\paragraph{Path Loss}
Due to the much short distance for chip-scale networks, the path loss is much different from that in macro-scale scenarios and depends on specific environments. Several studies have revealed the path loss characteristics in chip-scale environments. Firstly, for the chip-to-chip channel inside a desktop-size metal enclosure, \textit{Fu et al.} have proposed a path loss model consisting of four terms~\cite{fu2020modeling}, the mean path loss of traveling wave that depends on the frequency and distance, the received power variation contributed by resonating modes that is related to the height of the Rx, the antenna misalignment loss that is influenced by the departure direction and arrival direction, and a random shadow fading term. The comparison with measurement results verifies the effectiveness of the path loss model. Secondly, for wireless networks-on-chip (WiNoC) scenario, \textit{Yi~et al.} have found that the path loss exponents are 0.81, 1.04 and 0.84 at 100, 300 and 800~GHz respectively. Interestingly, the path loss is found to have oscillating and periodic behaviors along the frequency band with a period of 124~GHz, mainly due to the existence of surface waves and guided waves.
\paragraph{Multi-path Effect}
Since the distances among Tx, Rx as well as the scatterers for inter-chip communications are very short, the influence of multi-path effects need to be carefully analyzed. Based on measurement campaigns for chip-to-chip communications on a computer motherboard, \textit{Kim~et al.} find that the ground plane or parallel-plate structures on a computer motherboard introduce multipaths. Besides, the backside of the dual in-line memory modules (DIMM) can serve as an excellent reflecting surface, which enables the transmission for NLoS cases. In general, the antennas need to be carefully positioned with respect to the motherboard layout to achieve optimal communications. Besides, \textit {Fricke et al.} have derived a model for the specular reflection behavior of printed circuit boards in the Terahertz range applicable for chip-to-chip communication~\cite{fricke2018}. The model has been calibrated based on VNA measurements with a box using simulated annealing.
\par In general, chip-scale THz channels have not been fully investigated. Channel characteristics like delay spread and angular spread in chip-scale THz channels still remain unexplored, which is important for the system design. Besides, existing studies are dedicated to particular environments such as computer motherboards or metal enclosures, while extensive measurements and investigations of other practical scenarios are still needed.

\subsubsection{Channel Characteristics in Nano-scale Channels}
\par Nano-scale communication is known as wireless transmission between nano-sized devices, which is interesting to be explored in health monitoring, drug delivery, industry monitoring, etc.~\cite{akyildiz2010internet,yang2020comprehensive}. To enable the nano-devices with new functionality, new nano-materials, nano-particles as well as the new propagation environments, such as the human body for healthcare applications, are creating new characteristics for the propagation of THz wave. 
\paragraph{Path loss}
The path loss in the THz band for nano-scale channels consists of three frequency-dependent terms, including the spreading (free-space) loss, the scattering loss and the molecular absorption loss~\cite{elayan2017terahertz}. The spreading loss is due to the spherical propagation of the electromagnetic wave, which can be directly characterized by the Friss's law. Moreover, the scattering loss is attributed to the scattering by particles as well as cells and organelles inside the human body. The scattering caused by particles can be modeled as Rayleigh scattering, since the particle diameters are smaller than the wavelength of the THz wave. Besides, for in-vivo scenarios, the scattering by cells can be studied by applying the van de Hulst approximation. However, the numerical results in~\cite{elayan2018characterising} have revealed that the scattering loss is much smaller compared to molecular absorption loss and spreading loss, which makes little impact on the THz wave propagation and hence, can be neglected. 
\par Besides the spreading loss and scattering loss, the molecular absorption loss occurs due to the molecular vibration excited by the propagating THz wave, which can be characterized by the Beer-Lambert law~\cite{fuwa1963physical}. The molecular absorption loss is related to the composition of the medium and the dielectric properties. For communication scenarios that only involve molecular absorption in atmosphere, the HITRAN or ITU-R P.676 can be used, while for in-vivo applications, the characterization of the skin, cells, organelles and other composites are needed, in particular their refractive index, complex permittivity and absorption coefficient. Representative results for the electromagnetic properties of human skin can be found in~\cite{chopra2016fibroblasts,chopra2016thz}, e.g., refractive index and absorption coefficient are around 2 and 18.45 $\text{cm}^{-1}$, respectively.
\par Apart from the theoretical analysis mentioned above, an empirical path loss model for nano-communications inside the human skin is proposed in~\cite{abbasi2016terahertz} by curve-fitting the measurement results. The proposed model is dependent on frequency, distance and number of sweat ducts. Comparison with measurements has validated the effectiveness of this model within the considered scenario, while further studies are needed to verify its generality.
\paragraph{Noise Characteristics}
\par Unlike macro-scale scenarios, where the white Gaussian thermal noise dominates, the noise in nano-communications mainly consists of background noise and molecular absorption noise. On one hand, the background noise, also known as the black-body noise, is created by the temperature of the absorbing medium, which is independent of the transmitted signals. On the other hand, as investigated by \textit{Jornet et al.}~\cite{jornet2011channel}, the vibration of the molecules excited by the propagating electromagnetic waves would re-emit electromagnetic radiation at the vibration frequency, which forms the molecular absorption noise at the Rx. According to~\cite{papanikolaou2017channel}, the molecular absorption noise contributes most of the noise in THz nano-communication channels, which is thoroughly investigated in~\cite{jornet2011channel}, where a key parameter of emissivity is defined and calculated. Besides these two main noise terms, for intra-body communications, the Doppler-shift-induced noise due to cells' random and irregular velocity is also studied in~\cite{elayan2018end}, where the Doppler-shift-induced noise is found to be larger than the background noise and matters most for intra-body THz channels.
\par In general, existing studies on channel characteristics for nano-scale THz channels mainly rely on analytical methods, while experimental measurements should be conducted to validate the theoretical results.

\subsection{Non-stationary Properties}
\par Two general assumptions are normally made for channel models: i) the channel is wide-sense stationary uncorrelated scattering (WSSUS)~\cite{matz2005non}; ii) the transmission is from point to point, i.e., the antenna size is not large enough to have an impact on the channel. These assumptions underpin stationarity in temporal, frequency and spatial domains.
\par However, these assumptions may not be valid under particular scenarios. 
The first assumption may fail due to two reasons. On  one hand, the variation of the environment in time-varying channels can result in temporal non-stationarity. On the other hand, the large bandwidth used in the THz band brings a high delay resolution, where MPCs from the same objects, e.g., MPCs from the same wall, can be distinguished. Thus, the uncorrelated scattering assumption is violated, which further causes frequency non-stationarity. Furthermore, as UM-MIMO with thousands of antenna elements may be used in the THz band, the array size can be much larger relative to the wavelength, which may further invalidate the second assumption. Therefore, the variations of the power, delays, and angles of MPCs over different antenna elements result in the non-stationarity in the array spatial domain.
\par All these nonstationarities have been extensively studied at lower frequencies, and both generic modeling methods, and parameterizations have been developed, e.g.,~\cite{3gpp.38.901,liu2012cost,zwick2002stochastic,gao2013massive}, while the nonstationarities in the THz band is still under-explored. The nonstationarities can be evaluated by the stationarity region, defined as the temporal/frequency/spatial interval within which the channel can be assumed to be stationary. 
Due to the lack of efficient measurement equipment in non-stationary scenarios in the THz band, such as in time-varying channels and UM-MIMO channels, the evaluation of stationarity region in temporal and spatial domains remains as an open problem, while several studies investigated the stationarity bandwidth in the frequency domain. In~\cite{wang2020novel,wang2021general}, the stationarity bandwidth was investigated based on a theoretical model and found to linearly increase with the increases of the frequency, as 16~GHz at 300~GHz, 17.5~GHz at 325~GHz and 19~GHz at 350~GHz, respectively. These results indicate that the THz band exhibits large stationarity bandwidth, which is beneficial for the system design. However, further studies are needed to verify this conclusion. In general, more measurement campaigns are needed to further explore the stationarity region in the THz band.

%% file: v1/openproblems.tex
\section{Open Problems and Future Directions} \label{sec:openproblem}
Although there are increasing research studies on the THz channel these years, still many open problems remain, which motivate future research efforts to proceed along the directions including: i) High-performance THz channel measurement systems, ii) Extensive THz channel measurement campaigns, iii) Efficient 3D THz RT simulators, iv) Complete, accurate, and flexible THz channel models, including effects of UM-MIMO, and temporal-frequency-spatial non-stationary channels, v) intelligent surface in the THz band, vi) AI-powered THz channel analysis, ix) Standardization of THz channel models. These aspects will be dealt with in the following sub-sections.

\subsection{High-performance THz Channel Measurement Systems}
The unique propagation characteristics of the THz wave, the narrow beam, and the different application scenarios result in the distinct requirements of the THz channel measurement systems, compared with those for lower frequency bands. Specifically, we envision the requirements include: (i) high measurement frequency going beyond 1~THz and up to 10~THz, (ii) measurement bandwidth as large as tens of GHz, (iii) high dynamic range and sensitivity to detect multipaths due to large path loss, (iv) high measurement speed due to short coherence time (e.g., less than 0.6~ms) of the THz waves, (v) flexible transceiver distance of the measuring system according to different THz communication scenarios, ranging from centimeters to kilometers.
Currently, most channel measurement systems in the THz band are based on VNAs, which is evolved from mmWave channel measurement devices, with additional up- and down- converters to penetrate into the THz spectrum. The method of RFoF-extension is applied to overcome the limited transceiver distance~\cite{abbasi2020double,Nguyen2018Comparing, Nguyen2016Dual}. Besides, algorithms of noise floor estimation and cut-off margin selection are also under investigation for noise removal in the post-data processing~\cite{agarwal2017novel, nikonowicz2018noise, dupleich2021practical}. The other two sounding methods are based on sliding correlation and THz-TDS, which features, compared with VNA-based sounders, high measuring speed and large measuring bandwidth, respectively. However, a comprehensive high-performance THz channel measurement system is still subject to further research. In particular, phased arrays enabling fast beam sweeping for real-time directionally resolved measurements will need to be obtained and incorporated to achieve the above-mentioned goals.

\subsection{Extensive THz Channel Measurement Campaigns and Simulators}
For the variety of THz communication applications discussed in Sec.~\ref{sec:intro}, specific channel measurements are needed for their characteristic scenarios such as UAVs, ships, and vehicles, as well as additional measurements in the well-known scenarios of indoor, outdoor hotspots, inter-chip, etc., and cover the entire THz spectrum over 0.1-1~THz~\cite{tataria20216g,wang20206G}. 
The more difficult measurement setups, more time-consuming measurements, and high expense of the equipment will require collaborations between many institutions to establish a comprehensive measurement program, as is done, e.g., in the NextG channel alliance~\cite{NextG}.

RT simulators are built to efficiently reproduce channel characteristics as a supplement to channel measurement campaigns, and provide more, and thus statistically more significant results than can be obtained with measurements. Hence, accurate, stable, and efficient simulators are fundamental for subsequent research works concerned with THz channels. Before RT techniques can be widely applied in THz communications, RT techniques need to be fully justified and validated~\cite{han2018propagation}, which requires extensive measurements for characterizing and parameterizing the EM properties of materials covering the THz spectrum. Also, the support for intelligent reflecting surfaces is the new feature in the THz band to be incorporated in RT simulators.

\subsection{Evolution of Ray-Tracing-Statistical Hybrid Channel Modeling}
As discussed in Section~\ref{sec:modeling}, a
ray-tracing-statistical hybrid  channel model can achieve  balanced accuracy and  low complexity  simultaneously, as illustrated in Fig.~\ref{fig:axis_channel_modeling_methods}. For the deterministic part, propagation environments should be chosen according to specific communication scenarios. Among existing THz channel models, an accurate EM wave model for NLoS, particularly studying diffuse scattering, is still missing.
Additionally, while parameters in statistic models can be directly obtained from measurement results through channel estimation and approximation, for deterministic-stochastic hybrid modeling methods,  the extraction of model parameters from measurement is still an open issue. The reason for this is, that the process is more complex and requires a large amount of measurement campaigns~\cite{chen2021channel,liu2012cost}.

\subsection{Channel Modeling for THz UM-MIMO and Intelligent Reflective Surfaces Systems}
As discussed previously, UM-MIMO is realistic in THz because electrically large arrays still have reasonably small physical dimensions. Furthermore, large arrays also can be used for intelligent reflective surfaces (IRSs). An IRS~\cite{Ning2021Terahertz,liaskos2020end,nie2020beamforming,Peng2019} is an artificial surface consisting of metallic patches, which control the propagation of the radio waves impinging upon them by electronically changing the EM properties of the patches.
IRS shows potential in realizing pervasive networks by re-engineering and recycling EM waves, which is regarded as a promising technique to solve the LoS blockage and weak scattering problems in THz communications. 

However, by including very large antenna arrays at the transceivers for UM-MIMO and in the environment for IRS, challenges in channel modeling and characterization arise.

\subsubsection{Modeling of Mutual Coupling Effect}
Mutual coupling is crucial in modeling the very large array and analyzing the impact of array configurations of UM-MIMO and IRS performance. The analysis of the mutual coupling effect still needs to be tailored for different antenna technologies and THz transceivers with different materials. Moreover, the mutual coupling is additionally dependent on antenna space and frequency in the THz band. Remarkably, for a very large array, efficient coupling modeling among proximal antennas rather than all antennas is suggested~\cite{han2018propagation}. Hence, an efficient model of mutual coupling for the very large array in the THz band is required.

\subsubsection{Near-field Effects}
The Rayleigh distance of a large antenna array is increased for very large arrays in the THz band. When the distance between the Tx and the Rx is smaller than the Rayleigh distance, the near-field effect arises, which can be interpreted either as that the DoD and DoA of rays cannot be assumed constant~\cite{payami2012channel}, or that the waves should be modeled as spherical (instead of planar) waves.
Future work should include experimental characterization of near-field effects, since most investigations to date are based on modeling with simplifying assumptions. 

\subsection{Temporal-frequency-spatial Non-stationary Channel Properties}
The THz channels exhibit non-stationarity in temporal-frequency-spatial domain in certain scenarios:
\subsubsection{Temporal non-stationarity}
The temporal birth and death (BD) process of MPCs due to the variation of the visible state of scatterers, is essential for channel modeling and channel simulation. Whether the existing models capturing the temporal BD process proposed for microwave and mmWave bands, e.g., those in \cite{zwick2002stochastic,molisch2006cost259,3gpp.38.901}, are effective for THz time-varying channel needs further validation. Thorough measurement campaigns or simulations in THz time-varying channels, such as vehicular channels, railway channels, UAV channels, etc., should be conducted. Besides, the stationarity time and the coherence time need to be estimated to guide the system design in THz time-varying channels.

\subsubsection{Frequency Non-stationarity}
As ultra-wideband communication is promising in THz communications to enhance the channel capacity, the non-stationarity in the frequency domain led by the large bandwidth needs to be considered. The stationarity bandwidth, as well as the coherence bandwidth, are two key parameters to decide whether the channel is frequency-selective and how the physical layer mechanisms should be designed. To address this, ultra-wideband channel measurement campaigns in THz band needs to be further carried out in scenarios that require high data rate like indoor communications. 
\subsubsection{Spatial Non-stationarity}
As mentioned before, the utilization of UM-MIMO brings spatial non-stationarity.
Measurement campaigns with UM-MIMO are needed to investigate the spatial birth and death process of MPCs, where effective models are still missing to capture this. Besides, representative values for array   stationarity distance and coherence distance in typical scenarios such as indoor office room, outdoor urban scenario, etc., are necessary for utilization and design of UM-MIMO communication system in THz band.

\subsection{AI-powered THz Channel Analysis}
Artificial intelligence (AI) is capable of handling complex problems without explicit programming and designs. Recently, AI has drawn significant attention in the area of wireless communications, including antenna design, radio propagation study, multiple access, signal processing, resource allocation, etc.~\cite{wang20206G,zhang2020channel,aldossari2019machine,kato2020ten}. Pertaining to channel modeling, clustering algorithms enabled by unsupervised machine learning have been widely used in the analysis of channel parameters~\cite{he2018clustering}. Furthermore, path loss prediction powered by deep neural networks is reported to outperform the conventional Euclidean-distance-based empirical path loss models~\cite{ostlin2010macrocell}. Recent work has established the possibility of using AI for path loss prediction in the mmWave range based on geographical databases only~\cite{ratnam2020fadenet}. Coming to the THz band, unique propagation features bring many challenges to THz channel modeling, which require robust intelligent algorithms and models. Specifically, more efforts are expected to explore AI techniques in clustering multi-path components, extracting and classifying channel parameters, developing novel learning-based channel models, and others.

\subsection{Standardization of THz Channel Models}
As early as 2008, standardization of the future wireless communications systems in the THz band was initiated by ``Terahertz Interest Group (IGthz)''~\cite{igthz} in IEEE 802.15 (Wireless Specialty Networks - WSN).
Later in 2013, IEEE 802.15  Task Group 3d 100~Gbit/s Wireless (TG~3d (100G))\footnote{IEEE 802.15 WSN Task Group 3d 100~Gbit/s Wireless (TG~3d (100G)): http://www.ieee802.org/15/pub/SG100G.html} was established  in order to develop  the first standard for wireless 300 GHz wireless  communication, which was released in 2017 as IEEE Std. 802.15.3d-2017~\cite{ieee802153d} for 100~Gbps wireless communication operating in the frequency range 252-321~GHz~\cite{tekbiyik2019terahertz, petrov2020IEEE}. 
Among the released documents during the development of the standard, the Channel Modeling Document (CMD)~\cite{ieee802153d_CMD} summarizes channel propagation characteristics and proposes application-based channel models for target scenarios, including close proximity peer-to-peer communications, intra-device communications, wireless backhaul/fronthaul, and data center network.
CMD is used as a reference when submitting further technical contributions to the Task Group~\cite{petrov2016terahertz}.
However, the document provides limited channel models in specific scenarios, centered around 300~GHz only. To accelerate the standardization of the THz band, further application-specific studies of THz channels over the full THz band are still greatly demanded.

\par By reviewing the timeline of 5G standardization process, the channel model standardization for 6G possibly begins in 2023-2025, expanding the supporting frequency to the THz band. There are various expectations on THz channel models in the future 6G standardization. First, new application scenarios for THz communications are to be specified. Currently, 5G channel standardization is limited to the hot-spot scenarios, including macro and micro cells, and indoor office rooms. Possible new scenarios in 6G channel standardization are inter- and intra-device communication, on-desk communication, data center, vehicular communications, UAV, and space-ground communications, among others~\cite{akyildiz2014terahertz}. Second, the channel models will have to be able to describe channel interactions with new system components, such as IRS or massive and even ultra-massive MIMO, and also account for temporal-frequency-spatial non-stationarity. Third, new channel modeling methods, such as map-based hybrid channel models, which have already been adopted as an alternative methodology in 5G channel standardization, might be further expanded. Several research consortia, such as the NextG channel modeling alliance~\cite{NextG} are also working on THz channel models. 

%% file: v1/conclusion.tex
\section{Conclusion}
In this article, we presented a comprehensive overview and analysis of studies of THz wireless channels, including channel measurement, channel modeling and channel characterization, and discuss open problems for future research directions concerned with THz wireless channels. To be specific, three THz channel measurement methodologies, namely, frequency-domain channel measurement based on VNA, time-domain channel measurement based on correlation, and time-domain channel measurement based on THz pulses from THz-TDS, were introduced and compared. Since none of the techniques can meet all the requirements of a high-performance THz channel measurement system, more research on reliable THz channel measurement techniques is still required. At the same time, measurement campaigns are carried out, most of which apply the VNA-based method and focus on the sub-THz band under 300~GHz. Furthermore, channel simulators have been developed by research groups, which is also summarized in this article. In terms of channel modeling, deterministic, stochastic and hybrid channel modeling methodologies were introduced and categorized. Remaining open problems and potential research directions in channel modeling include hybrid channel modeling methods, and modeling of UM-MIMO channels. After that, we gave a review of THz channel properties, channel statistics and non-stationary properties, as obtained from measurements. Due to the lack of channel measurement campaigns over unexplored bands and scenarios, the corresponding characterization of these channels still has significant gaps. Further studies are also needed for preparing THz band standardization.

%% file: v1/1.tex
%Chong Han
(M'16) received Ph.D. degree in Electrical and Computer Engineering from Georgia Institute of Technology, USA in 2016. He is currently an Associate Professor with University of Michigan-Shanghai Jiao Tong University (UM- SJTU) Joint Institute, Shanghai Jiao Tong University, China, and founded the Terahertz Wireless Communications (TWC) Laboratory. Since 2021, he is also affiliated with Department of Electronic Engineering, Shanghai Jiao Tong University. He is the recipient of 2018 Elsevier NanoComNet (Nano Communication Network Journal) Young Investigator Award, 2017 Shanghai Sailing Program 2017, and 2018 Shanghai ChenGuang Program. He is a guest editor with IEEE Journal on Selected Topics in Signal Processing (JSTSP) and IEEE Transactions on Nanotechnology, an editor with IEEE Open Journal of Vehicular Technology since 2020, IEEE Access since 2017, Elsevier Nano Communication Network journal since 2016. He is a TPC chair to organize multiple IEEE and ACM conferences and workshops. He is a co-founder and vice-chair of IEEE ComSoc Special Interest Group (SIG) on Terahertz Communications, since 2021. His research interests include Terahertz communications, and intelligent sustainable power systems. He is a member of the IEEE and ACM.

%% file: v1/2.tex
%Yiqin Wang
received the B.E. degree in Electrical and Computer Engineering from the University of Michigan-Shanghai Jiao Tong University Joint Institute, Shanghai Jiao Tong University, Shanghai, China in 2020. She is currently working toward the Ph.D. degree with Terahertz Wireless Communication Laboratory, Shanghai Jiao Tong University, Shanghai, China. Her research interests include channel measurement and modeling in the Terahertz band.

%% file: v1/3.tex
%Yuanbo Li
received the B.E. degree in Communication Engineering from Harbin Institute of Technology, Harbin, China, in 2020. Since 2020, he has been working toward the Ph.D. degree with Terahertz Wireless Communication Laboratory, Shanghai Jiao Tong University, China. His research interests include Terahertz band channel modeling and time-varying channel modeling.

%% file: v1/4.tex
%Yi Chen
(S'18) received the Ph.D degree from the University of Michigan-Shanghai Jiao Tong University Joint Institute, Shanghai Jiao Tong University, Shanghai, China in 2022. His research interests include Terahertz band networks, wireless network-on-chip, channel measurements, and channel modeling.

%% file: v1/5.tex
%Naveed A. Abbasi
(M'18) received his bachelor's degree in electrical engineering from Air University, Islamabad, Pakistan in 2007. He then received his master's degree from Northwestern Polytechnical University, Xi'an, P.R. China in 2010 and received his Ph.D. degree from Koc University, Istanbul, Turkey in 2018.
He is currently a postdoctoral fellow at the University of Southern California and his current research interests include channel sounder design, measurements and modeling for various wireless communication bands including THz, mm-wave and Wi-fi as well as the applications of machine learning towards wireless communication. He is also interested in molecular and nano-scale communication.

%% file: v1/6.tex
%Thomas K{\"u}rner
(Fellow IEEE) received his Dipl.-Ing. degree in Electrical Engineering in 1990, and his Dr.-Ing. degree in 1993, both from University of Karlsruhe (Germany). From 1990 to 1994 he was with the Institut f{\"u}r H{\"o}chstfrequenztechnik und Elektronik (IHE) at the University of Karlsruhe working on wave propagation modelling, radio channel characterisation and radio network planning. From 1994 to 2003, he was with the radio network planning department at the headquarters of the GSM 1800 and UMTS operator E-Plus Mobilfunk GmbH \& Co KG, D{\"u}sseldorf, where he was team manager radio network planning support responsible for radio network planning tools, algorithms, processes and parameters from 1999 to 2003. Since 2003 he is Full University Professor for Mobile Radio Systems at the Technische Universit{\"a}t Braunschweig. In 2012 he was a guest lecturer at Dublin City University within the Telecommunications Graduate Initiative in Ireland. Currently he is chairing the IEEE 802.15 Standing Committee THz. He was also the chair of IEEE 802.15.3d TG 100G, which developed the worldwide first wireless communications standard operating at 300 GHz. He is also the project coordinator of the H2020-EU-Japan project ThoR (``TeraHertz end-to-end wireless systems supporting ultra-high data Rate applications'') and Coordinator of the German DFG-Research Unit FOR 2863 Meteracom (``Metrology for THz Communications''). In 2019 he received the Neal-Shephard Award of the IEEE Vehicular Technology Society (VTS). From 2016 to 2021 he was a member of the Board of Directors of the European Association on Antennas and Propagation (EurAAP) and since 2020 he is a Distinguished Lecturer of IEEE Vehicular Technology Society.

%% file: v1/7.tex
%Andy Molisch
received his degrees (Dipl.Ing. 1990, PhD 1994, Habilitation 1999) from the Technical University Vienna, Austria. He spent the next 10 years in industry, at FTW, AT\&T (Bell) Laboratories, and Mitsubishi Electric Research Labs (where he rose to Chief Wireless Standards Architect). In 2009 he joined the University of Southern California (USC) in Los Angeles, CA, as Professor, and founded the Wireless Devices and Systems (WiDeS) group. In 2017, he was appointed to the Solomon Golomb - Andrew and Erna Viterbi Chair. 
His research interests revolve around wireless propagation channels, wireless systems design, and their interaction. Recently, his main interests have been wireless channel measurement and modeling for 5G and beyond 5G systems, joint communication-caching-computation, hybrid beamforming, UWB/TOA based localization, and novel modulation/multiple access methods. Overall, he has published 5 books (among them the textbook ``Wireless Communications'', third edition in 2022), 21 book chapters, 280 journal papers, and 370 conference papers. He is also the inventor of 70 granted (and more than 10 pending) patents, and co-author of some 70 standards contributions. His work has been cited more than 59,000 times, his h-index is >100, and he is a Clarivate Highly Cited Researcher. 
Dr. Molisch has been an Editor of a number of journals and special issues, General Chair, Technical Program Committee Chair, or Symposium Chair of multiple international conferences, as well as Chairperson of various international standardization groups. He is a Fellow of the National Academy of Inventors, Fellow of the AAAS, Fellow of the IEEE, Fellow of the IET, an IEEE Distinguished Lecturer, and a member of the Austrian Academy of Sciences. He has received numerous awards, among them the IET Achievement Medal, the Technical Achievement Awards of IEEE Vehicular Technology Society (Evans Avant-Garde Award) and the IEEE Communications Society (Edwin Howard Armstrong Award), and the Technical Field Award of the IEEE for Communications, the Eric Sumner Award.